\documentclass[preprint,12pt]{elsarticle}

\usepackage{amsmath,amssymb}

\def\sub#1{_{\mathrm{#1}}}
\def\up#1{^{\mathrm{#1}}}
\def\Vec#1{\boldsymbol #1}
\def\dps{\displaystyle}
\def\arrayret{\vspace{5pt} \\}

\journal{Physics Reports}

\begin{document}

\begin{frontmatter}



\title{Quantum hydrodynamics}


\author[OCU]{Makoto Tsubota}
\author[UT]{Michikazu Kobayashi}
\author[HU]{Hiromitsu Takeuchi}

\address[OCU]{Department of Physics, Osaka City University, Sugimoto 3-3-138, Sumiyoshi-ku, Osaka 558-8585, Japan}
\address[UT]{Department of Basic Sciences, University of Tokyo, Komaba 3-8-1, Meguro-ku, Tokyo 153-8902, Japan}
\address[HU]{Graduate School of Integrated Arts and Sciences, Hiroshima University, Kagamiyama 1-7-1, Higashi-Hiroshima 739-8521, Japan}

\bibliographystyle{model1a-num-names}

\begin{abstract}
Quantum hydrodynamics in superfluid helium and atomic Bose--Einstein condensates (BECs) has been recently one of the most important topics in low temperature physics. 
 In these systems, a macroscopic wave function (order parameter) appears because of Bose--Einstein condensation, which creates quantized vortices.  
 Turbulence consisting of quantized vortices is called quantum turbulence (QT). 
 The study of quantized vortices and QT has increased in intensity for two reasons.  
The first is that recent studies of QT are considerably advanced over older studies, which were chiefly limited to thermal counterflow in $^4$He, which has no analogue with classical traditional turbulence,  
whereas new studies on QT are focused on a comparison between QT and classical turbulence.
The second reason is the realization of atomic BECs in 1995, 
for which modern optical techniques enable the direct control and visualization of the condensate and can even change the interaction; such direct control is impossible in other quantum condensates like superfluid helium and superconductors. 
Our group has made many important theoretical and numerical contributions to the field of quantum hydrodynamics of both superfluid helium and atomic BECs. 
In this article, we review some of the important topics in detail. 
The topics of quantum hydrodynamics are diverse, so we have not attempted to cover all these topics in this article. 
We also ensure that the scope of this article does not overlap with our recent review article (arXiv:1004.5458), "Quantized vortices in superfluid helium and atomic Bose--Einstein condensates", and other review articles. 
\end{abstract}


\end{frontmatter}


\section{Introduction}
\label{sec:introduction}

Quantum hydrodynamics (QHD) refers to the hydrodynamics of quantum condensed fluids,
such as superfluid helium and atomic Bose--Einstein condensates (BECs),
which are subject to quantum restrictions.
As a result of Bose--Einstein condensation, a system exhibits a macroscopic wave function
$\Psi(\Vec{r},t)=|\Psi(\Vec{r},t)| e^{i \phi(\Vec{r},t)}$ as an order parameter. 
The superfluid velocity field of a quantum condensed fluid is given by $\mathbf{v}_s=(\hbar/M) \nabla \phi$, with boson mass $M$,
representing the potential flow. 
Since the macroscopic wave function should be single-valued for the space coordinate $\Vec{r}$, the circulation $\oint \Vec{v} \cdot d\Vec{\ell}$ for an arbitrary closed loop in the fluid is quantized by the quantum $\kappa=h/M$.  A vortex with such quantized circulation is called a quantized vortex. Any rotational motion of a superfluid is sustained only
by quantized vortices.

The appearance of quantized vortices distinguishes the hydrodynamics of quantum condensed fluids from that of classical fluids.
A quantized vortex is a stable topological defect characteristic of a BEC and is different from a vortex in a classical viscous fluid. 
Firstly, the circulation is quantized, which contrasts with a classical vortex that can have any value of circulation. 
Secondly, a quantized vortex is a vortex of inviscid superflow 
and thus it cannot decay by the viscous diffusion of vorticity that occurs in a classical fluid.  
Thirdly, the core of a quantized vortex is very thin, on the order of the coherence length, which is only a few angstroms in superfluid $^4$He and sub $\mu$m even in atomic BECs. 
Since the vortex core is very thin and does not decay by diffusion, it is always possible to identify the position of a quantized vortex in the fluid.  

The turbulence of a superfluid velocity field is called superfluid turbulence or quantum turbulence (QT) \cite{Halperin:2009,Skrbek:2012}.
Since any rotational motion of a superfluid is sustained by quantized vortices, QT usually takes the form of a disordered tangle of quantized vortices.
QT is currently the most important subject of research of QHD in the field of low temperature physics.
The turbulence of classical fluids, called classical turbulence (CT), has been studied intensively in a number of fields, but it is still not yet well understood \cite{Frisch:1995}. 
This is chiefly because turbulence is a complicated dynamical phenomenon with strong nonlinearity. 
Vortices may be the key to understanding turbulence. 
For example, Leonardo da Vinci observed the turbulent flow of water and drew some sketches showing that turbulence had a structure comprised of vortices of different sizes. 
However, vortices are not well-defined for a classical viscous fluid.
They are unstable and appear and disappear repeatedly. The circulation is not conserved and not identical for each vortex.
Comparing QT and CT reveals definite differences, which demonstrates the importance of studying QT. 
QT consists of a tangle of quantized vortices that have the same conserved circulation. 
Thus, QT can be an easier system to study than CT and present a much simpler model of turbulence than CT.
 
QHD was first studied in superfluid $^4$He and more recently in atomic BECs.
Liquid $^4$He enters a superfluid state below the $\lambda$ point (2.17 K) with
Bose--Einstein condensation of the $^4$He atoms \cite{Tilley:1990}. 
The characteristic phenomena of superfluidity were discovered experimentally
in the 1930s by Kapitza {\it et al.}
The hydrodynamics of superfluid helium is well described by the two-fluid model,
for which the system consists of an inviscid superfluid (density $\rho_s$) and 
a viscous normal fluid (density $\rho_n$) with two independent velocity fields $\mathbf{v}_s$ and $\mathbf{v}_n$. 
The mixing ratio of the two fluids depends on temperature. 
As the temperature is reduced below the $\lambda$ point, the ratio of 
the superfluid component increases, and the entire fluid becomes a superfluid below approximately 1 K.
Early experimental studies on superfluid turbulence focused primarily on thermal counterflow, in which the normal fluid and superfluid flow in opposite directions. 
The flow is driven by an injected heat current and it was found that the superflow becomes dissipative when the relative velocity between the two fluids exceeds a critical value.
This was nothing but the appearance of QT. 
The interaction between the cores of tangled vortices and the normal fluid causes the dissipation. 
Considerable effort has been devoted to the study of thermal counterflow in superfluid $^4$He. 
However, since counterflow turbulence has no classical analog, the relationship between QT and CT has not been satisfactorily studied.

Research into QHD has tended toward new directions since the mid 1990s.  
One new direction is in the field of low temperature physics, studying superfluid helium. 
This field of study started with the attempt to understand the relationship between QT and CT \cite{Vinen:2002, Vinen:2006,Vinen:2010}.  
The energy spectrum of fully developed CT is known to obey the Kolmogorov law in the inertial range.  
Recent experimental and numerical studies support a Kolmogorov spectrum in QT. 
Following these studies, QT research on superfluid helium has progressed to important topics such as the dissipation process at very low temperatures, QT created by vibrating structures, and the visualization of QT \cite{Halperin:2009,Skrbek:2012}.
Another new direction is the realization of Bose--Einstein condensation in trapped atomic gases, first performed in 1995, which has stimulated intense experimental and theoretical activity \cite{Pethick:2008, Pitaevskii:2003}.  
As proof of the existence of superfluidity, quantized vortices have been created and observed in atomic BECs, and considerable effort has been devoted to a number of fascinating problems related to this \cite{Fetter:2009, Kasamatsu:2009}.  
Atomic BECs have several advantages over superfluid helium.  
The most important is that modern optical techniques enable the direct control of condensates and the visualization of quantized vortices. 

This article reviews the recent theoretical and numerical contributions of the Osaka City University group on QHD.
Section 2 describes the basics of QHD. 
Section 3 describes QT using the vortex filament model and the Gross--Pitaevskii model.
Section 4 describes hydrodynamic instability in two-component BECs. 
Section 5 is devoted to conclusions.

\section{Basics of quantum hydrodynamics}
\label{sec:Basics}
This section reviews briefly the background of low temperature physics necessary for understanding this article.

\subsection{Bose--Einstein condensation}
Quantum condensation appearing in quantum fluids is caused chiefly by Bose--Einstein condensation. 
Quantum mechanics is often thought to give the physical laws at microscopic scales, but quantum mechanics appears even at macroscopic scales through BECs. 

The essence of quantum mechanics is the duality of the particle picture and wave picture. 
Let us consider an ideal atomic gas. 
At relatively high temperatures, the statistics of the atoms obey the classical Maxwell--Boltzmann distribution and each atom behaves like a particle. 
As the temperature is reduced, however, the thermal de Broglie wavelength is increased to become comparable to the mean distance between atoms. 
Then each atom begins to behave like a wave and the statistics changes to the quantum Fermi--Dirac or Bose--Einstein distribution depending on whether the atom is a Fermion or a Boson. 
If the atoms are Bosons and the system is cooled below a critical temperature $T_{\rm BEC}$, Bose--Einstein condensation occurs in which the atoms occupy the same single-particle ground state. The critical temperature is given by
\begin{equation} T_{\rm BEC}=3.3\frac{\hbar^2n^{2/3}}{M k_B}, \label{TBEC} \end{equation}
where the relevant quantities are the particle mass $M$, the number density $n$, the Planck constant $h=2\pi \hbar$, and the Boltzmann constant $k_B$.
Then, matter waves of atoms become coherent to form a macroscopic wave function (the order parameter) $\Psi(\mathbf{r},t)=|\Psi(\mathbf{r},t)| e^{i \theta(\mathbf{r},t)}$  extending over the whole volume of the system, and the assemblage of these atoms is called a Bose--Einstein condensate (BEC).  

\subsection{Liquid helium, superfluidity and the two-fluid model}  
 Independent of studies of quantum statistical mechanics, the field of low temperature physics has developed since the beginning of 20th century. 
 Low temperature physics is generally believed to start with the first liquefaction of $^4$He at 4.2 K by Onnes in 1908. 
 Subsequently, Onnes observed superconductivity in mercury in 1911. 
 Onnes noticed an anomaly in the heat capacity of liquid helium at the $\lambda$ point $T_{\lambda}=2.17$ K. 
 In 1938 Kapitza {\it et al.} observed that liquid $^4$He becomes inviscid below the $\lambda$ point and called this striking phenomenon {\it superfluidity} \cite{Kapitza:1938, Allen:1938}.  
The superfluid transition of liquid $^4$He at 2.17 K is called the $\lambda$ transition.  

In order to explain the various hydrodynamic phenomena of superfluidity, Tisza \cite{Tisza:1938} and Landau \cite{Landau:1941} introduced the two-fluid model.
According to the two-fluid model,  the system consists of an inviscid superfluid (density $\rho_s$) and 
a viscous normal fluid (density $\rho_n$) with two independent velocity fields $\mathbf{v}_s$ and $\mathbf{v}_n$. 
The mixing ratio of the two fluids depends on temperature. 
As the temperature is reduced below the $\lambda$ point, the ratio of the superfluid component increases and the entire fluid becomes
a superfluid below approximately 1 K.
While the two-fluid model successfully explained the phenomena of superfluidity, it was discovered in the 1940s that superfluidity breaks down when a superfluid flows fast \cite{Gorter:1949} and this phenomenon could not be explained through the two-fluid model. 
This was later found to be caused by turbulence of the superfluid component due to random motion of quantized vortices.  

\subsection{Bose--Einstein condensation, macroscopic wave function, and quantized vortices}  
The $\lambda$ transition is closely related to the Bose--Einstein condensation of $^4$He atoms.
London proposed theoretically in 1938 that the $\lambda$ transition of liquid $^4$He is caused by Bose--Einstein condensation of $^4$He atoms \cite{London:1938}.  
When $T_{\rm BEC}$ of Eq. (\ref{TBEC}) is evaluated for the mass and density appropriate to liquid $^4$He at saturated vapor pressure, 
$T_{\rm BEC}$ of approximately 3.13 K is obtained, which is close to $T_{\lambda}=2.17$ K.

A Bose-condensed system exhibits the macroscopic wave function $\Psi(\Vec{r},t)=|\Psi(\Vec{r},t)| e^{i \phi(\Vec{r},t)}$ as an order parameter. 
The superfluid velocity field is given by $\Vec{v}_s=(\hbar/M) \nabla \phi$, representing the potential flow. 
Since the macroscopic wave function should be single-valued for the space coordinate $\Vec{r}$,
the circulation $\oint \Vec{v} \cdot d\Vec{\ell}$ for an arbitrary closed loop in the fluid is quantized by the quantum $\kappa=h/M$. 
A vortex with quantized circulation is called a quantized vortex. 
Any rotational motion of a superfluid is sustained only by quantized vortices.

\section{Quantum turbulence}
\label{sec:QT}

\subsection{Research history}
\label{history}
This subsection describes briefly the research history of quantized vortices and QT in superfluid $^4$He. 

The idea of quantized circulation was first proposed by Onsager for a series of annular rings in a rotating superfluid \cite{Onsager:1949}.
Feynman considered that a vortex in a superfluid can take the form of a vortex filament with quantized circulation $\kappa$ and a core of atomic dimensions \cite{Feynman:1955}.
Early experimental studies on superfluid hydrodynamics focused primarily on thermal counterflow. 
The flow is driven by an injected heat current, and the normal fluid and superfluid flow in opposite directions. 
The superflow was found to become dissipative when the relative velocity between the two fluids exceeds a critical value \cite{Gorter:1949}. 
Gorter and Mellink attributed the dissipation to mutual friction between two fluids and considered the possibility of superfluid turbulence. 
Feynman proposed a turbulent superfluid state consisting of a tangle of quantized vortices \cite{Feynman:1955}. 
Hall and Vinen performed experiments of second sound attenuation in rotating $^4$He, where second sound refers to the entropy wave in which superfluid and normal fluid oscillate oppositely, and its propagation and attenuation give information on the vortex density in the fluid. They found that mutual friction arises from the interaction between the normal fluid and quantized vortices \cite{Hall:1956a,Hall:1956b}.
Vinen confirmed Feynman's findings experimentally by showing that the dissipation in thermal counterflow arises from mutual friction between vortices and the normal flow \cite{Vinen:1957a,Vinen:1957b,Vinen:1957c,Vinen:1957d}. 
Vinen also succeeded in observing quantized circulation using vibrating wires in rotating superfluid $^4$He \cite{Vinen:1961}.
Subsequently, many experimental studies have examined superfluid turbulence (ST) in thermal counterflow systems and have revealed a variety of physical phenomena \cite{Tough:1982}. 
Since the dynamics of quantized vortices are nonlinear and non-local, it has not been easy to quantitatively understand these observations on the basis of vortex dynamics. 
Schwarz clarified the picture of ST based on tangled vortices by numerical simulation of the quantized vortex filament model in the thermal counterflow \cite{Schwarz:1985,Schwarz:1988}. 
However, since the thermal counterflow has no analogy in conventional fluid dynamics, this study was not helpful in clarifying the relationship between ST and classical turbulence (CT). 
ST is often called quantum turbulence (QT), which emphasizes the quantum effects.

QHD, including QT, is reduced to the motion of quantized vortices.
Hence, understanding the dynamics of quantized vortices is a key issue in QHD.
Two formulations are generally available for studying the dynamics of quantized vortices.  
One is the vortex filament model and the other is the Gross--Pitaevskii (GP) model.
This section describes the results first by the vortex filament model and then by the GP model.

\subsection{Vortex filament model}
\label{filament}
As described in Sec. 1., a quantized vortex has quantized circulation. 
The vortex core is extremely thin, usually much smaller than other characteristic scales of vortex motion. 
These properties allow a quantized vortex to be represented as a vortex filament. 
In classical fluid dynamics \cite{Saffman:1992}, the vortex filament model is just a convenient idealization; the vorticity in a realistic classical fluid flow rarely takes the form of clearly discrete vorticity filaments.
However, the vortex filament model is accurate and realistic for a quantized vortex in superfluid helium.

\subsubsection{Schwarz's  model}
In considering the velocity field created by a vortex filament in this subsection, we introduce Schwarz's model, which is useful for superfluid helium.

The incompressible velocity $\mathbf{v}(\mathbf{r})$ created by the vorticity source ${\boldsymbol \omega}(\mathbf{r})$ satisfies $\rm{div} \mathbf{v}=0$ and $\rm{rot} \mathbf{v}=\boldsymbol{\omega}$, whose solution is \cite{Saffman:1992}
\begin{equation}
\mathbf{v}(\mathbf{r})=\frac{1}{4\pi}\int \boldsymbol{\omega}(\mathbf{r}') \times \frac{\mathbf{r}-\mathbf{r}'}{|\mathbf{r}-\mathbf{r}'|^3} d\mathbf{r}'. \label{BSv}
\end{equation}
This representation is applied to the vortex filament formulation, which describes a quantized vortex as a filament passing through the fluid, having a definite direction corresponding to its vorticity. 
The three-dimensional configuration of vortex filaments can be represented by differential geometry. 
A point on a filament at a time $t$ is represented by the parametric form $\mathbf{s}(\varsigma, t)$ with the one-dimensional coordinate $\varsigma$ along the filament. 
The positive direction of $\varsigma$ is taken to be the direction of the vorticity.
Then, $\partial \mathbf{s}/\partial \varsigma \equiv \mathbf{s}'(\varsigma, t)$ is a unit tangential vector along the filament at $\mathbf{s}$, 
and $\partial^2 \mathbf{s}/\partial \varsigma^2 \equiv \mathbf{s}''(\varsigma, t)$ is the principal normal vector at $\mathbf{s}$ with magnitude $R^{-1}$, where $R$  is the local radius of curvature.
Except for the thin core region, the superflow velocity field has a classically well-defined meaning and can be described by ideal fluid dynamics. 
The vorticity with quantized circulation $\kappa$ is focused only on the filament, represented by
\begin{equation}
{\boldsymbol \omega}(\mathbf{r},t)=\kappa \int_{\mathcal{L}}\mathbf{s}'(\varsigma, t)\delta(\mathbf{r}-\mathbf{s}(\varsigma, t))d\varsigma, \label{omega}
\end{equation}
where the integration is taken along the filament.
Inserting Eq.  (\ref{omega}) into Eq. (\ref{BSv}) yields the Biot--Savart expression
\begin{equation}
\mathbf{v}_{s, BS} (\mathbf{r},t) = \frac{\kappa}{4\pi} \int_{\mathcal{L}} \frac{\mathbf{s}'(\varsigma, t) \times (\mathbf{r}-\mathbf{s}(\varsigma, t))}{|\mathbf{r}-\mathbf{s}(\varsigma, t)|^3} d\varsigma. \label{eq-Biot-Savart}
\end{equation}
Thus some configuration of vortex filaments gives the superfluid velocity field $\mathbf{v}_{s, BS} (\mathbf{r},t)$.

Considering the forces acting on the vortex filament, we derive the equation of motion, namely Schwarz's equation.
When a vortex filament moves in the superflow field $\mathbf{v}_{s}$, the effective forces are the Magnus force, the mutual friction force, and the inertial force.
The Magnus force refers to the lift force acting on a spinning object when it moves in a fluid.
The Magnus force for our vortex filament per unit length is written as
\begin{equation}
\mathbf{f}_M=\rho_s \kappa \mathbf{s}'\times (\dot{\mathbf{s}}-\mathbf{v}_{s}), \label{Magnus}
\end{equation}
where $\dot{\mathbf{s}}\equiv d \mathbf{s}/dt$ refers to the velocity of the filament at $\mathbf{s}$.
The Magnus force tends to move the vortex filament at $\mathbf{s}$ normal to both the vorticity $\kappa \mathbf{s}'$ and the relative velocity $\dot{\mathbf{s}}-\mathbf{v}_{s}$ 
between the vortex velocity and the superflow.
At finite temperatures mutual friction works through the interaction between the normal flow and the vortex core:
 \begin{equation}
\mathbf{f}_D=-\alpha \rho_s \kappa \mathbf{s}'\times [\mathbf{s}'\times (\mathbf{v}_{n}-\mathbf{v}_{s})]-\alpha' \rho_s \kappa \mathbf{s}'\times (\mathbf{v}_{n}-\mathbf{v}_{s}), \label{MF}
\end{equation}
where $\alpha$ and $\alpha'$ are coefficients dependent on temperature \cite{Schwarz:1985}.
The right hand side represents the force normal to $\mathbf{s}'$.
We can write the equation of motion of the vortex filament per unit length as
\begin{equation}
m_{\rm eff}\frac{d^2 \mathbf{s}}{dt^2}=\mathbf{f}_M+\mathbf{f}_D, \label{eq. VF}
\end{equation}
with the effective mass $m_{\rm eff}$ of the filament per unit length.
The effective mass should be of the order of $\rho_s a_0^2$, which is usually quite small compared with other terms because of the small core radius $a_0$.
Thus we can neglect the inertia term, so that Eq. (\ref{eq. VF}) is reduced to Schwarz's equation
\begin{equation}
\dot{\mathbf{s}}=\mathbf{v}_{s}+\alpha \mathbf{s}'\times (\mathbf{v}_{n}-\mathbf{v}_{s})-\alpha' \mathbf{s}'\times [\mathbf{s}'\times (\mathbf{v}_{n}-\mathbf{v}_{s})].  \label{Schwarz}
\end{equation} 

When we attempt to obtain $\mathbf{v}_{s}$ at a point $\mathbf{s}(\varsigma_0)$ along a filament from Eq. (\ref{eq-Biot-Savart}), the integral diverges as $\mathbf{s}(\varsigma) \rightarrow \mathbf{s}(\varsigma_0)$.
In order to treat this difficulty, we introduce the localized induction velocity proposed by Arms and Hama \cite{Arms:1965}.
By using a cutoff $R$, the integral of Eq. (\ref{eq-Biot-Savart}) is divided into the contribution within $R$ around $\varsigma_0$ and that from the other distant region.
The neighborhood of $\mathbf{s}(\varsigma_0)$ is represented by
\begin{equation}
\mathbf{s}(\varsigma)=\mathbf{s}(\varsigma_0)+(\varsigma-\varsigma_0)\mathbf{s}'(\varsigma_0)+\frac12 (\varsigma-\varsigma_0)^2\mathbf{s}''(\varsigma_0).
\end{equation}
This expression with Eq. (\ref{eq-Biot-Savart}) yields the local contribution
\begin{equation}
\mathbf{v}_{i}(\varsigma_0)=\frac{\kappa}{4\pi}\int_{a_0}^{R}\frac{d\varsigma}{\varsigma} \mathbf{s}'(\varsigma_0)\times \mathbf{s}''(\varsigma_0) =\beta \mathbf{s}'(\varsigma_0)\times \mathbf{s}''(\varsigma_0)
\end{equation} 
with $\beta=(\kappa/4\pi)\log (R/a_0)$. The parameter $a_0$ is the cutoff, corresponding to the core radius. 
This velocity is called the localized induction velocity or self-induced velocity.
Since the contribution from the outer region is obtained by the usual integration, $\mathbf{v}_{s}$ of Eq. (\ref{Schwarz}) is reduced to  \cite{Schwarz:1985}:
\begin{equation}
\mathbf{v}_{s}= \beta \mathbf{s}' \times \mathbf{s}''+ \frac{\kappa}{4\pi} \int_{\mathcal{L}}^\prime \frac{(\mathbf{s}_1 - \mathbf{r}) \times d\mathbf{s}_1}{|\mathbf{s}_1-\mathbf{r}|^3}. \label{eq-sdot}
\end{equation}
The second term represents the non-local field obtained by integrating the integral of Eq. (\ref{eq-Biot-Savart}) along the rest of the filament, except in the neighborhood of $\mathbf{s}$. 

A better understanding of vortices in a real system is obtained when boundaries are included in the analysis. 
For this purpose, a boundary-induced velocity field $\mathbf{v}_{s,b}$ is added to $\mathbf{v}_{s}$, so that the superflow can satisfy the boundary condition of an inviscid flow; 
that is, the normal component of the velocity should disappear at the boundaries. 
To allow for another, presently unspecified, applied field, we include $\mathbf{v}_{s,a}$.

Consequently, the total velocity $\dot{\mathbf{s}}_0$ of the vortex filament without dissipation is
\begin{equation}
\dot{\mathbf{s}}_0 =\beta \mathbf{s}' \times \mathbf{s}'' + \frac{\kappa}{4\pi} \int_{\mathcal{L}}^\prime \frac{(\mathbf{s}_1 - \mathbf{r}) \times d\mathbf{s}_1}{|\mathbf{s}_1-\mathbf{r}|^3} + \mathbf{v}_{s,b}(\mathbf{s}) + \mathbf{v}_{s,a}(\mathbf{s}). \label{eq-s0dot}
\end{equation}
\noindent
At finite temperatures, it is necessary to take into account the mutual friction between the vortex core and the normal flow $\mathbf{v}_{n}$. 
Including this term, the velocity of $\mathbf{s}$ is given by 
\begin{equation}
\dot{\mathbf{s}} =\dot{\mathbf{s}}_0 + \alpha \mathbf{s}^\prime \times (\mathbf{v}_{n} - \dot{\mathbf{s}}_0) - \alpha^\prime \mathbf{s}^\prime \times [\mathbf{s}^\prime \times (\mathbf{v}_{n} - \dot{\mathbf{s}}_0)], \label{eq-sdotmf}
\end{equation}
where $\dot{\mathbf{s}}_0$ is calculated from Eq. (\ref{eq-s0dot}). 

\subsubsection{Basic motion of vortex filaments}
In this subsection we discuss the simple motion of vortex filaments in order to develop a basic understanding.
The addressed equations of motion are Eqs. (\ref{eq-s0dot}) and (\ref{eq-sdotmf}).

The first term of Eq. (\ref{eq-s0dot}) refers to the localized induction field arising from a curved line element acting on itself. 
The mutually perpendicular vectors $\mathbf{s}'$, $\mathbf{s}''$, and $\mathbf{s}' \times \mathbf{s}''$ are directed along the tangent, 
the principal normal, and the binormal, respectively, at the point $\mathbf{s}$, and their respective magnitudes are 1, $R^{-1}$, and $R^{-1}$, where $R$ is the local radius of curvature. 
Thus, the first term represents the tendency to move the local point $\mathbf{s}$ in the binormal direction with a velocity inversely proportional to $R$. 
Neglecting the non-local terms is referred to as the localized induction approximation (LIA). 
This approximation is believed to be effective for analyzing isotropic dense tangles due to cancellations between non-local contributions \cite{Schwarz:1988}. 
However, the LIA lacks the interaction between vortices, and is not necessarily suitable for the description of a realistic vortex tangle, as shown in 3.3.

We consider vortex motion under mutual friction. The first example is the propagation of a vortex ring.
Without any other vortices or a velocity field at zero temperature,  a vortex ring with a radius $R$ just propagates with a velocity approximately equal to the self-induced velocity $\beta/R$ normal to the circular plane, keeping its shape.
We consider the motion of a vortex ring under the applied fields $\mathbf{v}_{n}$ and $\mathbf{v}_{s,a}$.
Using the LIA and neglecting the friction term $\alpha'$ for simplicity, Eqs. (\ref{eq-s0dot}) and (\ref{eq-sdotmf}) are reduced to
\begin{equation}
\dot{\mathbf{s}} =\beta \mathbf{s}' \times \mathbf{s}''+ \mathbf{v}_{s,a}+\alpha \mathbf{s}^\prime \times (\mathbf{v}_{n} - \mathbf{v}_{s,a}-\beta \mathbf{s}' \times \mathbf{s}''). \label{motion} 
\end{equation}
The self-induced velocity is supposed to be in the $z$ direction with $\mathbf{s}' \times \mathbf{s}''=\hat{z}/R$. 
If we take $\mathbf{v}_{n}=v_n \hat{z}$ and $\mathbf{v}_{s,a}=v_{s,a} \hat{z}$, Eq. (\ref{motion}) describes the time development of $R$ as 
\begin{equation}
\frac{dR}{dt}=\alpha \Bigl(v_n-v_{s,a}-\frac{\beta}{R} \Bigr).
\end{equation}
 The absence of any applied fields gives $dR/dt=-\alpha (\beta/R)$, whose solution is $R=\sqrt{R_0^2-2\alpha \beta t}$ with the initial radius $R_0$.
 Thus the mutual friction shrinks the ring. The presence of fields complicates the situation.
 For $v_n-v_{s,a}<0$, the ring always shrinks. 
 When $v_n-v_{s,a}>0$, the ring expands for $v_n-v_{s,a}>\beta/R$ and shrinks otherwise, which leads to a critical radius of curvature given by $R_c \simeq \beta/(v_n-v_{s,a})$.
 These simple considerations emphasize the important role of the mutual friction.
 The mutual friction can both shrink and expand a ring depending on the applied field and the radius.
 In a vortex tangle, fine structure with a small radius of curvature generally shrinks under the mutual friction.
 In other words, the mutual friction tends to make the vortex configuration smooth.
 
 The second example we consider is the motion of two parallel or antiparallel vortices.
 Suppose that two straight vortices with circulation $\kappa$ are placed in parallel with a distance $2r$.
 Each vortex moves through the velocity $\kappa/4\pi r$ from the other. 
 At zero temperature, they rotate around their middle point, maintaining their distance.
 At a finite temperature, Eq. (\ref{eq-sdotmf}), neglecting the $\alpha'$ term, yields $dr/dt=\alpha(\kappa/4\pi r)$, whose solution is $r=\sqrt{r_0^2+(\alpha \kappa/2\pi)t}$ with the initial distance $r_0$.
 Thus the two vortices spiral outward, which means that their interaction is effectively repulsive.
 On the other hand, two antiparallel straight vortices move normal to the line segment connecting them with velocity $\kappa/4\pi r$ at zero temperature.
 The mutual friction at a finite temperature reduces their distance as $r=\sqrt{r_0^2-(\alpha \kappa/2\pi)t}$.
 The interaction between them is effectively attractive, which eventually leads to pair-annihilation of two voritces.
 
\subsubsection{Numerical simulation}
The numerical simulation method based on this model has been described in detail elsewhere \cite{Schwarz:1985,Schwarz:1988,Tsubota:2000,Adachi:2010}. 
A vortex filament is represented by a single string of points separated by a distance $\Delta\varsigma$. 
The vortex configuration at a given time determines the velocity field in the fluid, thus moving the vortex filaments according to Eqs. (\ref{eq-s0dot}) and (\ref{eq-sdotmf}). 
When vortices move, they have chances to encounter other vortices.
Thus, vortex reconnection should be properly included when simulating vortex dynamics. 
 A numerical study of a classical fluid shows that the close interaction of two vortices leads to their reconnection, primarily because of viscous diffusion of the vorticity \cite{Boratav:1992}. 
 Schwarz assumed that two vortex filaments reconnect when they come within a critical distance of one another, 
 and showed that statistical quantities such as the vortex line density were not sensitive to how these reconnections occur \cite{Schwarz:1985,Schwarz:1988}.  
 Even after Schwarz's study, it remained unclear as to whether quantized vortices can actually reconnect. 
 However, Koplik and Levine directly solved the GP equation to show that two closely quantized vortices reconnect, even in an inviscid superfluid \cite{Koplik:1993}. 
 Therefore such an artificial procedure of vortex reconnection is currently thought to be allowed in the vortex filament model too.
 The more modern and reasonable procedure is to reconnect two vortices when they pass within the spatial resolution $\Delta \varsigma$ with unit probability. 
 Every vortex initially consists of a string of points at regular intervals of $\Delta\varsigma$. 
 When a point on a vortex approaches another point on another vortex more closely than the fixed space resolution $\Delta \varsigma$, we join these two points and reconnect the vortices. 
 This reconnection procedure is standard in the vortex filament model, but a different procedure is used in some studies \cite{Kondaurova:2008}.

\subsubsection{Some statistical quantities}
 Some important quantities that are useful for characterizing the vortex tangle are introduced below \cite{Schwarz:1988}. The vortex line density (VLD) is the total length of vortex lines per unit volume, defined by
\begin{equation}
L=\frac{1}{\Omega}\int_{\cal L}d\varsigma,
\end{equation}
where the integral is performed over all vortices in the sample volume $\Omega$. 
The anisotropy of the vortex tangle that is formed under the counterflow $\mathbf{v}_{ns}$ is represented by the dimensionless parameters
\begin{equation}
I_{\|}=\frac{1}{\Omega L} \displaystyle\int^{}_{\cal L}[1-({\bf s}'\cdot{\hat {\bf r}_{\|}})^2]d\varsigma,
\end{equation}
\begin{equation}
I_{\bot}=\frac{1}{\Omega L} \displaystyle\int^{}_{\cal L}[1-({\bf s}'\cdot{\hat {\bf r}_{\bot}})^2]d\varsigma,
\end{equation}
\begin{equation}
I_l{\hat {\bf r}_{\|}}=\frac{1}{\Omega L^{3/2}}\int^{}_{\cal L} {\bf s}'\times {\bf s}''d \varsigma.
\end{equation}
 Here, $\hat {\bf r}_{\|}$ and $\hat {\bf r}_{\bot}$ represent unit vectors parallel and perpendicular to the $\mathbf{v}_{ns}$ direction, respectively. 
 Symmetry generally yields the relation $I_{\|}/2+I_{\bot}=1$. 
 If the vortex tangle is isotropic, the averages of these parameters are ${\bar I_{\|}}={\bar I_{\bot}}=2/3$ and ${\bar I}_l=0$. 
 At the other extreme, if the tangle consists entirely of curves lying in planes normal to ${\bf v}_{ns}$, then ${\bar I_{\|}}=1$ and ${\bar I_{\bot}}=1/2$.

When we perform numerical simulations, we should deal with such statistical quantities as well as the dynamics of each vortex.
The characteristic behavior of the VLD $L$ in thermal counterflow was considered by Vinen.
 In order to describe amplification of a temperature difference at the ends of a capillary retaining thermal counterflow, Gorter and Mellink introduced some additional interactions (mutual friction) between the normal fluid and superfluid \cite{Gorter:1949}.
 Through experimental studies of the second-sound attenuation, Vinen considered this Gorter--Mellink mutual friction in relation to the macroscopic dynamics of the vortex tangle \cite{Vinen:1957a, Vinen:1957b, Vinen:1957c, Vinen:1957d}. 
 Assuming homogeneous superfluid turbulence, Vinen obtained an equation for the evolution of $L(t)$, which we call Vinen's equation:
\begin{equation}
\frac{dL}{dt}=\frac{\chi_1 B \rho_n}{2\rho} |{\bf v}_{ns}|L^{3/2}-\chi_2 \frac{\kappa}{2\pi}L^2,
\end{equation}
where $\chi_1$ is a constant, and $B$ and $\chi_2$ are temperature-dependent parameters. 
The first term represents the energy injection from the normal fluid to the vortices. 
The second term denotes the energy dissipation of vortices due to reconnection between vortices.
 The first and second terms indicate the growth and the degeneration of a vortex tangle, respectively.
 After the growth period of the VLD, the vortex tangle enters a statistically steady state.
 In the steady state, the VLD is obtained by setting $dL/dt$ equal to zero, which gives
\begin{equation}
L=\gamma^2v_{ns}^2,
\label{Lvns}
\end{equation}
where $\gamma=\pi B \rho_n \chi_1/ \kappa \rho \chi_2$ is a temperature-dependent parameter.
This relation can describe a large number of the observations of stationary cases \cite{Tough:1982}.
When we conduct a simulation of the counterflow, the confirmation of Eq. (\ref{Lvns}) is a touchstone.

\subsection{Thermal counterflow turbulence by the full Biot--Savart law}
\label{cf}
The difficulty in accounting for the nonlinear and nonlocal dynamics of vortices has long delayed progress in achieving a microscopic understanding of QT.
 It was Schwarz who made the breakthrough \cite{Schwarz:1988}.
 He investigated counterflow turbulence using the vortex filament model and dynamical scaling. 
The observable quantities obtained by his calculation agreed well with the experimental results for the steady state of vortex tangles. 
This study confirmed the idea proposed by Feynman that superfluid turbulence consists of a quantized vortex tangle. 
However, thermal counterflow turbulence was still far from being perfectly understood.
 The numerical simulation of Schwarz had serious defects.
 One is that the calculations were performed under the LIA neglecting interactions between vortices. 
 Schwarz reported that as a result the layer structure is constructed gradually when periodic boundary conditions are applied. 
 Of course, this behavior is not realistic.
 In order to address this, an unphysical, artificial mixing procedure was employed, in which half the vortices are randomly selected to be rotated by 90$^\circ$ around the axis defined by the flow velocity. 
 This method enables the steady state to be sustained under periodic boundary conditions.
 These defects cause us to conjecture that the LIA is unsuitable due to the absence of interactions between vortices.
 
Adachi {\it et al.} performed numerical simulations of counterflow turbulence using the full Biot--Savart law under periodic boundary conditions and succeeded in obtaining a statistically steady state without any unphysical procedures \cite{Adachi:2010}.
 Figure 1 shows a typical result of the time evolution of the vortices, whose VLD grows as shown in Fig. \ref{T19_line_t}. 
The initial configuration consists of six vortex rings.
 
\begin{figure}[h]
\begin{center}
\scalebox{0.4}{\includegraphics{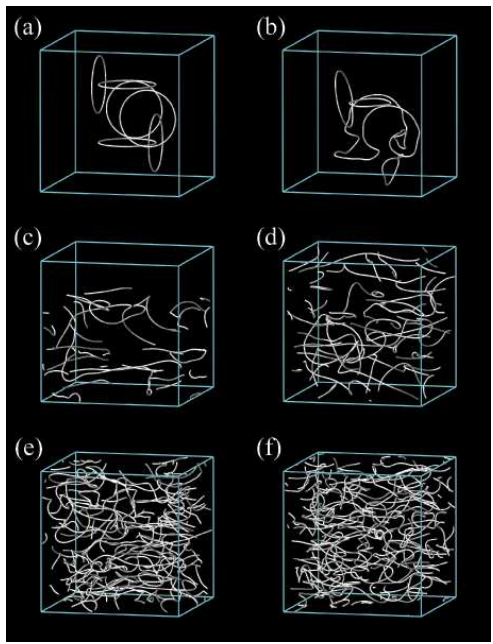}} 
\caption{Development of a vortex tangle by the full Biot--Savart calculation in a periodic box with a size of 0.1 cm. 
Here, the temperature is $T=1.9\,{\rm K}$ and the counterflow velocity $v_{ns}=0.572\,{\rm cm/s}$ is along the vertical axis. (a) $t=0\,{\rm s}$, (b) $t=0.05\,{\rm s}$, (c) $t=0.5\,{\rm s}$, (d) $t=1.0\,{\rm s}$, (e) $t=3.0\,{\rm s}$, (f) $t=4.0\,{\rm s}$. [Adachi, Fujiyama and Tsubota: Phys. Rev. B \textbf{81} (2010) 104511, reproduced with permission. Copyright 2010 the American Physical Society.]} 
\end{center}
\label{Adachi} 
\end{figure}

In the first stage ($0 \leq t \leq 0.4\, {\rm s}$), the critical radius $R_c$ determines the vortex destiny. 
Vortex ring sections in which the radius of curvature exceeds $R_c$ expand in the direction perpendicular to ${\bf v}_{ns}$ through mutual friction, while small vortex rings shrink. 
Thus, vortices evolve and become anisotropic. At the end of this stage, large vortices appear that are comparable to the system size under periodic boundary conditions.
These vortices survive with a large radius of curvature, and continuously generate small vortices by reconnections in the subsequent stages so that they function as ``vortex mills''\cite{Schwarz:1993}.
In the second stage ($0.4 < t \leq 2.0\, {\rm s}$), vortex tangles undergo continuous evolution despite the decreasing anisotropy.
As vortex rings expand, reconnections between vortices occur frequently. 
Reconnections generate vortices with various curvatures, resulting in them shrinking and expanding as discussed for the first stage. 
Local sections with a small radius of curvature formed by reconnections have an almost isotropic self-induced velocity, 
which prevents the vortices from lying perpendicular to ${\bf v}_{ns}$. 
In addition, as the VLD increases, vortex expansion becomes slower than in the first stage because the reconnection distorts vortices, which prevents a vortex from smoothly expanding.
In the third stage ($t > 2.0\,{\rm s}$), the statistically steady state is realized by the competition between the growth and decay of a vortex tangle.
The growth mechanism is still vortex expansion through mutual friction.
The decay mechanism either creates vortices with local radii of curvature smaller than $R_c$ or vortices with the self-induced velocity oriented in the opposite direction to ${\bf v}_{ns}$ after the reconnections. 
The increasing VLD causes more reconnections so that the decay mechanism becomes effective. 
When the VLD has increased sufficiently, the two mechanisms begin to compete so that the vortex tangle enters the statistically steady state. 
The LIA calculation cannot realize this competition, which shows that vortex interaction is essential for creating a steady state.

\begin{figure}[h]
\begin{center}
\includegraphics[width=0.45 \textwidth]{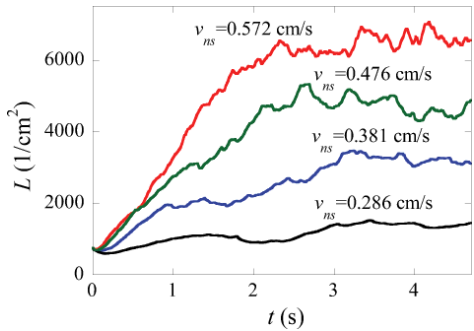}
\caption{Vortex line density as a function of time for four different counterflow velocities. [Adachi, Fujiyama and Tsubota: Phys. Rev. B \textbf{81} (2010) 104511, reproduced with permission. Copyright 2010 the American Physical Society.]\label{T19_line_t}} 
\end{center} 
\end{figure}

\begin{figure}[h]
\begin{center}
\scalebox{0.44}{\includegraphics{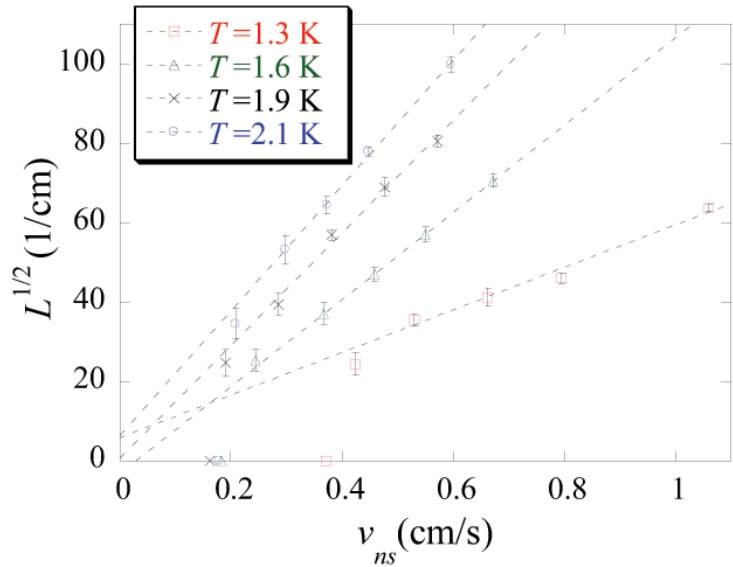}}
\caption{Steady state vortex line density $L(t)$ as a function of the counterflow velocity $v_{ns}$. The error bars represent the standard deviation.  [Adachi, Fujiyama and Tsubota: Phys. Rev. B \textbf{81} (2010) 104511, reproduced with permission. Copyright 2010 the American Physical Society.]\label{vinen}} 
\end{center}
\end{figure}

The obtained steady states almost satisfy the relation of Eq. (\ref{Lvns}) when $v_{ns}$ and $L$ are relatively large, as shown in Fig. \ref{vinen}. 
Table \ref{gamma} shows the parameter $\gamma$ as a function of $T$. 
The results quantitatively agree with the typical experimental observations of Childers and Tough \cite {Tough:1982,Childers:1976}. 
Additionally, there is a critical velocity of turbulence, below which vortices disappear. 
This critical velocity has been measured in many previous studies \cite{Tough:1982,Childers:1974,deHass:1976}; it is given by 
\begin{equation}
v_{ns,c}\approx \frac{2.5+1.44\sigma}{\gamma d},
\label{critical velocity}
\end{equation}
where $d$ is the channel size of the experimental system and $\sigma$ is a constant of order unity. 
In the simulation, the system size may be taken to be the size of the periodic box. 
Then, Eq. (\ref{critical velocity}) gives $v_{ns,c} \sim 0.1\, {\rm cm}$, which is almost consistent with the numerical results. 
However, the temperature dependence of $ v_{ns,c}$ should be considered. 
Equation (\ref{critical velocity}) states that $v_{ns,c}$ should decrease with $T$, which differs from the behavior in Fig. \ref{vinen}. 
The numerical results show that $v_{ns,c}$ decreases with $T$ below $1.9\, {\rm K}$ but increases slightly at $2.1\, {\rm K}$. 
This is because the strong mutual friction makes the vortices so anisotropic that they cannot form enough reconnections with other vortices, and so become degenerate.

Thus a full account of the intervortex interaction by the full Biot--Savart law enables us to obtain the statistical steady states without any unphysical procedures.
A detailed comparison between the numerical results of the full Biot--Savart law and the LIA is discussed in Ref. \cite{Adachi:2010}.

\begin{center}
\begin{table}
\begin{tabular}{cccc}
\hline
$T$ (K)& $\gamma_{num}({\rm s/cm^2})$ & $\gamma_{exp}({\rm s/cm^2})$ & $I_{\|}$ \\ \hline
        1.3 & 53.5    &   59  &    0.738 \\
        1.6 & 109.6   &   93  &    0.771 \\
        1.9 & 140.1   &   133 &    0.820 \\
        2.1 & 157.3   &   - &    0.901 \\
\hline
\end{tabular}
\caption{Line density coefficients $\gamma$ and anisotropy parameter $I_{\|}$.
 $\gamma_{num}$ and $\gamma_{exp}$ denote our numerical results and the experimental results of Childers and Tough \cite{Tough:1982,Childers:1976}, respectively. [Adachi, Fujiyama and Tsubota: Phys. Rev. B \textbf{81} (2010) 104511, reproduced with permission. Copyright 2010 the American Physical Society.]\label{gamma}}
\end{table}
\end{center}

However, the situation may be not so simple.
The above simulation was done under the assumption that the normal flow is laminar or uniform.
By the recent visualization experiments using metastable helium molecules, Guo {\it et al.} showed that the normal fluid could be turbulent too at relatively large velocities \cite{Guo:2010}.
In order to take account of the turbulent normal flow, we should address the coupled equations of the vortex dynamics and the Navier-Stokes equation describing the normal flow.
It would be a future work.  

\subsubsection{Velocity statistics}
\label{velocity statistics}
Velocity statistics, namely the probability density function (PDF) of the velocity field, is another important statistic in turbulence.
It is known that the PDF of classical viscous turbulence is Gaussian \cite{Vincent:1991,Noullez:1997}.
The question remains, what happens to the PDF in QT?

Paoletti {\it et al.}\cite{Paoletti:2008} performed visualization of quantized vortices in a relaxation process of counterflow using solid hydrogen particles and obtained the non-classical (non-Gaussian) PDF of the particle velocity. 
They reported that the non-classical statistics are due to the velocity induced by the reconnection of a quantized vortex because the PDF exhibits a power-law distribution of $v^{-3}$, which is derived from the vortex velocity before or after reconnection. 
However, they observed the velocity of particles, which is not necessary the velocity of the superflow. The non-classical velocity statistics were also confirmed by White {\it et al} \cite{White:2010}. 
They performed numerical simulations of QT in a trapped BEC by calculating the GP equation to obtain the PDFs of the superflow field. 
The PDFs do not show classical Gaussian distributions, but rather power-law distributions, due to the velocity field  $v=\kappa/(2\pi r)$ induced by the singular quantized vortex, where $r$ is the distance from the core of a quantized vortex.

\begin{figure}[h]
  \begin{center}
 \scalebox{0.28}{\includegraphics{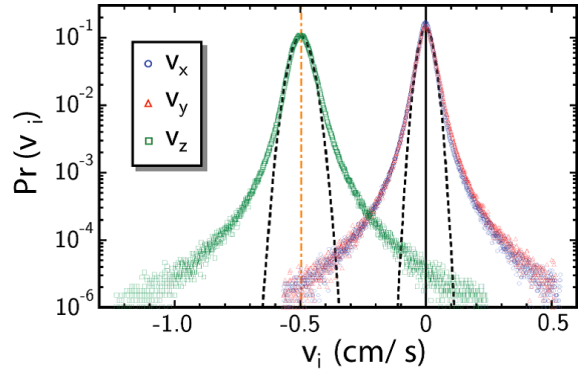}} 
 \caption{(Color online) Probability distribution of the velocity components $v_x$, $v_y$, and $v_z$ in steady counterflow turbulence of $v_{sa}=-0.496$ cm/s at $T=2.1{\rm K}$. 
 The vertical dot-dashed line indicates $v_{sa}$. [Adachi and Tsubota: Phys. Rev. B \textbf{83} (2011) 132503, reproduced with permission. Copyright 2011 the American Physical Society.]
 \label{pdf_vx_vy_vz}} 
  \end{center} 
 \end{figure}

Adachi {\it et al.} studied the PDF of superflow for the steady state of counterflow \cite{Adachi:2011}. 
Figure \ref{pdf_vx_vy_vz} shows the PDFs of $v_x$, $v_y$, and $v_z$, where the counterflow is applied along the $z$ direction. 
The PDFs exhibit a non-Gaussian distribution with a large tail in the high-velocity region. 
Since the vortex tangle of steady counterflow turbulence is isotropic in the direction perpendicular to the relative velocity ${\bf v}_{ns}$, the PDF of $v_{x}$ almost overlaps with that of $v_{y}$, with the peaks of two PDFs at $v_{x}=0$ and $v_{y}=0$.
 In contrast, since the superfluid velocity $v_{sa}=-0.496$ cm/s due to counterflow is applied in the $-z$ direction, the PDF of $v_z$ has a peak at $v_{sa}$. 

The PDF in the high-velocity region shows a power-law distribution of ${\rm Pr}(v_i)\propto v^{-3}_i$ ($i=x,y,z$). 
For the single, straight vortex case, the probability of separation occurring between $r$ and $r+dr$ is $2\pi r dr$, and the velocity scales as $1/r$, which leads to ${\rm Pr}(v)\sim {\rm Pr}(r(v))|dr/dv| \sim 1/v^3$. 
The PDF converges to a Gaussian distribution in the low-velocity region, probably because the vortex configuration is random in the tangle.
We can roughly estimate the transition velocity from the Gaussian distribution to the power-law distribution. 
In order to easily understand the velocity field induced by multiple vortices, we consider the simple case of two straight parallel vortices. 
Although the $1/r$ velocity primarily appears near each vortex, in the region halfway between vortices, the velocity becomes complicated because the velocities induced by the two vortices become comparable and interfere with each other. 
Hence, the statistics of velocity appear to change near the midpoint between vortices. 
In the vortex tangle, the mean inter-vortex distance is denoted by $l=1/L^{1/2}$, and so the midpoint between the vortices is located at $l/2$. 
Thus, the transition velocity of the statistics should be represented by 
\begin{equation}
v_t=\frac{\kappa}{2\pi(l/2)}.
\end{equation}
The transition in Fig.  \ref{pdf_vx_vy_vz} certainly occurs approximately at this scale.
Thus the PDF shows the classical behavior at the low-velocity region and quantum behavior at the high-velocity region. 

\subsection{Quantum turbulence created by vibrating structures}
\label{vibrating structure}
Recently, vibrating structures, such as discs, spheres, grids, and wires, have been widely used for research into QT \cite{Skrbek:2009}. 
Despite detailed differences between the structures considered, the experiments show some surprisingly common phenomena. 

This trend started with the pioneering observation of QT on an oscillating microsphere by J\"ager {\it et al.}\cite{Schoepe:1995}. 
The sphere used by J\"ager {\it et al.} had a radius of approximately 100 $\mu$m, 
and was made from a strongly ferromagnetic material with a very rough surface. 
The sphere was magnetically levitated in superfluid $^4$He and its response with respect to the alternating drive was observed.  
At low drives, the velocity response $v$ was proportional to the drive $F_D$, taking the "laminar" form $F_D=\lambda(T) v$, with the temperature-dependent coefficient $\lambda(T)$.  
At high drives, the response changed to the "turbulent" form $F_D=\gamma(T) (v^2-v_0^2)$ above the critical velocity $v_0$. 
At relatively low temperatures the transition from laminar to turbulent response was accompanied by significant hysteresis.
Subsequently, several groups have experimentally investigated the transition to turbulence in the superfluids $^4$He and $^3$He-B by using grids \cite{Nichol:2004a,Nichol:2004b,Charalambous:2006, Bradley:2005b, Bradley:2006,Bradley:2012}, wires \cite{Fisher:2001,Bradley:2004,Yano:2007,Hashimoto:2007, Goto:2008}, and tuning forks \cite{Blazkova:2007b,Blazkova:2009}.  
The details of the observations are described in a review article \cite{Skrbek:2009}. 
Here we shall briefly describe a few important points necessary for the current article.

\begin{figure}[htb] \centering \begin{minipage}[t]{0.7\linewidth} \begin{center} \includegraphics[width=.6\linewidth]{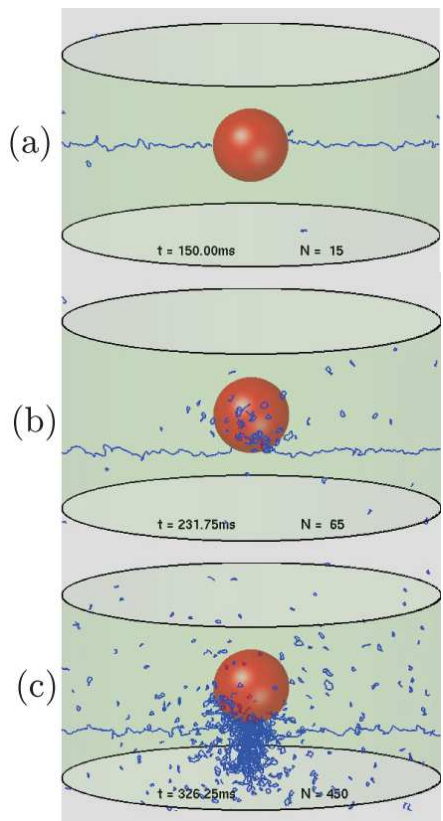} \end{center} \end{minipage} 
\caption{Evolution of the vortex line near a sphere of radius 100~$\mu$m in an oscillating superflow of 150 mm$^{-1}$ at 200~Hz.
[H\"anninen, Tsubota and Vinen: Phys. Rev. B \textbf{75} (2007) 064502, reproduced with permission. Copyright 2007 the American Physical Society.]} \label{Hanninen}\end{figure}

These experimental studies reported some common behavior independent of the details of the structures, such as the type, shape, and surface roughness. 
The observed critical velocities are in the range from 1 mm/s to approximately 200 mm/s.  
Since the velocity is usually much lower than the Landau critical velocity of approximately 50 m/s, the transition to turbulence should come not from intrinsic nucleation of vortices but from the extension or amplification of remnant vortices. 
Such behavior is shown in the numerical simulation by the vortex filament model \cite{Hanninen:2007}.  
Figure \ref{Hanninen} shows how the remnant vortices that are initially attached to a sphere develop into turbulence under an oscillating flow. 
Such behavior must be related to the essence of the observations.

\begin{figure}
	\begin{center}
		\begin{tabular}{cc}		
			\includegraphics[width=0.4\linewidth]{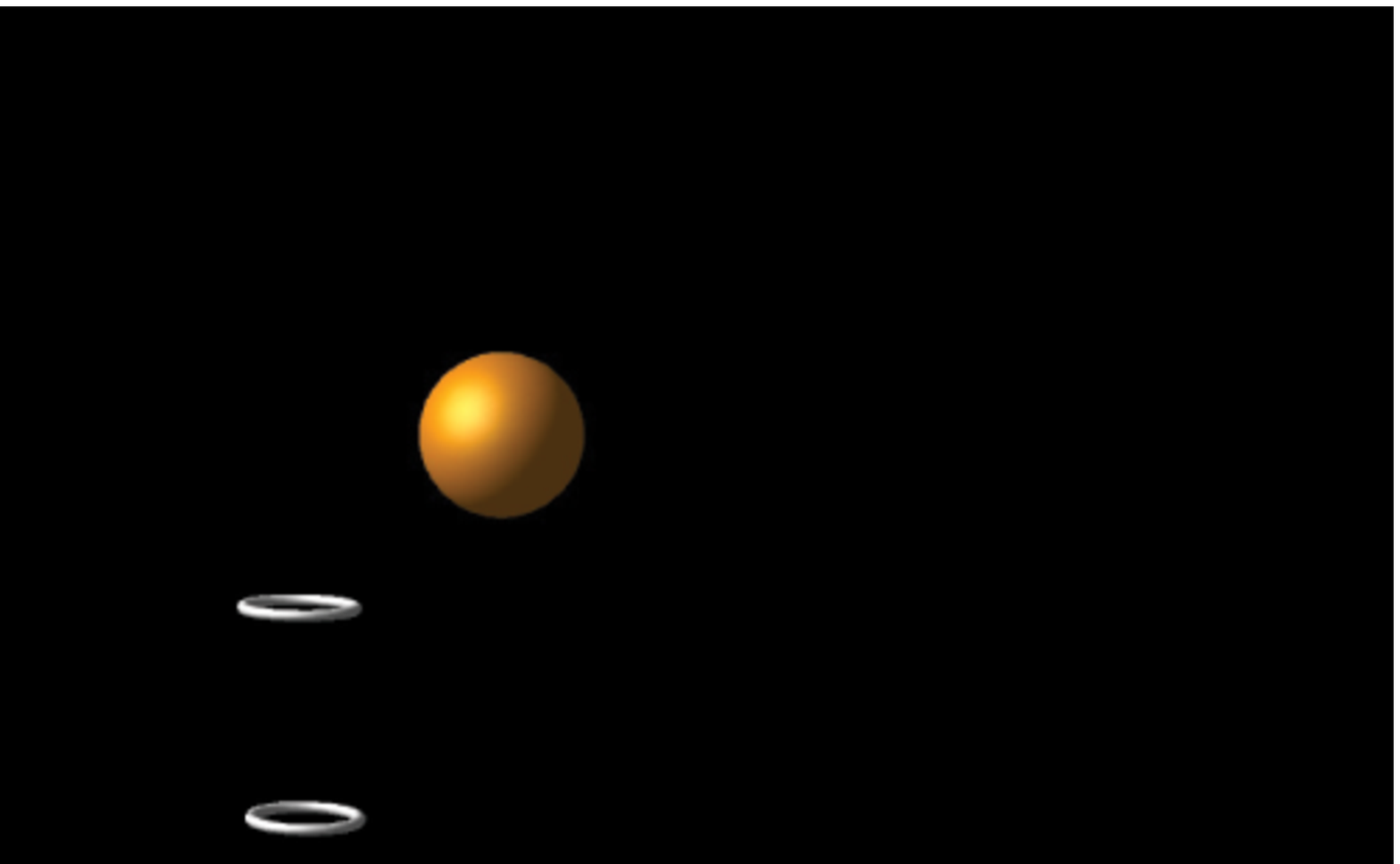}	&
			\includegraphics[width=0.4\linewidth]{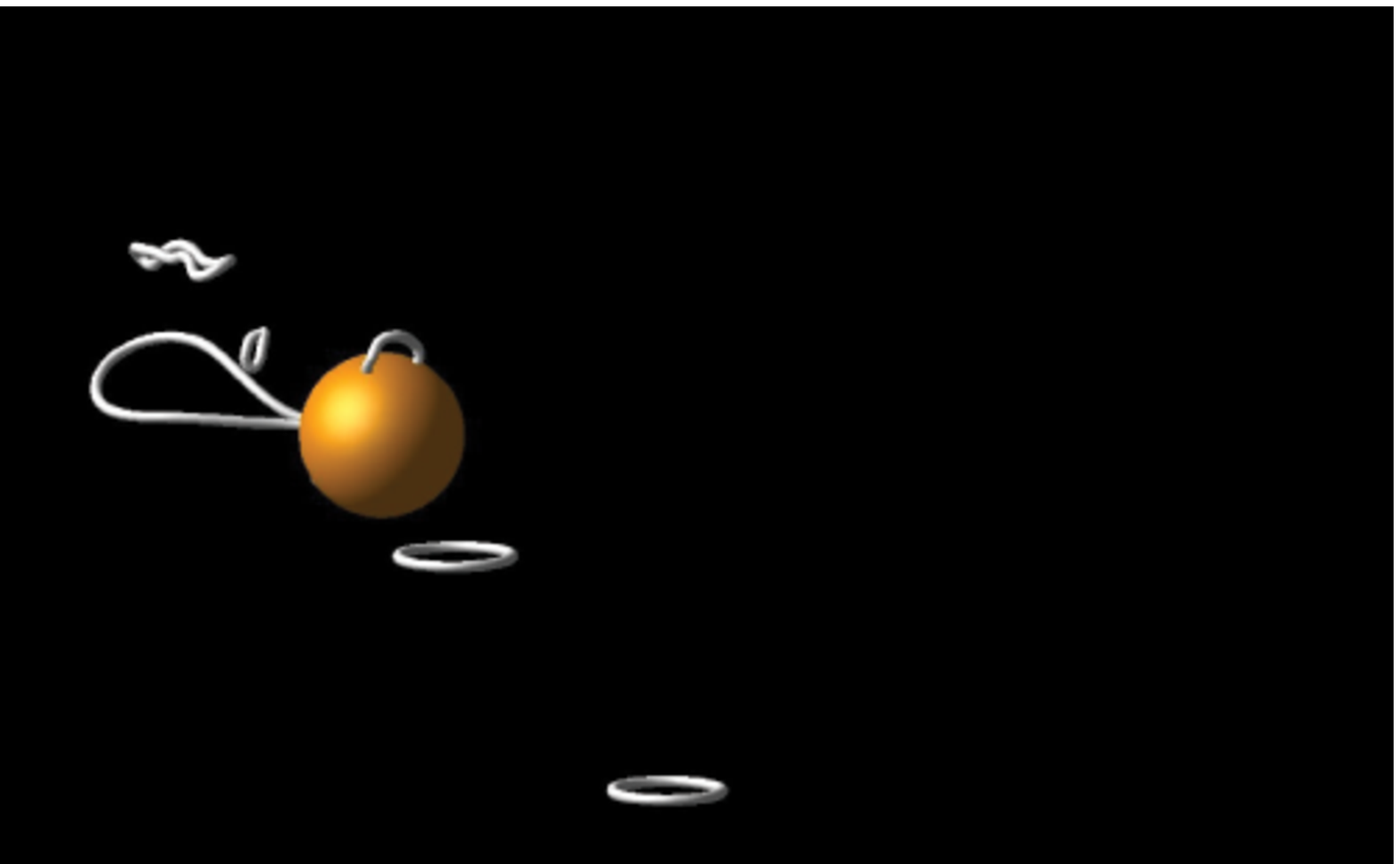}	\\
			(a) $t=19$ ms	&	(b) $t=40$ ms	\\
			\includegraphics[width=0.4\linewidth]{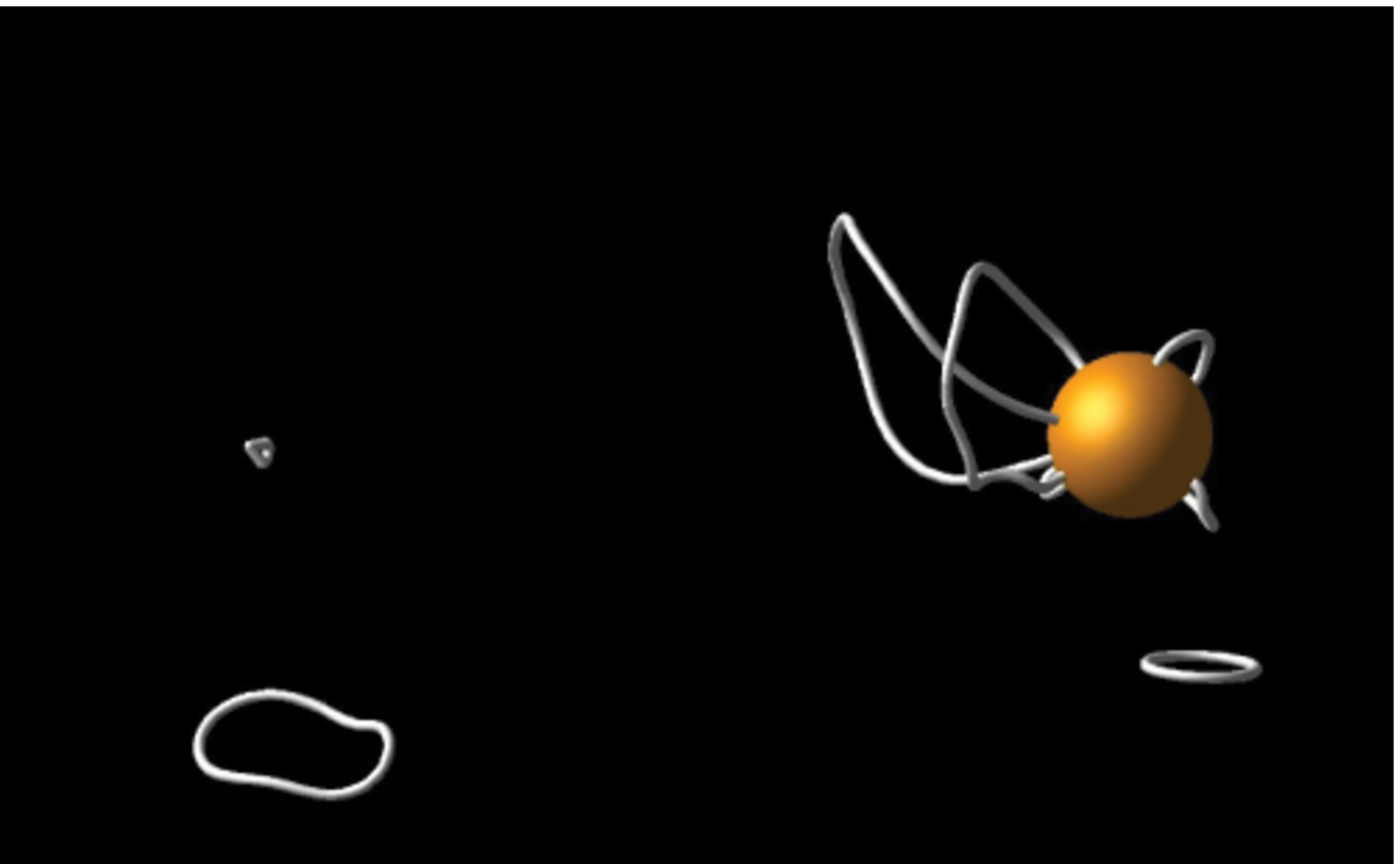}	&
			\includegraphics[width=0.4\linewidth]{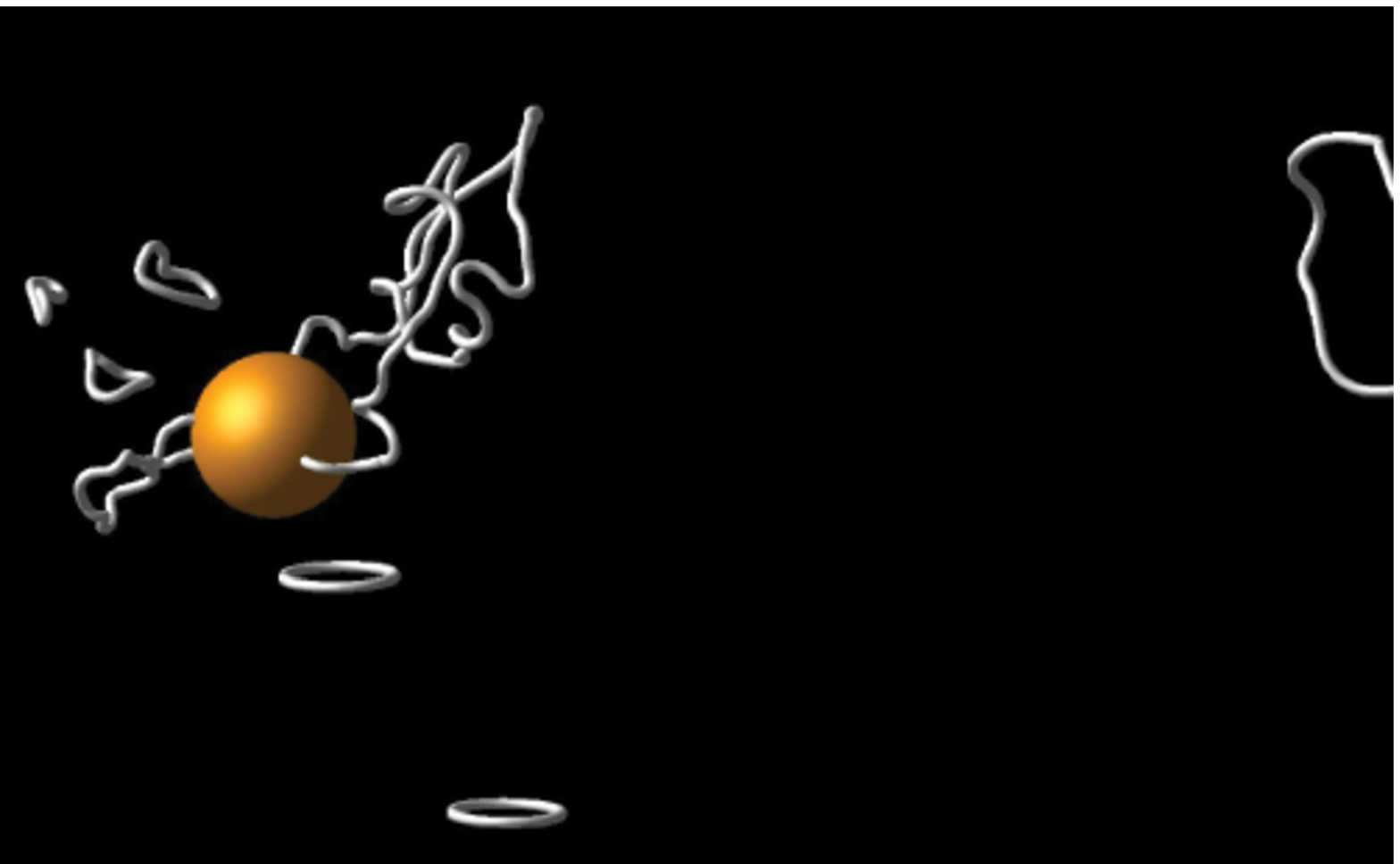}	\\
			(c) $t=58$ ms	&	(d) $t=100$ ms
		\end{tabular}
		\caption{Time evolution of turbulence generation for the case of a sphere oscillating with a velocity magnitude of 90 mm/s. See the text for details. [Fujiyama and Tsubota: Phys. Rev. B \textbf{79} (2009) 094513, reproduced with permission. Copyright 2009, the American Physical Society.]}
		\label{Fujiyama}
	\end{center}
\end{figure}

Generally it is not easy to control the remnant vortices in an actual experimental setup.
However, Goto {\it et al.} succeeded in preparing a vibrating wire free from remnant vortices \cite{Goto:2008}.
This wire never causes a transition to turbulence by itself; it can cause turbulence only when it receives seed vortices from another wire.
Such a simulation was performed by Fujiyama {\it et al.} as shown in Fig. \ref{Fujiyama}  \cite{Fujiyama:2009}.
The sphere oscillates horizontally; the diameter of the sphere is 3 $\mu$m, the frequency of the oscillation is 1590 Hz, while the oscillation velocity is chosen in the range of 30--90 mm/s.
Vortex rings of radius 1 $\mu$m are injected from the bottom of the medium [Fig. \ref{Fujiyama}(a)].
When the vortex rings collide with the sphere, reconnections occur and the vortices become attached to the sphere.
Then, the attached vortices are stretched as the sphere moves [Fig. \ref{Fujiyama}(b) and (c)].
Due to the successive injection of vortex rings the process is repeated and the stretched vortices form a tangle around the sphere  [Fig. \ref{Fujiyama}(d)].
The vortices grow in size and some then detach from the sphere.
In spite of the detachment, the oscillating sphere still sustains the vortex tangle when its velocity is relatively large. 
The vortex line length in a finite volume including the sphere was calculated at different oscillation velocities as a function of elapsed time.
The loss of the vortices escaping from the volume balances the injection and the growth of the vortices so that the line length saturates.
Only a slight increase in the line length can be observed  for a velocity magnitude of 30 mm/s, which means that the vortices are not stretched by the sphere.
The saturated value of the line length increases with the oscillation velocity magnitude.
For velocity magnitudes above 50 mm/s, the saturated line length value is much larger than the injection of vortices, which suggests that vortex tangles are formed around the sphere.
This behavior is qualitatively consistent with the observations \cite{Goto:2008}.

In order to characterize the transition to turbulence, Fujiyama {\it et al.} also studied the drag force \cite{Fujiyama:2009}.
The drag force acting on an object in a uniform flow is generally represented by
\begin{equation} F_D=\frac12 C_D \rho A U^2, \label{Drag} \end{equation}
where $C_D$ is the drag coefficient, $\rho$ is the fluid density, $A$ is the projection area of the object normal to the flow, and $U$ is the flow velocity.
It is known in classical fluid mechanics how $C_D$ depends on the properties of the flow.
At low Reynolds number, Stokes's drag force acts on the object, which is proportional to the magnitude of $U$, with the result that $C_D$ is inversely proportional to $U$.
When the flow becomes turbulent at high Reynolds number, $C_D$ is of order unity. 
Fujiyama {\it et al.} estimated the drag force for the cases such as those in Fig. \ref{Fujiyama}.
The amplified line length can be related to the increase in energy, which should be equivalent to the work by the sphere. 
The drag coefficient $C_D$ obtained by these considerations was of order unity.
Thus we could confirm an analogy between CT and QT for this problem too.

\begin{figure}
	\begin{center}
		\begin{tabular}{cc}		
			\includegraphics[width=0.5\linewidth]{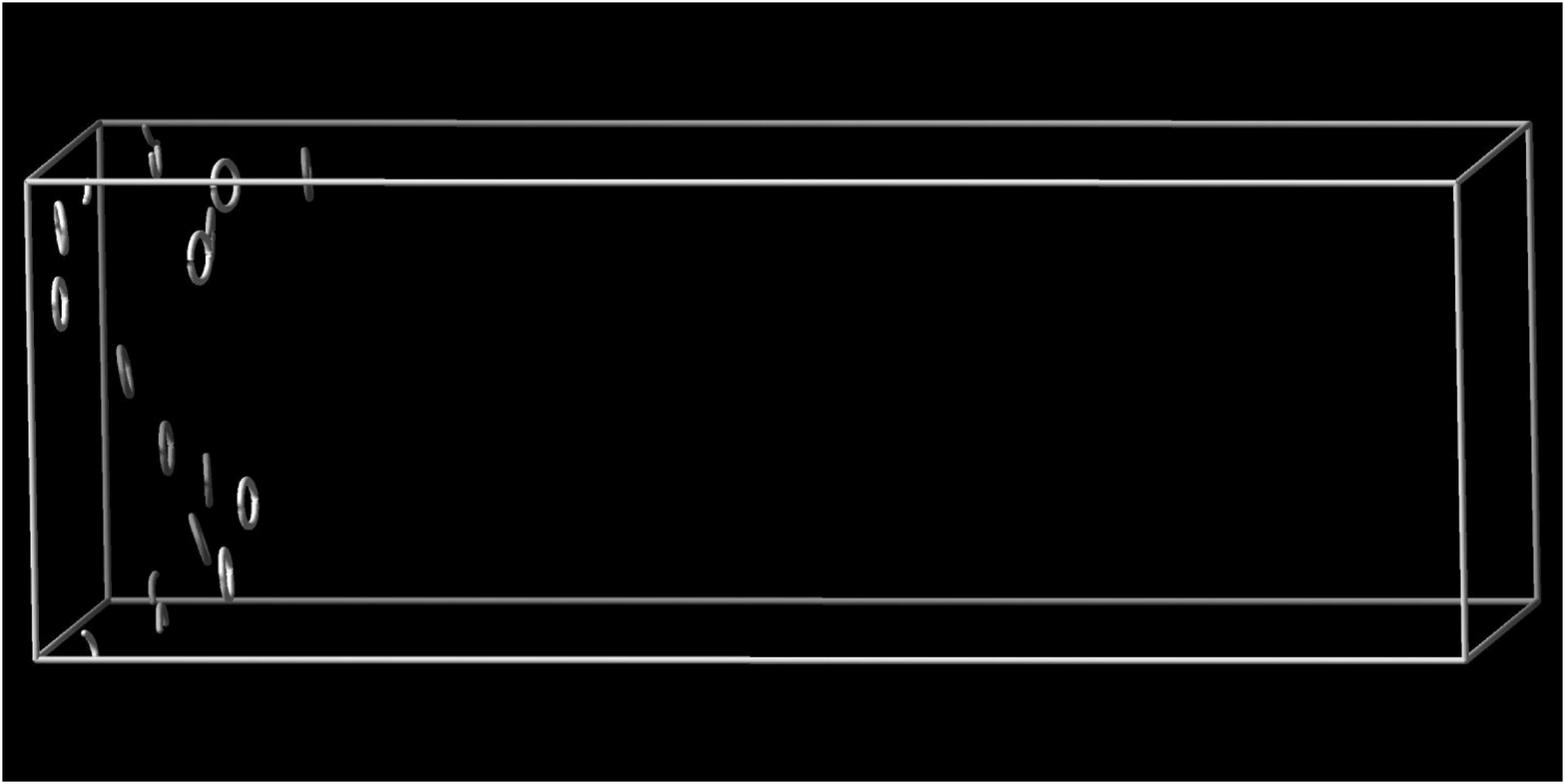}	&
			\includegraphics[width=0.5\linewidth]{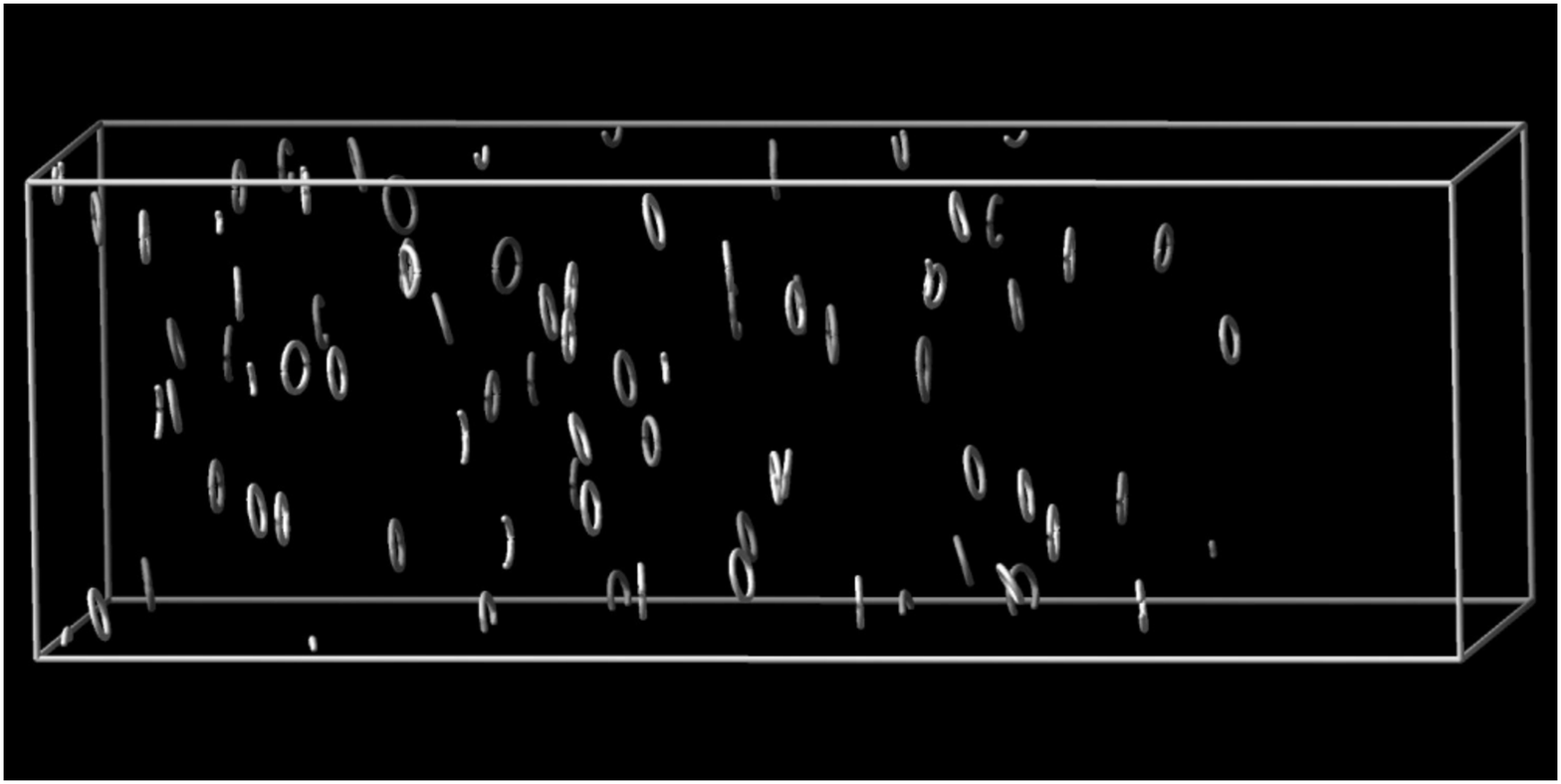}	\\
			(a) $t=20$ ms	&	(b) $t=100$ ms	\\
			\includegraphics[width=0.5\linewidth]{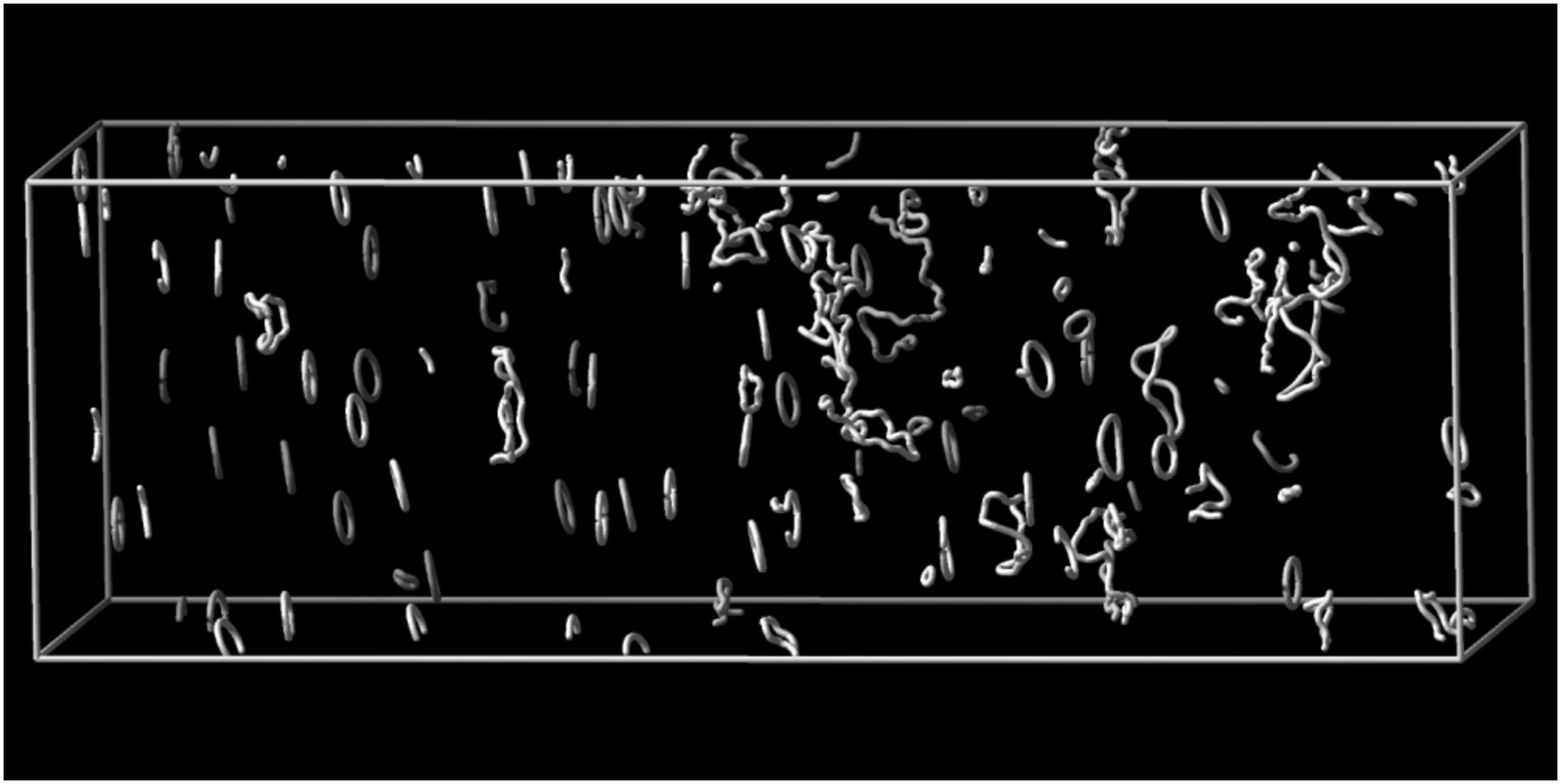}	&
			\includegraphics[width=0.5\linewidth]{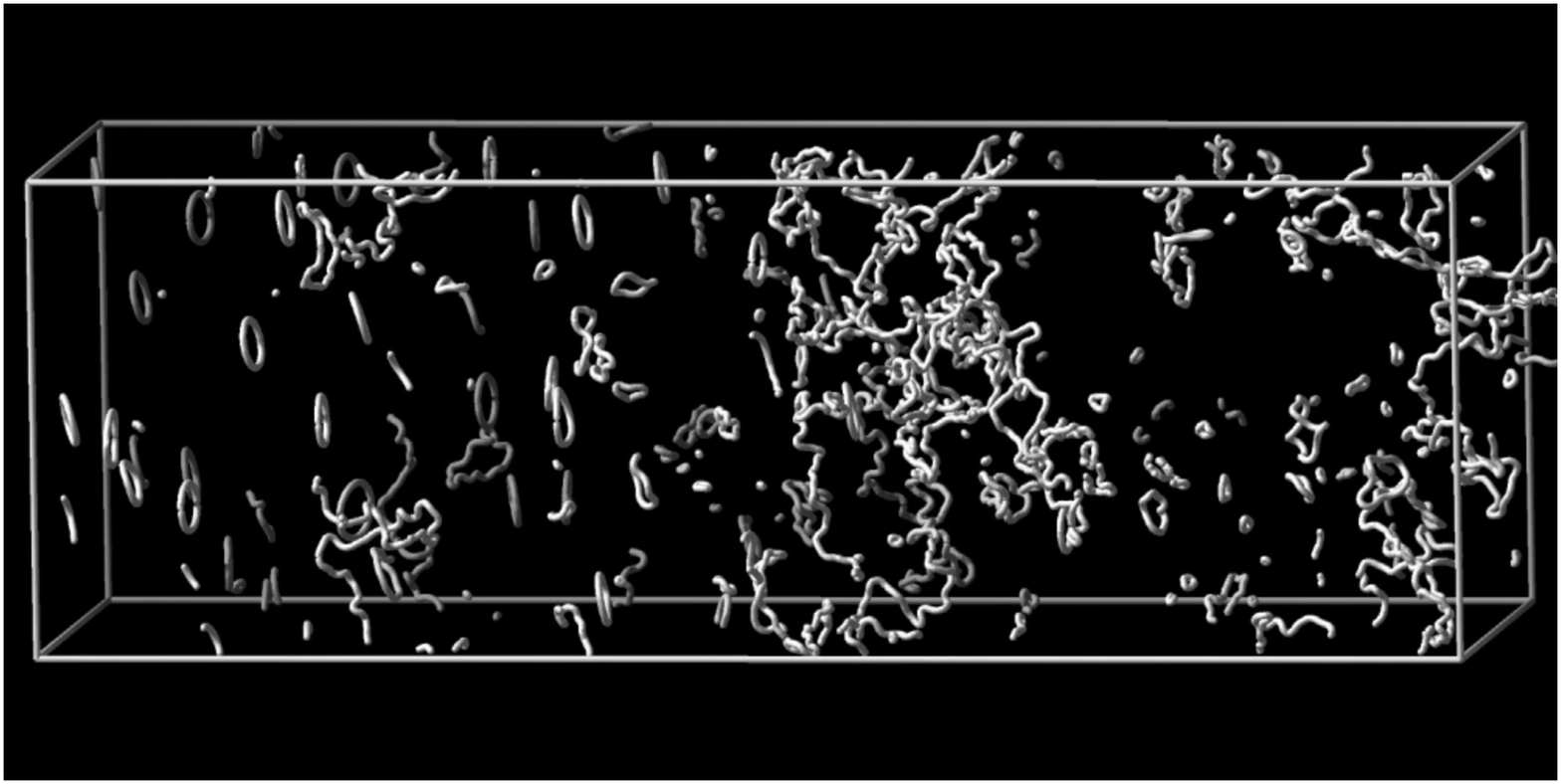}	\\
			(c) $t=300$ ms	&	(d) $t=500$ ms
		\end{tabular}
		\caption{Simulation of QT formation. Each frame shows the vortex configuration at the labeled time. Rings injected from the left quickly collide and recombine to produce a vortex tangle. See the text for details. }
		\label{Lancaster}
	\end{center}
\end{figure}

Another important simulation was performed for the current problem.
Bradley {\it et al.} studied experimentally the transition to QT in the B phase of superfluid $^3$He \cite{Vollhardt:1990} by vibrating a grid \cite{Bradley:2005b}. 
In superfluid $^3$He-B they set up a grid, 5.1 $\times$ 2.8 mm, composed of $\sim$10 $\mu$m square cross section wires 50 $\mu$m apart.
Directly in front of the grid were two vibrating wires, which could observe the vortices coming from the grid.
The observed behavior showed two distinct regimes.
At low grid velocities (below 3.5 mm/s) the two wires caught only vortex rings coming ballistically from the grid.
At high grid velocities, however,  the observation of two wires showed a signature of QT; the grid produced a vortex tangle.
Such behavior was confirmed through a simulation by Fujiyama {\it et al.} \cite{Fujiyama:2010}.
They followed the dynamics of vortex rings injected into a simulation "cell" such that the left-hand side of the cell represents the face of the grid.
The simulation cell was a box of cross section 200 $\mu$m $\times$ 200 $\mu$m and length 600 $\mu$m, as shown in Fig. \ref{Lancaster}.
In the transverse directions the cell had periodic boundary conditions.
Vortex rings of diameter 20 $\mu$m were injected at the left-hand side of the cell at a regular time interval $\tau_i$ but at random positions and random angles within a $\sim$20 deg. cone around the forward direction.
The rings traveled at a self-induced velocity of 4.6 mm/s.
At a low injection rate ($\tau_i=5$ms) the simulation confirms that the rings travel essentially independently.
At higher ring injection rates, however, corresponding to higher grid velocities, they found a very different behavior, as shown in Fig. \ref{Lancaster} for $\tau_i=1.5$ ms.
Here the rings immediately start to collide and reconnect, establishing a vortex tangle, which corresponds to the behavior at high grid velocities in Ref. \cite{Bradley:2005b}.

\subsection{Gross--Pitaevskii model} \label{subsec-GP}

In a weakly interacting Bose system, the macroscopic wave function $\Psi$ appears as the order parameter of Bose--Einstein condensation, obeying the Gross--Pitaevskii (GP) equation \cite{Gross:1961,Pitaevskii:1961}:
\begin{align}
i \hbar \frac{\partial \Psi}{\partial t} = \left( - \frac{\hbar^2 \nabla^2}{2M} + g \left| \Psi \right|^2 - \mu \right) \Psi. \label{eq-original-GP}
\end{align}
Here, $g = 4 \pi \hbar^2 a / M$ represents the coupling constant characterized by the $s$-wave scattering length $a$, $M$ is the particle mass, and $\mu$ is the chemical potential.
Expressing the order parameter with the amplitude and phase, {\it i.e.}, $\Psi = \sqrt{\rho} e^{i \phi}$, we obtain the condensate density $\rho$ and the superfluid velocity $\Vec{v}\sub{s} = (\hbar / M) \nabla \phi$.
The vorticity $\nabla \times \Vec{v}\sub{s}$ defined from the superfluid velocity vanishes everywhere in a singly-connected region of the order parameter, and all rotational flow is carried only by quantized vortices.
The quantized vortex is defined as a topological excitation in which $\rho$ vanishes at the core and $\phi$ rotates by $2 \pi$ around the core.
The only characteristic length scale of the GP equation is the healing length defined by $\xi = \hbar / \sqrt{2 M g \bar{\rho}}$ with mean condensate density $\bar{\rho}$; the vortex core size is given by $\xi$.

The GP model can explain not only the dynamics of vortex lines but also vortex core phenomena such as reconnection and nucleation of vortices.
However, strictly speaking, the GP equation is not applicable to superfluid helium, which is not a weakly interacting Bose system.
The GP equation is well applicable to dilute atomic BECs \cite{Pethick:2008, Pitaevskii:2003}

\subsubsection{Hydrodynamic properties and dynamics of a single quantum vortex}

Before investigating turbulent properties in the GP model, we briefly overview several hydrodynamic properties of the GP equation.
With condensate density $\rho$ and superfluid velocity $\Vec{v}\sub{s}$, the GP equation \eqref{eq-original-GP} can be rewritten as follows:
\begin{align}
\frac{\partial \rho}{\partial t} + \nabla \cdot \left( \rho \Vec{v}\sub{s} \right) = 0, \label{eq-continuity}
\end{align}
\begin{align}
\frac{\partial \Vec{v}\sub{s}}{\partial t} + \frac{\nabla \Vec{v}\sub{s}^2}{2} = - \frac{\nabla}{M \rho} \left( \frac{g \rho^2}{2} \right) + \frac{\hbar^2 \nabla}{2 M^2} \left(\frac{\nabla^2 \sqrt{\rho}}{\sqrt{\rho}} \right). \label{eq-quantum-Euler}
\end{align}
Equations \eqref{eq-continuity} and \eqref{eq-quantum-Euler} express the equations of conservation of mass and momentum for a compressible inviscid fluid.
The first term in the right-hand-side of Eq. \eqref{eq-quantum-Euler} corresponds to an effective pressure $p = g \rho^2 / 2$ due to the nonlinearity of the GP equation.
The second term, the so called quantum pressure, has no analog in standard fluid mechanics, and becomes especially important at small scales comparable to the healing length such as near vortex cores, where $\rho$ rapidly changes in the scale of $\xi$.
Another big difference between QHD described by the GP model and perfect classical fluid hydrodynamics is the existence of quantized vortices.
In the GP model, any rotational flow is carried by quantized vortices with the quantized circulation:
\begin{align}
\kappa = \oint \Vec{v}\sub{s} \cdot d\Vec{\ell}= \frac{h}{M},
\end{align}
and it is well known that quantized vortices behave simply as vortex filaments in a perfect fluid.

To study QT, we introduce a dissipation into the GP equation \cite{Kobayashi:2005a,Kobayashi:2005b}.
In atomic BECs, the main origin of the dissipation is considered to be interaction between the condensate cloud and the thermal component.
In this section, we introduce the dissipation term in the following simple way \cite{2003KasamatsuPRA}:
We assume that the system is described by $\Psi$ such that energy and particles are exchanged with a particle reservoir.
The particle reservoir is a thermodynamic environment that lets the chemical potential of the system equal to that of the reservoir.
The interaction with the particle reservoir is provided by an imaginary term in the GP equation:
\begin{align}
i \hbar \frac{\partial \Psi_\mu}{\partial t} = \left( - \frac{\hbar^2 \nabla^2}{2M} + g \left| \Psi_\mu \right|^2 - i \Gamma \right) \Psi_\mu. \label{eq-GP-with-Gamma}
\end{align}
Here the wave function $\Psi_\mu$ includes the chemical potential $\mu$ through the gauge transformation $\Psi_\mu = \Psi e^{- i \mu t}$.
Since $\Gamma$ arises from the difference of the chemical potential, we can write $\Gamma = \gamma (\mu - \mu\sub{pr})$ with the chemical potential $\mu\sub{pr}$ of the particle reservoir.
We assume that the system is nearly in equilibrium with the particle reservoir and that $\Gamma$ is proportional to the difference in the chemical potentials.
Using the approximation
\begin{align}
i \hbar \frac{\partial \Psi_\mu}{\partial t} \simeq \mu \Psi_\mu,
\end{align}
and $\mu \simeq \mu\sub{pr}$, Eq. \eqref{eq-GP-with-Gamma} becomes
\begin{align}
\left( i - \gamma \right) \hbar \frac{\partial \Psi_\mu}{\partial t} = \left( - \frac{\hbar^2 \nabla^2}{2 M} + g \left| \Psi_\mu \right|^2 + i \mu \gamma \right) \Psi_\mu. \label{eq-GP-gamma-gamma}
\end{align}
Substituting $\Psi_\mu = \Psi e^{- i \mu t}$ into Eq. \eqref{eq-GP-gamma-gamma}, we finally obtain the modified GP equation with the effective dissipation $\gamma$:
\begin{align}
\left( i - \gamma \right) \hbar \frac{\partial \Psi}{\partial t} = \left( - \frac{\hbar^2 \nabla^2}{2 M} + g \left| \Psi \right|^2 - \mu \right) \Psi. \label{eq-dissipated-GP}
\end{align}
Introducing $\gamma$ conserves neither the energy nor the number of particles.
For studying the hydrodynamics, however, it is sometimes realistic to assume that the number of particles is conserved.
Hence, we can introduce the time dependence of the chemical potential so that the total number of particles $N = \int d \Vec{r} \: |\Psi|^2$ is conserved.

When $\gamma \ll 1$, Eq. \eqref{eq-dissipated-GP} becomes approximately
\begin{align}
i \hbar \frac{\partial \Psi}{\partial t} = (1 - i \gamma) \left( - \frac{\hbar^2 \nabla^2}{2 M} + g \left| \Psi \right|^2 - \mu \right) \Psi, \label{eq-complex-Ginzburg-Landau}
\end{align}
which is widely known as complex-Ginzburg--Landau equation for the case when $\gamma$ is uniform \cite{Aranson:2002}.
By using the density $\rho$ and the superfluid velocity $\Vec{v}\sub{s}$, Eq. \eqref{eq-complex-Ginzburg-Landau} can be rewritten as
\begin{align}
\frac{\partial \rho}{\partial t} + \nabla \cdot (\rho \Vec{v}\sub{s}) = \gamma \left\{ \frac{\hbar \sqrt{\rho} \nabla^2 \sqrt{\rho}}{M} - \frac{M \rho \Vec{v}\sub{s}^2}{\hbar} - \frac{2 (g \rho - mu) \rho}{\hbar} \right\},
\end{align}
\begin{align}
\frac{\partial \Vec{v}\sub{s}}{\partial t} + \frac{\nabla \Vec{v}\sub{s}^2}{2} = - \frac{\nabla}{M \rho} \left( \frac{g \rho^2}{2} \right) + \frac{\hbar^2 \nabla}{2 M^2} \left(\frac{\nabla^2 \sqrt{\rho}}{\sqrt{\rho}} \right) + \frac{\hbar \nabla}{2 M} \left\{ \frac{\gamma}{\rho} \nabla \cdot (\rho \Vec{v}\sub{s}) \right\}. \label{eq-quantum-Navier-Stokes}
\end{align}
The third term in Eq. \eqref{eq-quantum-Navier-Stokes} works as a viscous term with the kinematic viscosity $\hbar \gamma / (2 M)$.
We can therefore define the effective Reynolds number
\begin{align}
R\sub{QT} = \frac{M \bar{v}\sub{s} D}{\hbar \bar{\gamma}},
\end{align}
when we consider QT in the GP model.
Here $\bar{v}\sub{s}$ and $\bar{\gamma}$ are the averaged values of $\Vec{v}\sub{s}$ and $\gamma$ over the space defined as
\begin{align}
\bar{v}\sub{s} = \frac{\dps \int d\Vec{r} \: |\rho \Vec{v}\sub{s}|}{\dps \int d\Vec{r} \: \rho}, \quad
\bar{\gamma} = \frac{\dps \int d\Vec{r} \: \rho \gamma}{\dps \int d\Vec{r} \: \rho},
\end{align}
and $D$ is the system size.

Next, we investigate the dynamics of a single vortex line \cite{Kobayashi:2005a}.
The steady solution of the GP equation with a single vortex line along the $z$ axis is represented by
\begin{align}
\Psi(\Vec{r}) = \sqrt{\rho(r)} e^{\pm i \varphi}, \label{eq-straight-vortex}
\end{align}
where $r = \sqrt{x^2 + y^2}$ and $\varphi = \tan^{-1}(y/ x)$ are the radius and angle in cylindrical coordinates.
$\rho(r)$ follows the equation
\begin{align}
\frac{\hbar^2}{2 M} \left( \frac{d^2 \sqrt{\rho}}{d r^2} + \frac{1}{r} \frac{d \sqrt{\rho}}{d r} - \frac{\sqrt{\rho}}{r^2} \right) - g \sqrt{\rho^3} + \mu \sqrt{\rho} = 0 \label{eq-straight-vortex-solution}.
\end{align}
Starting from the solution of Eq. \eqref{eq-straight-vortex-solution}, we calculate the GP equation under an external flow $\Vec{v}\sub{e}$ as
\begin{align}
\left( i - \gamma \right) \hbar \frac{\partial}{\partial t} \Psi = \left( - \frac{\hbar^2}{2 M} \nabla^2 + g \left| \Psi \right|^2 - \mu - i \hbar \Vec{v}\sub{e} \cdot \nabla \right) \Psi,
\end{align}
and obtain the time development of the density $\rho$:
\begin{align}
\frac{\partial \sqrt{\rho}}{\partial t} = - v_{\mathrm{e}r} \frac{d \sqrt{\rho}}{d r} \mp \gamma v_{\mathrm{e}\varphi} + \gamma^2 v_{\mathrm{e}r} \frac{d \sqrt{\rho}}{d r}, \label{eq-dissipated-density-movement}
\end{align}
which is second order in $\gamma$.
Here $\Vec{v}\sub{e} = v_{\mathrm{e}r} \hat{\Vec{r}} + v_{\mathrm{e}\varphi} \hat{\Vec{\varphi}}$.
The vortex moves from the positive to negative sides of $\partial \sqrt{\rho} / \partial t$ in Eq. \eqref{eq-dissipated-density-movement}.
Assuming $\sqrt{\rho} \propto r$ around the core, we obtain the time development of the vortex position $\Vec{r}_0$:
\begin{align}
\frac{d \Vec{r}_0}{d t} = \Vec{v}\sub{e} \pm \gamma \Vec{v}\sub{e} \times \hat{\Vec{z}} - \gamma^2 \Vec{v}\sub{e} \label{eq-dissipated-vortex-movement}
\end{align}
The first term states that the vortex moves with the velocity field $\Vec{v}\sub{e}$ \cite{Fetter:1966}.
The second and third terms describe the drag forces perpendicular and parallel to $\Vec{v}\sub{e}$ \cite{Peach:1950,Kawasaki:1984}.

Recognizing $\Vec{v}\sub{e}$ in Eq. \eqref{eq-dissipated-vortex-movement} as the velocity field induced by other vortices, we can derive the effective vortex dynamics for a system with many vortices.
In that case, the second and third terms work as the mutual friction forces by comparing Eq. \eqref{eq-dissipated-vortex-movement} with Eq. \eqref{Schwarz} of the vortex-filament model \cite{Hall:1956a,Hall:1956b,Schwarz:1985}, with the corresponding coefficients $\alpha = \gamma$ and $\alpha^\prime = \gamma^2$.

\begin{figure}[htb]
\centering
\includegraphics[width=0.9\linewidth]{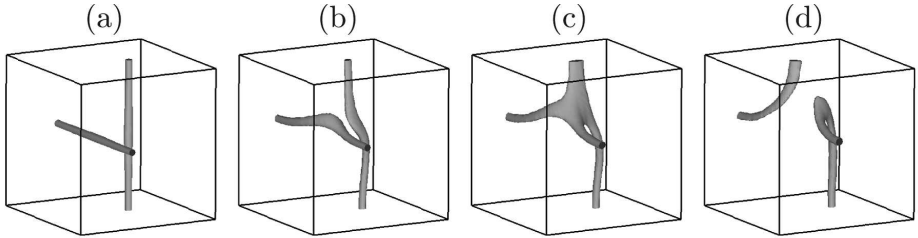}
\caption{\label{fig-reconnection} Reconnection of two vortices starting from a skewed position of two straight vortex lines. Simulation of the GP model of Eq. \eqref{eq-original-GP}. (a) Initial state. (b) State just before the two vortices connect. (c) Connection of two vortex lines. (d) State after the vortex lines separate in a newly connected arrangement, also called reconnection. The contours in all the figures show the point of low density (10\% of maximum density). [Kobayashi and Tsubota: J. Phys. Soc. Jpn. {\bf 74} (2005) 3248, reproduced with permission. Copyright 2005 the Physical Society of Japan.]}
\end{figure}

Furthermore, there is another important vortex dynamics which is not directly described in Eq. \eqref{eq-dissipated-vortex-movement}: the reconnection of two vortices \cite{Koplik:1993,Koplik:1996,Leadbeater:2001,Ogawa:2002}.
When two vortices are close to each other, they approach and become locally anti-parallel, and then reconnection occurs.
Figure \ref{fig-reconnection} shows an example of vortex reconnection given by the numerical calculation of Eq. \eqref{eq-original-GP} starting from two straight vortex lines in a skewed orientation.
Reconnection occurs even for the original GP equation \eqref{eq-original-GP} without dissipation.
We show that this process does not violate Kelvin's circulation theorem; moreover, the process becomes irreversible due to the emission of compressible excitations that have a wavelength smaller than $\xi$ \cite{Ogawa:2002}.

\subsubsection{Quantum turbulence at zero temperature} \label{subsubsec-turbulence-zero-temperature}

In this section, we overview our study of QT at zero temperature through the analysis of the GP equation.
QT is defined as the turbulent state of a quantum fluid with highly tangled quantized vortices, and the term is often used to emphasize that the state is dominated by the behavior of quantized vortices at very low temperatures at which the thermal effect is negligible \cite{Halperin:2009}.

For the analysis of turbulence, the energy spectrum is one of the most important statistical quantities.
The energy spectrum can be obtained from the Fourier transformation of the equal-time two-point velocity correlation function:
\begin{align}
F(\Vec{k},t) = \frac{1}{2} \int d\Vec{r} \: e^{i \Vec{k} \cdot \Vec{r}} \int d\Vec{r}^\prime \: \left\langle \Vec{v}(\Vec{r}^\prime,t) \cdot \Vec{v}(\Vec{r} + \Vec{r}^\prime,t) \right\rangle.
\end{align}
Here $\Vec{k}$ is the wavenumber from the Fourier transformation and $\langle \cdots \rangle$ shows the ensemble average over statistically equivalent states.
When studying the energy spectrum of QT, $\Vec{v}$ is regarded as the superfluid velocity $\Vec{v}\sub{s} = (\hbar / M) \nabla \phi$.
The integration of $F(\Vec{k},t)$ over the angle in wavenumber space is defined as the energy spectrum:
\begin{align}
E(k,t) = \frac{1}{(2 \pi)^3} \int d\varphi_{\Vec{k}} \: d\theta_{\Vec{k}} \: k^2 \sin\theta_{\Vec{k}} F(\Vec{k},t), \label{eq-energy-spectrum-definition}
\end{align}
where $\theta_{\Vec{k}}$ and $\varphi_{\Vec{k}}$ refer to the polar and the azimuthal angles in wave number space.
The energy spectrum holds the following relation with the spatial integration of the kinetic energy:
\begin{align}
\int dk\: E(k,t) = \frac{1}{2 (2 \pi)^3} \int d\Vec{k} \: \left\langle |\tilde{\Vec{v}}|^2 \right\rangle = \frac{1}{2} \int d\Vec{r} \: \left\langle |\Vec{v}|^2 \right\rangle = E(t).
\end{align}
Here $E(t)$ is the total kinetic energy per unit mass and $\tilde{\Vec{v}}$ is the Fourier transformation of $\Vec{v}$.

For turbulence of a typical viscous fluid, referred to as classical turbulence (CT) in this section, Kolmogorov proposed a statistically steady state of fully developed turbulence \cite{Kolmogorov:1941a,Kolmogorov:1941b} in which energy is injected into the fluid at scales comparable to the system size $D$ in the energy-containing range.
In the inertial range, this energy is transferred to smaller scales without being dissipated, supporting a statistical law for the energy spectrum known as the Kolmogorov law:
\begin{align}
E(k) = C \varepsilon^{2/3} k^{-5/3}. \label{eq-Kolmogorov-energy-spectrum}
\end{align}
The energy transferred to smaller scales in the energy-dissipated range is eventually dissipated at the Kolmogorov wavenumber $k\sub{K} = (\varepsilon / \nu)^{1/4}$ through the kinematic viscosity $\nu$ of the fluid, at the dissipation rate $\varepsilon$.
The Kolmogorov constant $C$ is a dimensionless parameter of order unity.

The inertial range is thought to be sustained by the self-similar Richardson cascade in which large eddies are broken up into smaller ones \cite{Richardson:2007}.
In CT, however, the Richardson cascade is not completely understood because it is impossible to definitively identify each individual eddy.
In contrast, quantized vortices in QT are definite and stable topological defects.
Therefore, QT gives the real Richardson cascade of definite quantized vortices, and thus is an ideal prototype for studying statistics such as the Kolmogorov law and the Richardson cascade in the inertial range of turbulence.
Quantized vortices at finite temperatures can decay through mutual friction with the normal fluid at any scale.
At very low temperatures, on the other hand, vortices can decay by the emission of compressible excitations and Kelvin waves through vortex reconnections.
Therefore, dissipation occurs only at small scales; for large scales, we can obtain the turbulent state at high Reynolds number.

We now consider the energy spectrum of the GP equation.
The total energy of the GP equation per unit mass is
\begin{align}
E = \frac{1}{M N} \int d\Vec{r} \: \left( \frac{\hbar^2}{2 M} |\nabla \Psi|^2 + \frac{g}{2} |\Psi|^4 \right). \label{eq-GP-energy}
\end{align}
Here $N = \int d\Vec{r} \: |\Psi|^2$ is the total number of particles.
$E$ can be separated into the gradient energy $E\sub{grad}$ and the interaction energy $E\sub{int}$.
$E\sub{grad}$ is further separated into the kinetic energy $E\sub{kin}$ and the quantum energy $E\sub{q}$ as in the following:
\begin{align}
\begin{array}{c}
\dps E\sub{grad} = E\sub{kin} + E\sub{q} = \frac{\hbar^2}{2 M^2 N} \int d\Vec{r} \: |\nabla \Psi|^2, \quad
E\sub{int} = \frac{g}{2 M N} \int d\Vec{r} \: |\Psi|^4, \arrayret
\dps E\sub{kin} = \frac{1}{2 N} \int d\Vec{r} \:  \Vec{p}^2, \quad
E\sub{q} = \frac{\hbar^2}{2 M^2 N} \int d\Vec{r} \: \left( \nabla \sqrt{\rho} \right)^2, \label{eq-GP-various-energy}
\end{array}
\end{align}
with $\Vec{p} = \sqrt{\rho} \Vec{v}\sub{s}$.
$E\sub{kin}$ can be further divided into a compressible part $E\sub{kin}\up{c}$ due to compressible excitations and an incompressible part $E\sub{kin}\up{i}$ due to quantized vortices \cite{Nore:1997a,Nore:1997b}:
\begin{align}
E\sub{kin}\up{c,i} = \frac{1}{2 N} \int d\Vec{r} \: \left( \left[ \Vec{p} \right]\up{c,i} \right)^2. \label{eq-GP-kinetic-energy}
\end{align}
Here $[\cdots]\up{c}$ denotes the compressible part, {\it i.e.}, $\nabla \times [\cdots]\up{c} = 0$, and $[\cdots]\up{i}$ denotes the incompressible part, {\it i.e.}, $\nabla \cdot [\cdots]\up{c} = 0$.
The compressible part $\Vec{A}\up{c}$ and the incompressible part$\Vec{A}\up{i}$ of an arbitrary vector field $\Vec{A}$ are given by
\begin{align}
\Vec{A}\up{c} = \sum_{\Vec{k}} \frac{\Vec{k} \cdot \tilde{\Vec{A}}}{k^2} k e^{i \Vec{k} \cdot \Vec{r}}, \quad
\Vec{A}\up{i} = \Vec{A} - \Vec{A}\up{c},
\end{align}
where $\tilde{\Vec{A}}$ is the Fourier component of $\Vec{A}$.
Corresponding to each energy, there are several kinds of energy spectra.
The most important is the energy spectrum of the incompressible kinetic energy:
\begin{align}
E\sub{kin}\up{i}(k,t) = \frac{1}{2 (2 \pi)^3 N} \int d\varphi_{\Vec{k}} \: d\theta_{\Vec{k}} \: k^2 \sin\theta_{\Vec{k}} \left( \tilde{\Vec{p}}\up{i} \right)^2,
\end{align}
because it should obey the Kolmogorov law with the Richardson cascade of quantized vortices.
Here $\tilde{\Vec{p}}\up{i}$ is the Fourier transformation of the incompressible momentum density $\left[\Vec{p} \right]\up{i}$.

Starting from a Taylor--Green vortex, Nore {\it et al.} simulated decaying turbulence by numerically solving the GP equation \eqref{eq-original-GP} \cite{Nore:1997a,Nore:1997b}.
After some time, the initial vortices became tangled and the calculated energy spectrum $E\sub{kin}\up{i}$ obeyed the power-law behavior $E\sub{kin}\up{i}(k,t) \propto k^{-\eta}$.
When vortices formed a tangle, the exponent $\eta$ was about $5/3$, but this value did not hold for long because the turbulence was decaying with the conservation of total energy $E$ in Eq. \eqref{eq-GP-energy}.
Because of energy conservation, the energy of the vortices $E\sub{kin}\up{i}$ was transferred to compressible excitations $E\sub{kin}\up{c}$ with wavelength comparable to $\xi$ through repeated reconnections.
Therefore, the dynamics of QT are affected by many compressible excitations and we cannot see the proper dynamics of quantized vortices, which causes $E\sub{kin}\up{i}(k,t)$ to deviate from the Kolmogorov law.

To obtain QT free from compressible excitations, we use the modified GP equation \eqref{eq-dissipated-GP} discussed in the previous section \cite{Kobayashi:2005a,Kobayashi:2005b}.
The space-independent constant $\gamma$, however, acts on vortices as the mutual friction at finite temperatures, as discussed in Eq. \eqref{eq-dissipated-vortex-movement}, and is inappropriate to study QT at zero temperature.
Thus, we now consider the space dependence of $\gamma$ as follows.
The Fourier transformation of Eq. \eqref{eq-dissipated-GP} is
\begin{align}
(i - \tilde{\gamma}) \hbar \frac{\partial \tilde{\Psi}}{\partial t} = \left( \frac{\hbar^2 k^2}{2 M} - \mu \right) \tilde{\Psi} + g \tilde{Y}. \label{eq-dissipated-GP-Fourier}
\end{align}
Here $\tilde{\gamma}$ and $\tilde{Y}$ are defined to satisfy the relations:
\begin{align}
\int d\Vec{r} \: e^{- i \Vec{k} \cdot \Vec{r}} \gamma \frac{\partial \Psi}{\partial t} = \tilde{\gamma} \frac{\partial \tilde{\Psi}}{\partial t}, \quad
\int d\Vec{r} \: e^{- i \Vec{k} \cdot \Vec{r}} |\Psi|^2 \Psi = \tilde{Y}, \label{eq-Fourier-gamma-interaction-definition}
\end{align}
where both depend on $\Psi$.
Especially, $\tilde{\gamma}(\Vec{k})$ directly acts on and dissipates the $\Vec{k}$-component of $\tilde{\Psi}$.
When we choose $\tilde{\gamma}$ as the step-function form:
\begin{align}
\tilde{\gamma} = \gamma_0 \theta(k - 2 \pi / \xi), \label{eq-step-function-dissipation}
\end{align}
we can expect that only compressible short-wavelength excitations produced via reconnections are dissipated.
Because the system is dissipationless at scales exceeding $\xi$, we can investigate proper vortex dynamics at zero temperature in large scales without dissipation.
We investigated the vortex dynamics in the modified GP equation \eqref{eq-dissipated-GP-Fourier} with the step function type dissipation \eqref{eq-step-function-dissipation} and found that this kind of dissipation does not work as the mutual friction as described in Eq. \eqref{eq-dissipated-vortex-movement} and just removes the short-wavelength compressible excitations.

\begin{figure}[htb]
\centering
\includegraphics[width=0.4\linewidth]{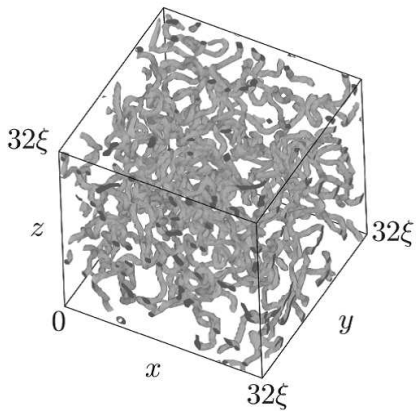}
\caption{\label{fig-turbulence-snapshot} Snapshot of a vortex configuration in the turbulent state.
The simulation box was set to $(32 \xi)^3$. Visualization of quantized vortices can be done using the following method.
For four numerical grid points $( x \; y \; z )$, $( x + \Delta x \; y \; z )$, $( x + \Delta x \; y + \Delta y \; z )$, and $( x \; y + \Delta y \; z )$, we can calculate the phase shift.
This value takes the values $0$, $2 \pi$, or $- 2 \pi$, with the two latter values indicating that a vortex line pierces this plaquette.
Therefore, by calculating the phase shift at six plaquettes for each unit cube and taking the isosurface of its absolute value, we can visualize vortices. [Kobayashi and Tsubota: J. Phys. Soc. Jpn. {\bf 74} (2005) 3248, reproduced with permission. Copyright 2005 the Physical Society of Japan.]}
\end{figure}

We next discuss simulations of QT described by Eq. \eqref{eq-dissipated-GP-Fourier} with the dissipation of Eq. \eqref{eq-step-function-dissipation}.
Here we consider two kinds of turbulence: decaying turbulence \cite{Kobayashi:2005a} and steady turbulence \cite{Kobayashi:2005b}.
For decaying turbulence, the initial configuration was set to have a uniform density $\rho = 1$ and the phase $\phi$ had a random spatial distribution.
The random phase $\phi$ is generated by placing random numbers between $-\pi$ to $\pi$ at every distance $\lambda$ and connecting them smoothly, which represents an energy injection at the scale of $\lambda$.
Because the initial superfluid velocity $\Vec{v}\sub{s} = (\hbar / m) \nabla \phi$ given by the initial random phase is random, the initial wave function is dynamically unstable and soon produces turbulence with many vortices (see Fig. \ref{fig-turbulence-snapshot}).
We confirm that only the compressible kinetic energy $E\sub{kin}\up{c}$ is decreased by the dissipation term and that the incompressible kinetic energy $E\sub{kin}\up{i}$ dominates the total kinetic energy $E\sub{kin}$, demonstrating that only compressible excitations are effectively dissipated by the dissipation term $\tilde{\gamma}$.
In the middle stage of the decay, the dissipation rate of the incompressible kinetic energy $\varepsilon\sub{kin}\up{i} = - \partial E\sub{kin}\up{i} / \partial t$ takes an almost constant value, showing the quasi-steady state of QT.
In this range, the energy spectrum $E\sub{kin}\up{i}(k,t)$ is consistent with the Kolmogorov law:
\begin{align}
E\sub{kin}\up{i}(k,t) \cong C \left(\varepsilon\sub{kin}\up{i} \right)^{2/3} k^{-5/3},
\end{align}
with the Kolmogorov constant $C \cong 0.32$, and smaller than that in CT which is estimated to be $1.4 \lesssim C \lesssim 1.8$.
Araki {\it et al.} obtained a Kolmogorov constant $C \simeq 0.7$ in their numerical simulation of the vortex-filament model and this is also smaller than that in CT.
This small Kolmogorov constant may therefore be characteristic of QT \cite{Araki:2002}.

We also considered steady QT, which can be modeled by introducing energy injection to the system.
The advantages of steady QT over decaying QT are the following.
First, steady turbulence gives a clearer correspondence with the Kolmogorov law: this is because the original statistics have been developed for steady turbulence.
Second, it enables us to confirm the presence of the energy-containing range, the inertial and the energy-dissipative range of QT.
Third, in all ranges we can obtain the time-independent energy flux in wavenumber space.
Consequently, it is possible to reveal the cascade process of QT, which occurs by quantized vortices.

\begin{figure}[htb]
\centering
\includegraphics[width=0.7\linewidth]{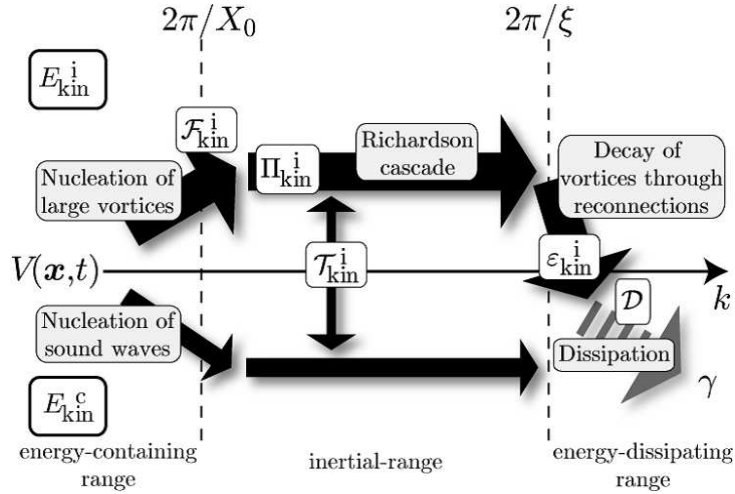}
\caption{\label{fig-cascade-image} Flow of the incompressible kinetic energy $E\sub{kin}\up{i}$ (upper half of diagram) and compressible kinetic energy $E\sub{kin}\up{c}$ (lower half) in wavenumber space. [Kobayashi and Tsubota: J. Phys. Soc. Jpn. {\bf 74} (2005) 3248, reproduced with permission. Copyright 2005 the Physical Society of Japan.]}
\end{figure}

For energy injection at large scales, we introduce the moving random potential $V(\Vec{r},t)$ in the GP equation:
\begin{align}
\left( i - \gamma \right) \hbar \frac{\partial \Psi}{\partial t} = \left( - \frac{\hbar^2 \nabla^2}{2 M} + g \left| \Psi \right|^2 - \mu + V \right) \Psi, \label{eq-dissipated-potential-GP}
\end{align}
or its Fourier-transformed form:
\begin{align}
(i - \tilde{\gamma}) \hbar \frac{\partial \tilde{\Psi}}{\partial t} = \left( \frac{\hbar^2 k^2}{2 M} - \mu \right) \tilde{\Psi} + g \tilde{Y} + \tilde{V}. \label{eq-dissipated-potential-GP-Fourier}
\end{align}
Here $\tilde{V}$ is defined to satisfy the relation:
\begin{align}
\int d\Vec{r} \: e^{- i \Vec{k} \cdot \Vec{r}} V \Psi = \tilde{V}. \label{eq-Fourier-potential-definition}
\end{align}
We set the statistical properties of $V(\Vec{r},t)$ to obey the Gaussian two-point correlation:
\begin{align}
\left\langle V(\Vec{r},t) V(\Vec{r}^\prime,t^\prime) \right\rangle = V_0^2 \exp\left[- \frac{(x - x^\prime)}{2 X_0^2} - \frac{(t - t^\prime)^2}{2 T_0^2} \right]. \label{eq-moving-random-potential}
\end{align}
This moving random potential has the characteristic spatial scale $X_0$ and thus quantized vortices of scale $X_0$ are nucleated when $V_0$ is strong and $T_0$ is short enough.
We define the wavenumber separating the energy-containing range and the inertial range as $2 \pi / X_0$.
The wavenumber $2 \pi / \xi$ between the inertial range and the energy-dissipative range is defined by the dissipation term $\tilde{\gamma}$.
Therefore, our steady QT has an energy-containing range of $k < 2 \pi / X_0$, inertial range of $2 \pi / X_0 < k < 2 \pi / \xi$, and energy-dissipative range of $2 \pi / \xi < k$ (see Fig. \ref{fig-cascade-image}).
Starting from the uniform state $\rho = 1$ and $\phi = 0$, we develop the GP equation \eqref{eq-dissipated-potential-GP-Fourier} with potential \eqref{eq-moving-random-potential}, and obtain steady QT in which $E$, $E\sub{q}$, $E\sub{kin}\up{c}$, and $E\sub{kin}\up{i}$ are statistically steady.
By choosing the appropriate parameters, the incompressible kinetic energy $E\sub{kin}\up{i}$ is always dominant in the total kinetic energy $E\sub{kin}$; the introduced potential contributes to the nucleation of vortices rather than that of compressible excitation with long wavelength.
The obtained energy spectrum was consistent with the Kolmogorov law in the inertial range.

We further calculate two other important values: the energy dissipation rate $\varepsilon\sub{kin}\up{i}$ and the flux $\Pi\sub{kin}\up{i}$ of the incompressible kinetic energy from small to large wavenumbers.
$\varepsilon\sub{kin}\up{i}$ in steady turbulence can be equated to $\varepsilon\sub{kin}\up{i} = - \partial E\sub{kin}\up{i} / \partial t$ after switching off the moving random potential.
This is because the incompressible kinetic energy $E\sub{kin}\up{i}$ decays to the energy of compressible short-wavelength excitations.
On the other hand, the energy flux $\Pi\sub{kin}\up{i}$ can be calculated by considering the scale-by-scale energy budget equation, which can be obtained by the time development of the cumulative incompressible kinetic energy:
\begin{align}
\mathcal{E}\sub{kin}\up{i} = \frac{1}{2 N} \int d\Vec{r}\: \left( L_k [\Vec{p}]\up{i} \right)^2.
\end{align}
Here $L_k$ is the operator for the low-pass filter:
\begin{align}
L_k[s(\Vec{r})] = \frac{1}{(2 \pi)^3} \int_{\Vec{k} < k} d\Vec{k} \: \int d\Vec{r}^\prime \: e^{i \Vec{k} \cdot (\Vec{r} - \Vec{r}^\prime)} s(\Vec{r})
\end{align}
The time derivative of $\mathcal{E}\sub{kin}\up{i}$ gives the energy budget equation:
\begin{align}
\frac{\partial \mathcal{E}\sub{kin}\up{i}}{\partial t} + \Pi\sub{kin}\up{i} = \mathcal{F}\sub{kin}\up{i} + \mathcal{T}\sub{kin}\up{i} - \mathcal{D}\sub{kin}\up{i} \label{eq-energy-budget}.
\end{align}
Here we introduce the cumulative energy injection $\mathcal{F}\sub{kin}\up{i}$
\begin{align}
\mathcal{F}\sub{kin}\up{i} = - \frac{1}{M N} \int d\Vec{r} \: L_k [\Vec{p}]\up{i} \cdot L_k \left[ \sqrt{\rho} \nabla V \right]\up{i},
\end{align}
the cumulative energy transfer $\mathcal{T}\sub{kin}\up{i}$
\begin{align}
\mathcal{T}\sub{kin}\up{i} = \frac{1}{N} \int d\Vec{r} \: L_k [\Vec{p}]\up{i} \cdot L_k \left[ \frac{\sqrt{\rho}}{M} \nabla \left( \frac{\hbar^2 \nabla^2 \sqrt{\rho}}{2 M \sqrt{\rho}} - g \rho\right) - \left\{ \frac{\Vec{v}\sub{s} \Vec{v}\sub{s} \cdot \nabla \rho }{2 \sqrt{\rho}} \right\} \right]\up{i},
\end{align}
the cumulative energy dissipation $\mathcal{D}\sub{kin}\up{i}$
\begin{align}
\begin{split}
& \mathcal{D}\sub{kin}\up{i} = - \frac{\hbar}{M N} \int d\Vec{r} \: L_k [\Vec{p}]\up{i} \cdot L_k \Bigg[ \frac{\sqrt{\rho}}{2} \nabla \left\{ \frac{\gamma}{\rho} \nabla \cdot \left( \rho \Vec{v}\sub{s} \right) \right\} \\
&\quad + \gamma \sqrt{\rho} \left\{ \frac{\Vec{v}\sub{s}}{2} \left( \frac{\nabla^2 \sqrt{\rho}}{\sqrt{\rho}} - \frac{M^2 \Vec{v}\sub{s}^2}{\hbar^2} \right) - \frac{ M \Vec{v}\sub{s} (V + g \rho - \mu)}{\hbar^2} \right\} \Bigg]\up{i} + \mathcal{O}(\gamma^2),
\end{split}
\end{align}
and the energy flux $\Pi\sub{kin}\up{i}$
\begin{align}
\Pi\sub{kin}\up{i} = \frac{1}{2 N} \int d\Vec{r} \: L_k [\Vec{p}]\up{i} \cdot L_k \left[ \Vec{p} \nabla \cdot \Vec{v}\sub{s} + \sqrt{\rho} \nabla \Vec{v}\sub{s}^2 \right]\up{i}.
\end{align}
Equation \eqref{eq-energy-budget} can be interpreted as follows: at a given scale $k$, the rate of change of the incompressible kinetic energy is equal to the energy injection by the force $\mathcal{F}\sub{kin}\up{i}$ plus the energy transfer between vortices and density fluctuations $\mathcal{T}\sub{kin}\up{i}$ minus the energy dissipation $\mathcal{D}\sub{kin}\up{i}$ minus the energy flux $\Pi\sub{kin}\up{i}$ to smaller scales.

\begin{figure}[htb]
\centering
\includegraphics[width=0.8\linewidth]{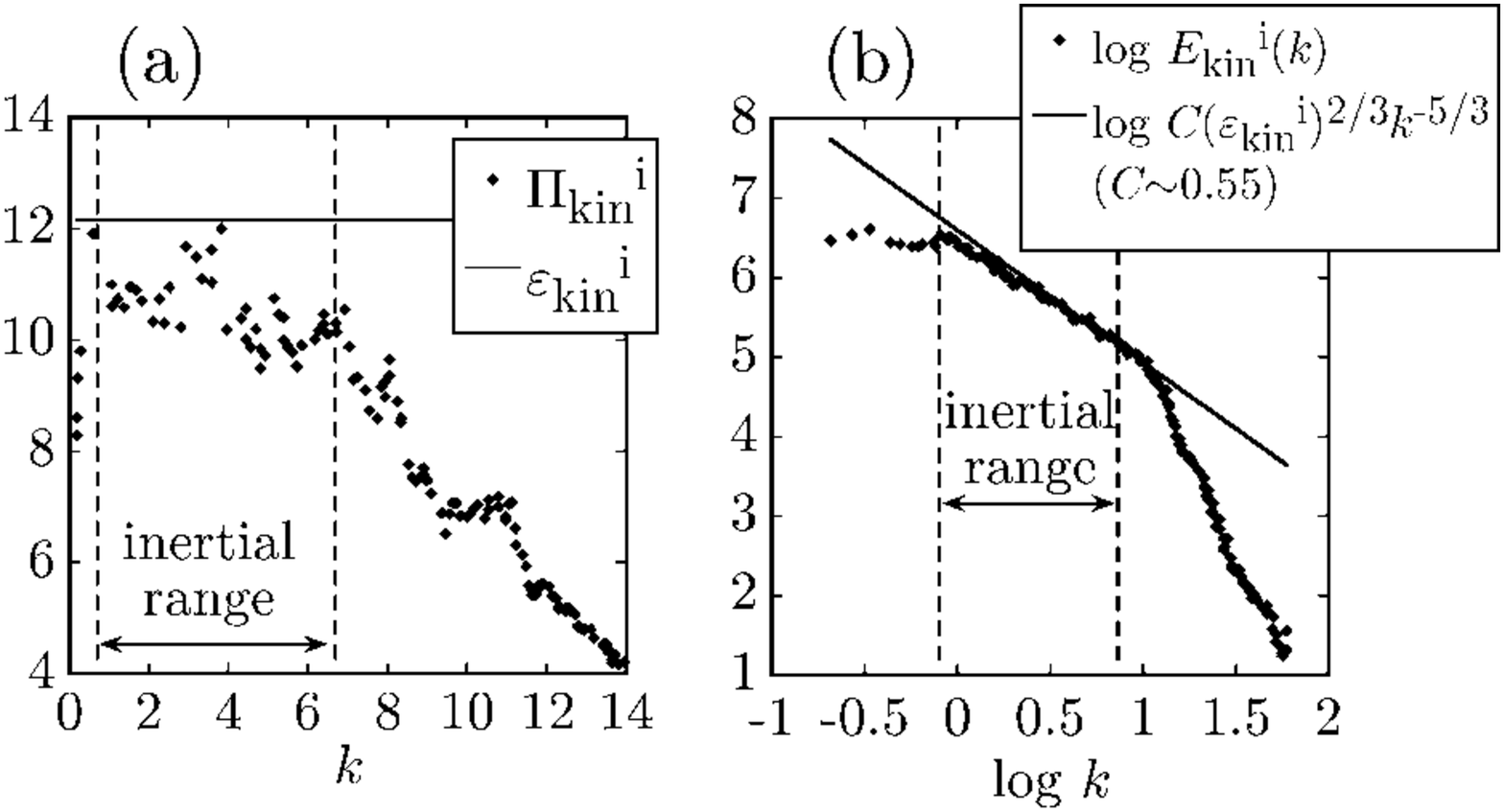}
\caption{\label{fig-steady-turbulence} (a) Dependence of the energy flux $\Pi\sub{kin}\up{i}$ on the wavenumber $k$ and the energy dissipation rate obtained from $\varepsilon\sub{kin}\up{i} = - \partial E\sub{kin}\up{i} / \partial t$. (b) Energy spectrum $E\sub{kin}\up{i}(k)$. The solid line is the Kolmogorov law $C (\varepsilon\sub{kin}\up{i})^{2/3} k^{-5/3}$. A simulation was performed in a periodic box of size $32$ with parameters $\gamma_0 = 1$, $V_0 = 50$, $X_0 = 4$, and $T_0 = 6.4 \times 10^{-2}$. Here, length, wavenumber, energy, and time are normalized by $\xi$, $1 / \xi$, $\hbar^2 / (2 M \xi^2)$, and $(2 M \xi^2) / \hbar$ respectively. With these parameters, the system enters steady turbulence after $t \simeq 25$. $\Pi\sub{kin}\up{i}$, $\varepsilon\sub{kin}\up{i}$ and $E\sub{kin}\up{i}(k)$ were obtained from an ensemble average of 50 randomly selected states at $t > 25$. [Kobayashi and Tsubota: J. Phys. Soc. Jpn. {\bf 74} (2005) 3248, reproduced with permission. Copyright 2005 the Physical Society of Japan.]}
\end{figure}

Figures \ref{fig-steady-turbulence} (a) and (b) shows the energy flux $\Pi\sub{kin}\up{i}$ and the energy dissipation rate $\varepsilon\sub{kin}\up{i}$, and the energy spectrum $E\sub{kin}\up{i}(k)$ respectively for numerically obtained steady QT.
The energy flux $\Pi\sub{kin}\up{i}$ is nearly constant and is consistent with the energy dissipation rate $\varepsilon\sub{kin}\up{i}$ in the inertial range, which indicates that the incompressible kinetic energy steadily flows in wavenumber space through the Richardson cascade at the constant rate $\Pi\sub{kin}\up{i}$, and finally dissipates to compressible excitations at the rate $\varepsilon\sub{kin}\up{i} \simeq \Pi\sub{kin}\up{i}$.
This energy flow is shown in the diagram of Fig. \ref{fig-cascade-image}.
The energy spectrum $E\sub{kin}\up{i}(k)$ shown in Fig. \ref{fig-steady-turbulence} (b) is quantitatively consistent with the Kolmogorov law in the inertial range.
The resulting Kolmogorov constant is $C \simeq 0.55$, which is smaller than that in CT as well as for decaying turbulence.

\subsubsection{Quantum turbulence at finite temperatures}

Experimental studies of superfluid $^4$He measured the energy spectrum of QT at finite temperatures, and supported the Kolmogorov spectrum directly or indirectly \cite{Smith:1993,Maurer:1998,Stalp:1999,Skrbek:2000a,Skrbek:2000b,Stalp:2002,Skrbek:2003}.
These experiments were also consistent in that they showed similarities between QT and CT.
Vinen theoretically considered this similarity and proposed that the superfluid and the normal fluid are likely to be coupled by mutual friction at scales larger than the intervortex spacing $l$ and would thus behave like a classical fluid \cite{Vinen:2000}.
This idea was confirmed by Kivotides {\it et al.} through numerical simulation of coupled dynamics of a vortex filament and a normal fluid \cite{Kivotides:2007} and by L'vov {\it et al.} through theoretical analysis of the two-fluid model \cite{Lvov:2006}.

Using the GP model, the easiest way to discuss QT at finite temperature is to solve the modified GP equation \eqref{eq-dissipated-GP} with a constant $\gamma$, because, as we discussed earlier, the constant $\gamma$ behaves like the mutual friction with relation to the coefficients $\gamma = \alpha$ and $\gamma^2 = \alpha^\prime$ (see Eq. \eqref{eq-dissipated-vortex-movement}).
As other methods to simulate atomic BECs at finite temperatures, the projected GP equation \cite{Davis:2001,Blakie:2007}, the stochastic GP equation \cite{Gardiner:2002}, and a coupled formalism involving the GP equation for condensate atoms and the Boltzmann equation for thermal cloud \cite{Zaremba:1999} are proposed.
All methods include not only dissipation but also thermal fluctuation, which becomes more important at high temperatures.

Before investigating QT at finite temperatures by using the above methods, we have to clarify the origin of the dissipation term $\gamma$ in Eq. \eqref{eq-dissipated-GP} from the microscopic point of view and its temperature dependence \cite{Kobayashi:2006}.
In quantum fluids, dissipation comes from the interaction between the condensate and its excitation such as quantum and thermal fluctuations.
For atomic BECs, the dynamics of the condensate and excitations can be described by the GP and Bogoliubov--de Gennes (BdG) equations, respectively, \cite{Bogoliubov:1947,deGennes:1966}.
Our goal in this section is to microscopically clarify the dissipation mechanism of a quantum fluid with quantized vortices by numerically solving the coupled equations involving the GP and BdG equations.
To do this, we start from the many-body Hamiltonian for bosons:
\begin{align}
\hat{H} = \int d\Vec{r} \: \hat{\Psi}^\dagger \left( - \frac{\hbar^2 \nabla^2}{2 M} - \mu + \frac{g}{2} \hat{\Psi}^\dagger \hat{\Psi} \right) \hat{\Psi}. \label{eq-field-operator-Hamiltonian}
\end{align}
Here $\hat{\Psi}$ is the boson field operator.
The time development of $\hat{\Psi}$ can be described by
\begin{align}
i \hbar \frac{\partial \hat{\Psi}}{\partial t} = \left( - \frac{\hbar^2 \nabla^2}{2 M} - \mu + g \hat{\Psi}^\dagger \hat{\Psi} \right) \hat{\Psi}. \label{eq-field-operator-equation}
\end{align}
In the Bose condensed system, the field operator $\hat{\Psi}$ can be separated in terms of the mean-field ansatz \cite{Girardeau:1959,Castin:1998}
\begin{align}
\hat{\Psi} = \Psi + \hat{\chi} + \hat{\zeta}. \label{eq-field-operator-separation}
\end{align}
In addition to the macroscopic wave function $\Psi = \mathcal{O}(\sqrt{N_0 / V})$, we define the first-order excitations $\hat{\chi} = \mathcal{O}(1 / \sqrt{V})$ and the higher-order excitations $\hat{\zeta} = \mathcal{O}(1 / \sqrt{N_0 V})$ with the number of condensate particles $N_0$ and the volume of the system $V$.
Substituting Eq. \eqref{eq-field-operator-separation} into Eq. \eqref{eq-field-operator-equation} and neglecting $\hat{\zeta}$, we obtain the GP equation:
\begin{align}
i \hbar \frac{\partial \Psi}{\partial t} = \left\{ - \frac{\hbar^2 \nabla^2}{2 M} - \mu + g \left(|\Psi|^2 + 2 \langle \hat{\chi}^\dagger \hat{\chi} \rangle \right) \right\} \Psi + g \langle \hat{\chi}^2 \rangle \Psi^\ast, \label{eq-GP-BdG-GP}
\end{align}
and the BdG equation:
\begin{align}
i \hbar \frac{\partial \hat{\chi}}{\partial t} = \left( - \frac{\hbar^2 \nabla^2}{2 M} - \mu + 2 g |\Psi|^2 \right) \hat{\chi} + g \Psi^2 \hat{\chi}^\dagger. \label{eq-GP-BdG-BdG}
\end{align}
When the GP equation \eqref{eq-GP-BdG-GP} is expressed as $i \hbar \partial \Psi / \partial t = H\sub{GP} \Psi$, the corresponding Hamiltonian $H\sub{GP}$ has the following imaginary term:
\begin{align}
\frac{\mathrm{Im}[H\sub{GP}]}{g} \equiv - \gamma = \mathrm{Im} \left[\frac{\left\langle \hat{\chi}^2 \right\rangle \Psi^\ast}{\Psi} \right] \label{eq-fix-dissipation-from-BdG}
\end{align}
This defines the dissipation $\gamma$ of the condensate caused by the interaction with the noncondensed particles.
We can calculate the dissipation $\gamma$ from Eq. \eqref{eq-fix-dissipation-from-BdG} by numerically solving the coupled GP \eqref{eq-GP-BdG-GP} and BdG \eqref{eq-GP-BdG-BdG} equations.
The BdG equation \eqref{eq-GP-BdG-BdG} can be solved by using the Bogoliubov transformation:
\begin{align}
\chi = \frac{1}{\sqrt{V}} \sum_j \left[u_j \hat{\alpha}_j + v_j^\ast \hat{\alpha}_j^\dagger \right], \label{eq-Bogoliubov-transformation-chi}
\end{align}
where $u_j$ and $v_j$ are the Bogoliubov coefficients and $\hat{\alpha}_j$ and $\hat{\alpha}_j^\dagger$ are the annihilation and creation operators of a quasiparticle, respectively, for the $j$th energy level.
Here, we assume that the quasiparticles are coupled with a heat bath at temperature $T$ and they are in a locally equilibrium state:
\begin{align}
\left\langle \hat{\alpha}_j \hat{\alpha}_j^\dagger \right \rangle = \frac{1}{e^{E_j / T} - 1} \equiv N_j, \label{eq-local-equilibrium-quasiparticle}
\end{align}
with the excitation spectrum $E_j$ of quasiparticles.
From these assumptions, we can expect that the energy of vortices or compressible excitations formed in the condensate is transferred to quasiparticles and finally dissipated to the heat bath.
Using Eqs. \eqref{eq-Bogoliubov-transformation-chi} and \eqref{eq-local-equilibrium-quasiparticle}, we can obtain the final form of the coupled GP and BdG equations:
\begin{align}
\begin{array}{c}
\dps i \hbar \frac{\partial \Psi}{\partial t} = \left\{ - \frac{\hbar^2 \nabla^2}{2 M} - \mu + g (|\Psi|^2 + 2 n\sub{e}) \right\} \Psi + g m\sub{e} \Psi^\ast, \arrayret
\dps i \hbar \frac{\partial u_j}{\partial t} = \left( - \frac{\hbar^2 \nabla^2}{2 M} - \mu + 2 g |\Psi|^2 \right) u_j - g \Psi^2 v_j \equiv A_j, \arrayret
\dps i \hbar \frac{\partial v_j}{\partial t} = - \left( - \frac{\hbar^2 \nabla^2}{2 M} - \mu + 2 g |\Psi|^2 \right) v_j + g (\Psi^\ast)^2 u_j \equiv B_j, \arrayret
\dps n\sub{e} = \sum_j \left\{ |u_j|^2 N_j + |v_j|^2 (N_j + 1) \right\}, \quad
\dps m\sub{e} = - \sum_j \left\{ u_j v_j^\ast (2 N_j + 1 ) \right\}, \arrayret
\dps E_j = \int d\Vec{r} \: \mathrm{Re} \left( u_j^\ast A_j + v_j ^\ast B_j \right), \quad
\gamma = \mathrm{Im} \left[\frac{m\sub{e} \Psi^\ast}{\Psi} \right]. \label{eq-numerical-GP-BdG}
\end{array}
\end{align}
For a given initial condition of $\Psi$, we adopt the uniform excitations given by
\begin{align}
\begin{array}{c}
\dps u_j = e^{i (\Vec{k}_j \cdot \Vec{r})} \sqrt{\frac{1}{2 V} \frac{\hbar^2 k_j^2 / (2 M) + g |\Psi|^2}{E_j} + 1}, \arrayret 
\dps v_j = e^{- i (\Vec{k}_j \cdot \Vec{r})} \sqrt{\frac{1}{2 V} \frac{\hbar^2 k_j^2 / (2 M) + g |\Psi|^2}{E_j} - 1},
\end{array}
\end{align}
as the initial condition for $u_j$ and $v_j$.
Here $\Vec{k}_j = 2 \pi ( j_x ,  j_y , j_z) / \sqrt[3]{V}$, with nonzero integers $j_x$, $j_y$ and $j_z$.

\begin{figure}[htb]
\centering
\includegraphics[width=0.45\linewidth]{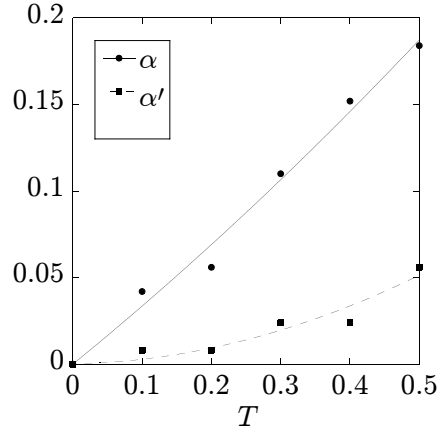}
\caption{\label{fig-BEC-mutual-friction} Temperature dependence of the mutual friction coefficients $\alpha$ and $\alpha^\prime$. The plots represent the numerical results and the lines indicate fitting. A simulation was performed in a periodic box of size $4$ with the velocity field $v\sub{e} = 0.1$, where length, velocity, and temperature are normalized by $\xi$, $\hbar / (M \xi)$, and the critical temperature for the BEC for free bosons, respectively. [Kobayashi and Tsubota: Phys. Rev. Lett. {\bf 97} (2006) 145301, reproduced with permission. Copyright 2006 the American Physical Society.]}
\end{figure}

First, we attempt to calculate the coefficients of mutual friction as functions of temperature \cite{Hall:1956a,Hall:1956b,Schwarz:1985}.
When one straight vortex along the $z$ axis is placed under the velocity field $\Vec{v}\sub{e} = (v\sub{e}, 0,  0) $, the dynamics of the vortex position $\Vec{s}(t) = (s_x(t), s_y(t),  0) $ are described through Eq. \eqref{Schwarz} by
\begin{align}
\Vec{s}(t) = (s_x(0) + (1 - \alpha^\prime) v\sub{e},  \quad s_y(0) - \alpha v\sub{e},  \quad 0) \label{eq-mutual-friction-vortex-movement}
\end{align}
Starting from the state $\Psi$ in Eq. \eqref{eq-straight-vortex} with one straight vortex line, we numerically solve the coupled GP and BdG equations under the velocity field $\Vec{v}\sub{e}$:
\begin{align}
\begin{array}{c}
\dps i \hbar \frac{\partial \Psi}{\partial t} = \left\{ - \frac{\hbar^2 \nabla^2}{2 M} - \mu + g (|\Psi|^2 + 2 n\sub{e}) - i \Vec{v}\sub{e} \cdot \nabla \right\} \Psi + g m\sub{e} \Psi^\ast, \arrayret
\dps i \hbar \frac{\partial u_j}{\partial t} = \left( - \frac{\hbar^2 \nabla^2}{2 M} - \mu + 2 g |\Psi|^2 - i \Vec{v}\sub{e} \cdot \nabla \right) u_j - g \Psi^2 v_j, \arrayret
\dps i \hbar \frac{\partial v_j}{\partial t} = - \left( - \frac{\hbar^2 \nabla^2}{2 M} - \mu + 2 g |\Psi|^2 - i \Vec{v}\sub{e} \cdot \nabla \right) v_j + g (\Psi^\ast)^2 u_j.
\end{array}
\end{align}
We can obtain $\alpha$ and $\alpha^\prime$ by comparing the position of the vortex with Eq. \eqref{eq-mutual-friction-vortex-movement} and find their monotonic increase with temperature as shown in Fig. \ref{fig-BEC-mutual-friction}, which is qualitatively consistent with mutual friction in superfluid $^4$He at temperatures much lower than the superfluid critical temperature.
This temperature dependence of $\alpha$ and $\alpha^\prime$ may be a standard scale for measuring the temperature in atomic BECs with quantized vortices.

\begin{figure}[htb]
\centering
\includegraphics[width=0.8\linewidth]{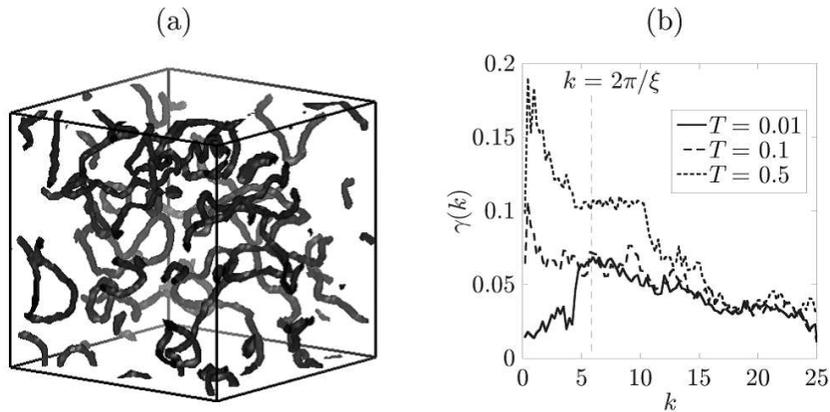}
\caption{\label{fig-GP-BdG-gamma} (a) Example of the configurations of quantized vortices at $t = 0$. (b) Wave number dependence of the Fourier transformed dissipation term $\tilde{\gamma}$ at $t = 1$. A simulation was performed in a periodic box of size $4$. Length, wavenumber, time, and temperature are normalized by $\xi$, $1 / \xi$, $(2 M \xi^2) / \hbar$, and the critical temperature for the BEC for free bosons, respectively. $\tilde{\gamma}$ is obtained from an ensemble average of 25 different initial states. [Kobayashi and Tsubota: Phys. Rev. Lett. {\bf 97} (2006) 145301, reproduced with permission. Copyright 2006 the American Physical Society.]}
\end{figure}

Next, we calculate the dissipation term $\gamma$ for the turbulent state.
We begin with $\Psi$ that includes several randomly placed vortices, as shown in Fig. \ref{fig-GP-BdG-gamma} (a).
After some time evolution, we calculate the dissipation term $\gamma$ in Eq. \eqref{eq-numerical-GP-BdG}.
Figure \ref{fig-GP-BdG-gamma} (b) shows the Fourier-transformed dissipation $\tilde{\gamma}$ at several temperatures.
At low temperature, dissipation works only at wavenumbers greater than $2 \pi / \xi$, which is consistent with the dissipation term $\tilde{\gamma}(k) = \gamma_0 \theta(k - 2 \pi / \xi)$ used in the previous section for the simulation of QT at zero temperature.
From this result, we expect that only short-wavelength excitations emitted during vortex reconnections or by high frequency Kelvin waves become dissipated at scales smaller than $\xi$.
On the other hand, as the temperature increases, dissipation works at small wavenumbers as well, which is consistent with the above simulation of a single vortex, because dissipation at small wavenumbers acts as the mutual friction, as discussed in Eq. \eqref{eq-dissipated-vortex-movement}

\subsubsection{Two-dimensional turbulence} \label{subsubsec-2D-turbulence}

In the previous section, we numerically verified that QT shows similar or the same statistical properties characterized by the Kolmogorov law.
The physical explanation for this similarity is that quantization of vortices is not essential at scales larger than the mean intervortex spacing $l$, and vortices behave like eddies in CT forming vortex bundle structures.
Considering this situation, a new question arises: what happens during QT in a 2D system, in which quantized vortices take a point structure.

Before discussing 2D QT, we consider 2D turbulence of a classical incompressible fluid obeying the Navier--Stokes equations \cite{Frisch:1995}:
\begin{align}
\begin{array}{c}
\dps \nabla \cdot \Vec{v} = 0, \arrayret
\dps \frac{\partial \Vec{v}}{\partial t} + (\Vec{v} \cdot \nabla) \Vec{v} = - \frac{1}{\rho} \nabla p + \nu \nabla^2 \Vec{v}, \label{eq-Navier-Stokes}
\end{array}
\end{align}
where $p$ is the pressure and $\nu$ is the kinematic viscosity.
In the dissipationless limit $\nu \to 0$, the kinetic energy
\begin{align}
E = \frac{1}{2} \int d\Vec{r} \: \Vec{v}^2,
\end{align}
becomes motion invariant, and this plays a key role in the energy cascade because viscosity does not apply and the energy is not dissipated but is transferred from larger to smaller scales.
In 2D system, the enstrophy $\Omega$ is defined as
\begin{align}
\Omega = \int d\Vec{r} \: \left( \nabla \times \Vec{v} \right)^2
\end{align}
and also becomes motion invariant in the dissipationless limit, giving further cascade physics.
Kraichnan recognized that the motion invariance of $\Omega$ drastically modifies the physics of 2D turbulence \cite{Kraichnan:1967}.
There are two inertial ranges, the first for the cascade of the kinetic energy and the second for the enstrophy.
In the energy cascading region, the direction of the energy flux is different from that in 3D turbulence, {\it i.e.}, an energy cascade from small to large scales.
The energy spectrum takes $E(k) \propto \varepsilon^{2/3} k^{-5/3}$ and $E(k) \propto \eta^{3/2} k^{-3}$ (plus certain logarithmic corrections, unimportant in the current discussion) in the energy and enstrophy cascading regions, respectively.
If energy is injected into a fluid in a wavenumber scale of $k_i$, the inertial ranges for the energy cascade and the enstrophy cascade are formed in the wavenumber regions of $k < k_i$ and $k_i < k < k_\nu$, respectively, as the steady turbulent state, where $k_\nu$ is the viscosity cutoff.
These predictions have been confirmed in laboratory experiments and using large-scale direct numerical simulations of Eq. \eqref{eq-Navier-Stokes} \cite{Herring:1985,Boffetta:2000,Laval:2004}. 

Our question is whether the inverse energy cascade and the enstrophy cascade are features of 2D QT described in the 2D GP equation.
The qualitative answer is that this is not necessarily the case.
In a Bose condensed system described by the GP equation, the enstrophy $\Omega$ coincides (up to a prefactor) with the total number of vortex points.
In the GP dynamics, however, the total number of vortices is not conserved because vortices appear and disappear through pair creation and annihilation.
Therefore, Kraichnan's arguments become, generally speaking, irrelevant for 2D QT.
Thus, for a certain range of parameters, where the creation or annihilation of vortex pairs becomes dynamically important, we can expect a direct energy cascade in 2D QT, exactly as in 3D turbulence.
The same argument can be made based on the Euler equations.
As shown in Eq. \eqref{eq-quantum-Euler}, the GP equation is reduced to a compressible Euler equation with an effective pressure function $P = g \rho - \hbar^2 (\nabla^2 \sqrt{\rho}) / (2 M \sqrt{\rho})$.
However, a compressible Euler equation does not preserve the enstrophy.
Therefore, at some level of compressibility (characterized by the Mach number, the ratio of the turbulent velocity fluctuations to the sound velocity), the direction of the energy flux can change its sign and, instead of an inverse energy cascade, we expect a direct cascade typical for 3D turbulence.

Experimentally, we can assume a 2D BEC that is dynamically frozen in the transversal direction, {\it i.e.}, a pancake-shaped BEC that is strongly trapped along the $z$ axis.
Let us consider a trapped BEC in the harmonic oscillator potential described by the GP equation
\begin{align}
i \hbar \frac{\partial \Psi}{\partial t} = \left\{ - \frac{\hbar^2 \nabla^2}{2M} - \mu + \frac{M}{2} (\omega_x^2 x^2 + \omega_y^2 y^2 + \omega_z^2 z^2) + g |\Psi|^2 \right\} \Psi. \label{eq-2D-trap-GP}
\end{align}
Here, $\omega_{x,y,z}$ are the oscillator frequencies on the $x$, $y$, and $z$ axes.
For simplicity, we here assume $\omega_x = \omega_y = \omega_\perp$.
The strength of the trap along the $z$ axis is determined by the ratio $\omega_z / \omega_\perp$, and a pancake-shaped potential gives $\omega_z / \omega_\perp \gg 1$.
If $\hbar \omega_z$ is much larger than the chemical potential $\mu$, the wave function can be separated into $\Psi_\perp(x,y)$ in the $x$--$y$ plane and $\Psi_z(z)$ along the $z$ axis approximated by a Gaussian function as
\begin{align}
\Psi = \Psi_z \Psi_\perp = \frac{\dps e^{- z^2 / (2 a_z^2)}}{\sqrt{a_z \sqrt{\pi}}} \Psi_\perp. \label{eq-2D-wave-approximation}
\end{align}
Here, $a_z = \sqrt{\hbar / (M \omega_z)}$ is the trap length along the $z$ direction.
Substituting Eq. \eqref{eq-2D-wave-approximation} into Eq. \eqref{eq-2D-trap-GP}, we obtain
\begin{align}
i \hbar \frac{\partial \Psi_\perp}{\partial t} = \left\{ - \frac{\hbar^2 \nabla_\perp^2}{2M}  - \mu_\perp + \frac{M}{2} (\omega_x^2 x^2 + \omega_y^2 y^2) + g_\perp |\Psi_\perp|^2 \right\} \Psi_\perp.
\end{align}
Here, $\nabla_\perp^2 = \partial^2 / \partial x^2 + \partial^2 / \partial y^2$, $\mu_\perp = \mu - \hbar \omega_z$, and $g_\perp = g / (\sqrt{2} \pi a_z)$.
As in the 3D case, we can define the 2D density $\rho_\perp = |\Psi_\perp|^2$ and 2D healing length $\xi_\perp = \hbar / \sqrt{2 M g_\perp \bar{\rho}_\perp}$.

For simplicity, in the work \cite{Numasato:2010} Numasato {\it et al.} use the uniform 2D GP equation for a small $\omega_\perp$ limit:
\begin{align}
i \hbar \frac{\partial \Psi_\perp}{\partial t} = \left\{ - \frac{\hbar^2 \nabla_\perp^2}{2M}  - \mu_\perp + g_\perp |\Psi_\perp|^2 \right\} \Psi_\perp.
\end{align}
To confirm whether the cascade is direct or inverse, the thermalization process of decaying turbulence in an isolated system is an effective indicator.
If the cascade is direct, an essential part of the energy reaches the smallest scales available in the simulation and the system quickly evolves toward thermodynamic equilibrium filled with short scale excitations of fluid.
On the other hand, if the cascade is inverse, fluid motions of the system size are strongly excited even at later stages.
To check this thermalization process, Numasato {\it et al.} introduce neither dissipation nor energy injection in this work.

The method to generate turbulence is almost the same as that used for 3D decaying turbulence \cite{Kobayashi:2005a}: the initial condition of $\Psi$ is set to constant density $\rho_\perp$ and random phase $\phi$, which varies at large scales.
From this initial condition, we obtain, during time evolution, a 2D QT composed of a random configuration of quantized vortices.
A simulation was performed in a periodic box with a size of $64 \xi$ for all the results shown below.

Figure \ref{fig-2Denergy} shows the evolution of the total energy $E$, the kinetic energy $E\sub{kin}$, and its compressible and incompressible parts $E\sub{kin}\up{c}$ and $E\sub{kin}\up{i}$, and the number of vortices.
$E$, $E\sub{kin}$, $E\sub{kin}\up{c}$, and $E\sub{kin}\up{i}$ are obtained from Eqs. \eqref{eq-GP-energy}, \eqref{eq-GP-various-energy}, and \eqref{eq-GP-kinetic-energy} by replacing $\Psi$ and $\rho$ with $\Psi_\perp$ and $\rho_\perp$.
Because the system has no dissipation, $E$ is time independent.
At the initial stage ($t \lesssim 5$), an intensive process of vortex creation leads to fast transformation of $E\sub{kin}\up{c}$ into $E\sub{kin}\up{i}$, which is larger than $E\sub{kin}\up{c}$ for the time interval $2 \lesssim t \lesssim 4$.
At later stages ($t \gtrsim 5$), $E\sub{kin}$ is practically independent of time.
The largest maximal value of the vortex number is achieved at the crossover time $t \sim 5$.
Its decay is faster for larger $g_\perp$ (not shown), because the probability of the dominant nonlinear process of vortex pair annihilation is larger for larger $g_\perp$.
A decay of the vortex number leads to an increase in the flow compressibility.
The compressibility measured by $E\sub{kin}\up{c} / E\sub{kin}\up{i}$ is larger for larger $g_\perp$ (not shown).
The system finally reaches its equilibrium state with no quantized vortex.
It takes longer to reach its thermodynamic equilibrium state for smaller $g_\perp$.
\begin{figure}[htb]
\centering
\includegraphics[width=0.99\linewidth]{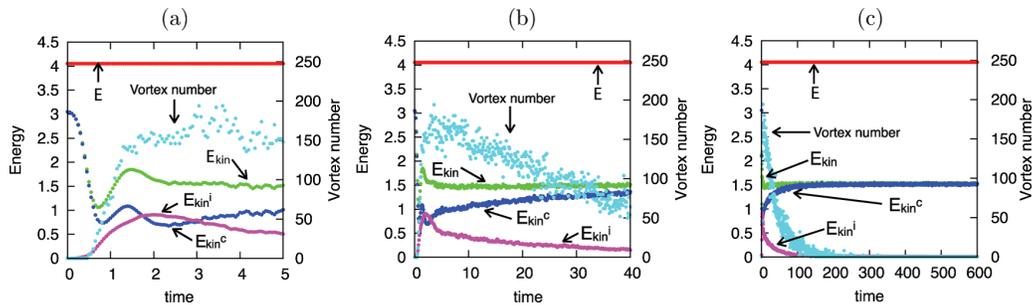}
\caption{\label{fig-2Denergy} Time evolution of $E$, $E\sub{kin}$, $E\sub{kin}\up{c}$, $E\sub{kin}\up{i}$, and the number of vortices.
A simulation was performed with $g_\perp = 4$, and with length, energy, and time normalized by $\xi_\perp$, $\hbar^2 / (2 M \xi_\perp^2)$, and $(M \xi_\perp^2) / \hbar$, respectively. [Numasato, Tsubota and L'vov: Phys. Rev. A {\bf 81} (2010) 063630, reproduced with permission. Copyright 2010 the American Physical Society.]}
\end{figure}

\begin{figure}[htb]
\centering
\includegraphics[width=0.99\linewidth]{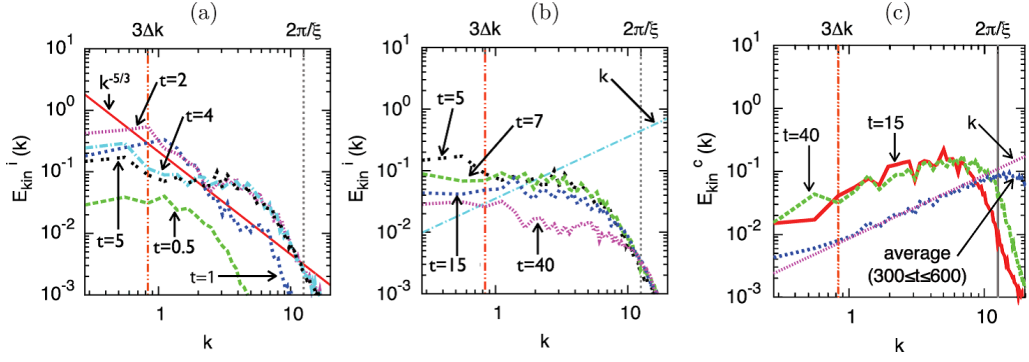}
\caption{\label{fig-2DE-spectrum} (a) $E\sub{kin}\up{i}(k)$ at an earlier moment of time. (b) $E\sub{kin}\up{i}(k)$ at a later moment of time. (c) $E\sub{kin}\up{c}(k)$ at a later moment of time. In all figures, $g_\perp = 4$. [Numasato, Tsubota and L'vov: Phys. Rev. A {\bf 81} (2010) 063630, reproduced with permission. Copyright 2010 the American Physical Society.]}
\end{figure}

The energy spectra $E\sub{kin}\up{i}(k)$ and $E\sub{kin}\up{c}(k)$ at different moments of time are shown in Fig. \ref{fig-2DE-spectrum}.
Here, $E\sub{kin}\up{i}(k)$ and $E\sub{kin}\up{c}(k)$ are defined by
\begin{align}
E\sub{kin}\up{c,i}(k) = \frac{1}{2 (2 \pi)^2 N_\perp} \int d \varphi_{\Vec{k}} k \left( \tilde{\Vec{p}}_\perp\up{c,i} \right)^2,
\end{align}
for $[\Vec{p}_\perp]\up{c,i} = [\sqrt{\rho_\perp} \Vec{v}\sub{s}]\up{c,i}$ and its Fourier transformation $\tilde{\Vec{p}}\up{c,i}_\perp$, and the 2D total number of particles $N_\perp = \int d \Vec{r} |\Psi_\perp|^2$.
At around $t \sim 3$, $E\sub{kin}\up{i}(k)$ are close to the Kolmogorov law $E\sub{kin}\up{i}(k) \propto k^{-5/3}$ as shown in Fig. \ref{fig-2DE-spectrum} (a).
This behavior is related to the energy cascade and will be discussed later.
At later times, the energy begins to accumulate at large $k$ and the $E\sub{kin}\up{i}(k)$ asymptotically approach the quasistationary state.
$E\sub{kin}\up{c}(k)$ at this stage are close to the thermodynamic equilibrium with energy equipartition between degrees of freedom.
In 2D systems, this gives $E\sub{kin}\up{i}(k) \propto k$ as shown in Fig. \ref{fig-2DE-spectrum} (c).
This will be also discussed later.
This distribution, however, does not correspond to that under thermodynamic equilibrium.
This is related to the fact that the system does not achieve full equilibrium at these times.
Indeed, Fig. \ref{fig-2Denergy} (c) shows that $E\sub{kin}\up{i}$ continues to converge to $E\sub{kin}\up{c}$.
We interpret this stage as a kind of flux equilibrium, when the $E\sub{kin}\up{c}(k)$ is determined by the energy flux from $E\sub{kin}\up{i}$ to $E\sub{kin}\up{c}$.
The exchange between $E\sub{kin}\up{i}$ and $E\sub{kin}\up{c}$ plays a subdominant role at these times.
After a long time evolution, most of $E\sub{kin}$ finally comes to comprise $E\sub{kin}\up{c}$.
This is the full thermodynamic state.
As expected, $E\sub{kin}\up{c}(k)$ is proportional to $k$ as shown in Fig. \ref{fig-2DE-spectrum} (c).

\begin{figure}[htb]
\centering
\includegraphics[width=0.99\linewidth]{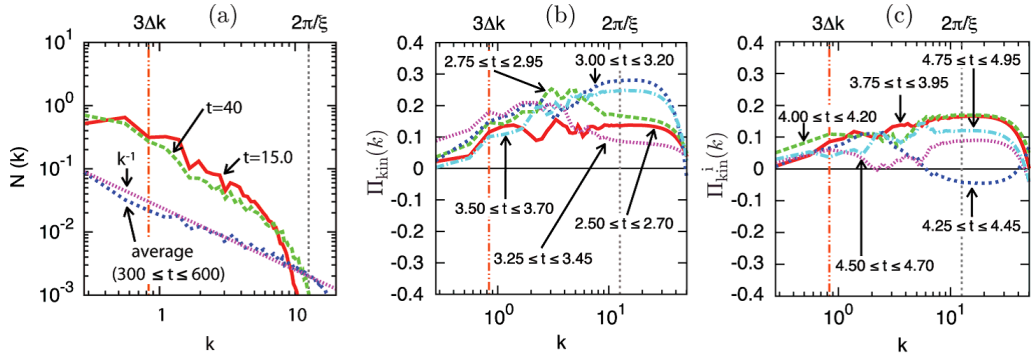}
\caption{\label{fig-2Dflux} (a) Particle number spectra. (b)--(c) Averaged incompressible kinetic energy flux in a short time period $\tau = 0.20$ in the time range $2.50 \leq t \leq 4.95$.
In all figures, $g_\perp = 4$. [Numasato, Tsubota and L'vov: Phys. Rev. A {\bf 81} (2010) 063630, reproduced with permission. Copyright 2010 the American Physical Society.]}
\end{figure}

Numasato {\it et al.} next consider the particle number spectrum $N(k) = |\tilde{\Psi}_\perp|^2 \equiv \tilde{\rho}_\perp$, where $\tilde{\Psi}_\perp$ is the Fourier transformation of $\Psi_\perp$.
Figure \ref{fig-2Dflux} (a) shows $N(k)$ for different time intervals.
Through the turbulent state, the spectrum asymptotically approaches the form $N(k) \propto k^{-1}$.
To rationalize this behavior, we note that this dependence should be related to $E\sub{kin}\up{c}(k) \propto k$ in the full thermodynamic equilibrium.
Indeed, in this state, the interaction energy can be neglected in comparison with the kinetic and quantum energy.
In addition, the quantum energy corresponds to zero-point motion and does not give rise to particle currents.
Thus, only the kinetic energy spectrum is related to the particle spectrum.
Moreover, almost all $E\sub{kin}$ becomes $E\sub{kin}\up{c}$ and there are no vortices, and as a result, $\phi$ has no singularity and becomes of order unity.
Thus the kinetic energy density can be written as follows:
\begin{align}
\begin{split}
E\sub{kin} &\simeq E\sub{kin}\up{c} = \frac{\hbar^2}{2 N_\perp M^2} \int d\Vec{r} \: \rho_\perp |\nabla \phi|^2 \sim \frac{\hbar^2}{2 (2 \pi)^2 N_\perp M^2} \int d\Vec{k} \: k^3 \tilde{\rho}_\perp |\phi|^2 \\
&\sim \frac{\hbar^2}{2 (2 \pi)^2 N_\perp M^2} \int d\Vec{k} \: k^3 \tilde{\rho}_\perp.
\end{split}
\end{align}
As a result,
\begin{align}
E\sub{kin}\up{c}(k) \simeq \frac{\hbar^2 k^2 N(k)}{2 (2 \pi)^2 N_\perp M^2}.
\end{align}
In this way, the two relations $E\sub{kin}\up{c}(k) \propto k$ and $N(k) \propto k^{-1}$ hold consistently.
This relation is similar to the relation between enstrophy $\Omega$ and kinetic energy $E$ in 2D CT.

Next, Numasato {\it et al.} consider the value and direction of the compressible and incompressible energy flux $\Pi\sub{kin}\up{c}$ and $\Pi\sub{kin}\up{i}$.
Numerical results for $\Pi\sub{kin}\up{c}$ and $\Pi\sub{kin}\up{i}$ are shown in Fig. \ref{fig-2Dflux} (b) and (c).
$\Pi\sub{kin}\up{i}$ takes positive values for $3 (2 \pi / L) \lesssim k \lesssim 2 \pi / \xi_\perp$ at least for $2.50 \lesssim t \lesssim 4.95$.
This strongly supports the idea that $E\sub{kin}\up{i}$ propagates from small $k$ to large $k$ and we conclude that at some region of the system parameters Numasato {\it et al.} can observe a 2D-direct energy cascade.

\begin{figure}[htb]
\centering
\includegraphics[width=0.75\linewidth]{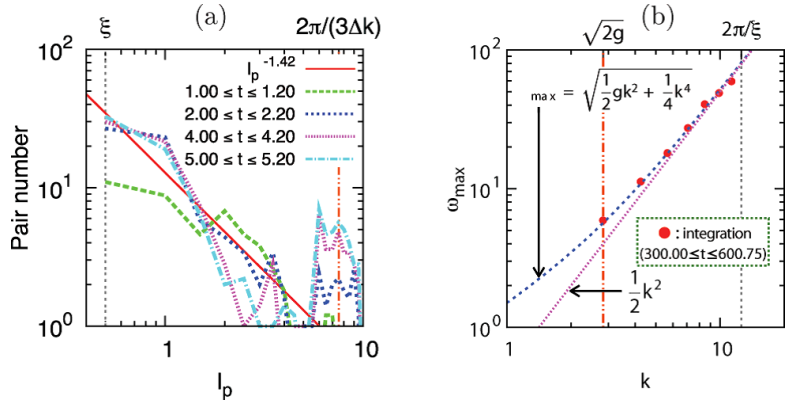}
\caption{\label{fig-2Dfrequency} (a) Averaged vortex pair numbers as a function of intervortex length $l\sub{p}$. (b) Position of the maximum frequency power spectra of the compressible velocity component for different wave vectors, averaged over a long time in the state of full thermodynamic equilibrium (full dots).
Bogoliubov's frequency spectrum $\omega\sub{max}$, Eq. \eqref{eq-Bogoliubov-spectrum}, and its large-$k$ asymptotic form of $\omega\sub{max} \propto k^2$, where the wavenumber, energy, and frequency are normalized by $1 / \xi_\perp$, $\hbar^2 / (2 M \xi_\perp^2)$, and $(M \xi_\perp^2) / \hbar$.
In both figures, $g_\perp=4$. [Numasato, Tsubota and L'vov: Phys. Rev. A {\bf 81} (2010) 063630, reproduced with permission. Copyright 2010 the American Physical Society.]}
\end{figure}

When the Kolmogorov spectrum is formed, a 2D Richardson cascade can be seen.
Numasato {\it et al.} choose the shortest intervortex pair length $l\sub{p}$ for all vortices.
In Fig. \ref{fig-2Dfrequency} (a), the averaged vortex pair number is shown and is proportional to $l\sub{p}^{-n}$, where $n$ depends on $g_\perp$ and $1.30 \leq n \leq 2.12$.
This power law suggests a self-similar spatial structure.
For 2D vortices, one of the most effective lengths, corresponding to the length of 3D vortex ring, is the intervortex length.

Important information about the motion can be extracted from the frequency power spectrum, which is the Fourier transformation of the different-time pair correlation function.
Numasato {\it et al.} observe this frequency power spectrum for the compressible velocity component:
\begin{align}
\int dt \: e^{i \omega t} \tilde{\Vec{v}}\sub{kin}\up{c}(\Vec{k}, t^\prime) \cdot \left\{ \tilde{\Vec{v}}\sub{kin}\up{c}(\Vec{k}, t^\prime + t) \right\}^\ast,
\end{align}
at later times.
As the system evolves to the thermodynamic equilibrium state, the power spectrum forms sharp peaks.
In Fig. \ref{fig-2Dfrequency} (b), Numasato {\it et al.} plot the position of the maxima integrated over a long time interval for $t$ in the thermodynamic equilibrium state.
The eigenfrequencies of the thermal fluctuations have been determined by Bogoliubov to be
\begin{align}
\omega\sub{max} = \sqrt{\frac{\hbar^2 k^4}{4 M^2} + \frac{g_\perp \bar{\rho}_\perp k^2}{M}}. \label{eq-Bogoliubov-spectrum}
\end{align}
The excellent agreement between the theoretical and numerical results indicates that the observed thermal fluctuations of the compressible velocity component do indeed correspond to Bogoliubov's elementary excitations.
The relatively small but finite width of the peak characterizes the finiteness of the lifetimes of these fluctuations, caused by interaction of the fluctuations with different $\Vec{k}$.

Numasato {\it et al.} finally note that $E\sub{kin}\up{i}$ approaches $0$ and there is no vortex in the thermodynamic equilibrium state.
The temperature of this state is, therefore, below the Kosterlitz--Thouless transition temperature $T\sub{KT}$.
If we increase the initial energy injection, we expect the final equilibrium state with many randomly paired nucleated and annihilated vortices, the temperature of which is above $T\sub{KT}$ \cite{Kosterlitz:1973}.
We also note that 2D QT is not restricted to theory.
For atomic BECs, the initial condition can be prepared by the phase imprinting technique.
By using experimentally realistic parameters $M = 1.46 \times 10^{-25}$kg, $a = 5.77$nm, $N = 10^3$, $a_z = 1.34$$\mu$m, $a_\perp \equiv \sqrt{\hbar / (M \omega_\perp)} = 4.25$$\mu$m, the healing length is estimated to be $\xi_\perp = 1.22$$\mu$m.
As a result, the size of the system in $x$--$y$ space must be $L \sim 2 a_\perp \sim 7 \xi$ to see the effects discussed in this section.

\subsubsection{Quantum turbulence in atomic Bose--Einstein condensates}

The study of the turbulent state in quantum fluids and its relation to CT is an intriguing physical problem.
Although the study of QT has a long history, only superfluid $^4$He and $^3$He systems have been used to realize QT until recently.
Recently, atomic BECs have become another candidate for QT research, since a turbulent state was realized in this system \cite{Weiler:2008,Henn:2009a,Henn:2009b,Seman:2011}.

Compared with a helium system, the characteristics of trapped BECs are: (i) a BEC system is weakly interacting and can be easily treated theoretically, (ii) many physical parameters of BECs are experimentally controllable, and (iii) various physical quantities such as the density and phase of BECs can be directly observed.
Quantized vortices can be considered to be holes of density and singularities of phase.
Shortly after trapped BECs were first realized, experimental groups reported vortex lattice structures, as well as the crystallization dynamics of these structures under rotation \cite{Madison:2000,AboShaeer:2001}.
These dynamics have been successfully confirmed quantitatively by numerical simulations using the GP equation \cite{2003KasamatsuPRA,Kasamatsu:2005}.
However, in experimental research on trapped BECs, another important phenomenon of quantized vortices, namely QT, has not been adequately studied until recently.
Noting that quantized vortices are observable and that almost all physical parameters of trapped BECs are controllable, such systems are an ideal prototype for truly controllable QT.
QT in trapped BECs is, therefore, used to determine several details of the system, such as the distribution of vortex length, details on the cascade of vortices, the isotropy or anisotropy of vortex configuration, and details on correlations among vortices related to eddy viscosity, as already considered for CT \cite{Frisch:1995}.
Clarifying any of these will lead to the detailed understandings of the transition to QT and its universality.
Therefore, research into QT offers the promise of greater advances in understanding turbulence than has been possible in past studies of turbulence.

There is a disadvantage in using trapped BECs to study QT: to generate turbulence, we cannot apply a velocity field, which is widely used for research on CT and QT of superfluid helium, because BECs are trapped.
To realize turbulence in this system, several ways have been theoretically proposed such as relaxation from a strongly degenerate nonequilibrium gas across the BEC critical temperature \cite{Svistunov:2001} or crystallization of an isolated BEC from the vortex-free to vortex lattice state under rotation \cite{Parker:2005}.

Here, we present a precession rotation which has two rotation axes, one of which rotates around the other \cite{Kobayashi:2007}.
We start from the GP equation under the rotating field $\Vec{\Omega}$:
\begin{align}
\left( i - \gamma \right) \hbar \frac{\partial \Psi}{\partial t} = \left( - \frac{\hbar^2 \nabla^2}{2 M} + g \left| \Psi \right|^2 - \mu + V - i \hbar \Vec{\Omega} \cdot \Vec{r} \times \nabla \right) \Psi.
\end{align}
Here $V$ is the trapping potential satisfying
\begin{align}
V = \frac{M \omega^2}{2} \left\{ (1 - \delta_z) (1 - \delta_y) x^2 + (1 + \delta_z) (1 - \delta_y) y^2 + (1 + \delta_y) z^2 \right\}, \label{eq-GP-trap-rotation}
\end{align}
with the trapping frequency $\omega$ and elliptical deformation parameters $\delta_z$ and $\delta_y$ in the $x$--$y$ and $z$--$x$ planes.
To develop the BEC to a turbulent state rather than a vortex lattice state, we use precession rotation along the $z$ and $x$ axes; the first rotation along the $z$ axis rotates around the second rotation along the  $x$ axis.
The resulting rotation field becomes $\Vec{\Omega} = (\Omega_x,  \Omega_z \sin \Omega_x t,  \Omega_z \cos \Omega_x t)$, where $\Omega_z$ and $\Omega_x$ are the frequencies of the first and second rotation, respectively.
The advantage of using this precession rotation to study turbulence is high controllability of the state from a nonturbulent vortex lattice to fully developed turbulence by changing the ratio $\Omega_x / \Omega_z$.

Now we consider a system at very low temperatures.
To apply the result shown in Fig. \ref{fig-GP-BdG-gamma} to the dissipation term $\gamma$, we consider the Fourier transformed form of Eq. \eqref{eq-GP-trap-rotation}
\begin{align}
\begin{array}{c}
\dps (i - \tilde{\gamma}) \hbar \frac{\partial \tilde{\Psi}}{\partial t} = \left( \frac{\hbar^2 k^2}{2 M} - \mu \right) \tilde{\Psi} + g \tilde{Y} + \tilde{V} - i \hbar \Vec{\Omega} \cdot \tilde{\Vec{R}}, \arrayret
\dps \tilde{\Vec{R}} = \int d\Vec{r} \: e^{- i \Vec{k} \cdot \Vec{r}} \Vec{r} \times \nabla \Psi, \label{eq-trap-GP-Fourier}
\end{array}
\end{align}
where $\tilde{\gamma}$, $\tilde{Y}$, and $\tilde{V}$ are defined in Eqs. \eqref{eq-Fourier-gamma-interaction-definition} and \eqref{eq-Fourier-potential-definition}.
In this work, we substitute the result at $T = 0.01$ shown in Fig. \ref{fig-GP-BdG-gamma} into $\tilde{\gamma}$ with the healing length at the trap center: $\xi = \hbar / \sqrt{2 M g \rho(\Vec{r} = 0)}$ at $t = 0$.
For other numerical parameters, we use the following, taken from experiments on $^{87}$Rb atoms: $M = 1.46 \times 10^{-25}$kg, $a = 5.61$ nm, $N = 2.50 \times 10^5$, and $\omega = 150 \times 2\pi$ Hz.

We start from a stationary solution without rotation and elliptical deformation.
At $t = 0$, we turn on the rotation $\Omega_x = \Omega_z = 0.6 \omega$ and elliptical deformation $\delta_z = \delta_y = 0.025$, and numerically calculate the time development of the GP equation \eqref{eq-trap-GP-Fourier}.
In the initial stage, vortices start to enter the BEC making the system quite anisotropic.
After $t \omega \simeq 150$, the BEC recovers isotropy and the system enters a statistically steady state.
The steady turbulence is sustained by the balance between the large-scale energy injection due to the rotation and the small-scale dissipation.
Furthermore, in all stages of the dynamics, $E\sub{kin}\up{i}$ is always much larger than $E\sub{kin}\up{c}$ and the dynamics of the BEC are dominated by vortices rather than compressible excitations.

\begin{figure}[htb]
\centering
\includegraphics[width=0.8\linewidth]{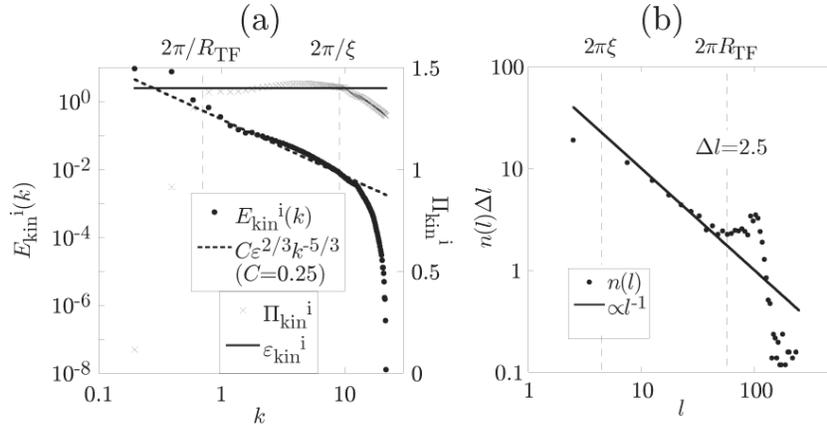}
\caption{\label{fig-BEC-spectrum} (a) $E\sub{kin}\up{i}$ and $\Pi\sub{kin}\up{i}$. (b) Vortex length distribution $n(l) \Delta l$ inside the Thomas--Fermi radius $R\sub{TF}$.
In both figures, length, wavenumber, energy, and time are normalized by $a\sub{h} = \sqrt{\hbar / (M \omega)}$, $1 / a\sub{h}$, $\hbar \omega$, and $1 / \omega$ respectively.
$E\sub{kin}\up{i}$, $\Pi\sub{kin}\up{i}$, and $n(l) \Delta l$ are obtained from an ensemble average of 25 randomly selected states at $t > 300$. [Kobayashi and Tsubota: Phys. Rev. A {\bf 76} (2007) 045603, reproduced with permission. Copyright 2007 the American Physical Society.]}
\end{figure}

Figure \ref{fig-BEC-spectrum} (a) shows $E\sub{kin}\up{i}(k)$ and $\Pi\sub{kin}\up{i}$ for the steady turbulent state.
$E\sub{kin}\up{i}(k)$ satisfies the Kolmogorov law in the inertial range $2 \pi / R\sub{TF} < k < 2 \pi / \xi$, where $R\sub{TF} = \sqrt{2 \mu(t = 0) / (M \omega)}$ is the Thomas--Fermi radius and represents the largest scale in the BEC.
Furthermore, the energy flux is nearly constant $\Pi\sub{kin}\up{i} \simeq 1.4 \hbar \omega^2 / (N M)$ in the inertial range, supporting the fact that the incompressible kinetic energy steadily flows in wavenumber space through the Richardson cascade at the constant energy transportation rate $\varepsilon\sub{kin}\up{i} = \Pi\sub{kin}\up{i}$.
Using this $\varepsilon\sub{kin}\up{i}$, we can estimate the Kolmogorov constant $C \simeq 0.25$, which is smaller than that in CT and consistent with the work for the uniform system in Sec. \ref{subsubsec-turbulence-zero-temperature}.

To investigate the relation between the Kolmogorov law and the Richardson cascade, we calculate the vortex length distribution $n(l) \Delta l$ inside the condensate, where $n(l) \Delta l$ represents the number of vortices with length from $l$ to $l + \Delta l$.
As shown in Fig. \ref{fig-BEC-spectrum} (b), $n(l) \Delta l$ obeys the scaling property $n(l) \propto l^{-\alpha}$ for $2 \pi \xi < l < 2 \pi R\sub{TF}$.
This reflects the self-similar Richardson cascade in which large vortices entering the condensate from the surface \cite{Kasamatsu:2005} are divided into smaller vortices.
The scaling exponent $\alpha$ is close to unity, which is consistent with those given by Araki {\it et al.} ($\alpha \simeq 1.34$) \cite{Araki:2002} and Mitani {\it et al.} ($\alpha \simeq 1$) \cite{Mitani:2006}.

\begin{figure}[htb]
\centering
\includegraphics[width=0.9\linewidth]{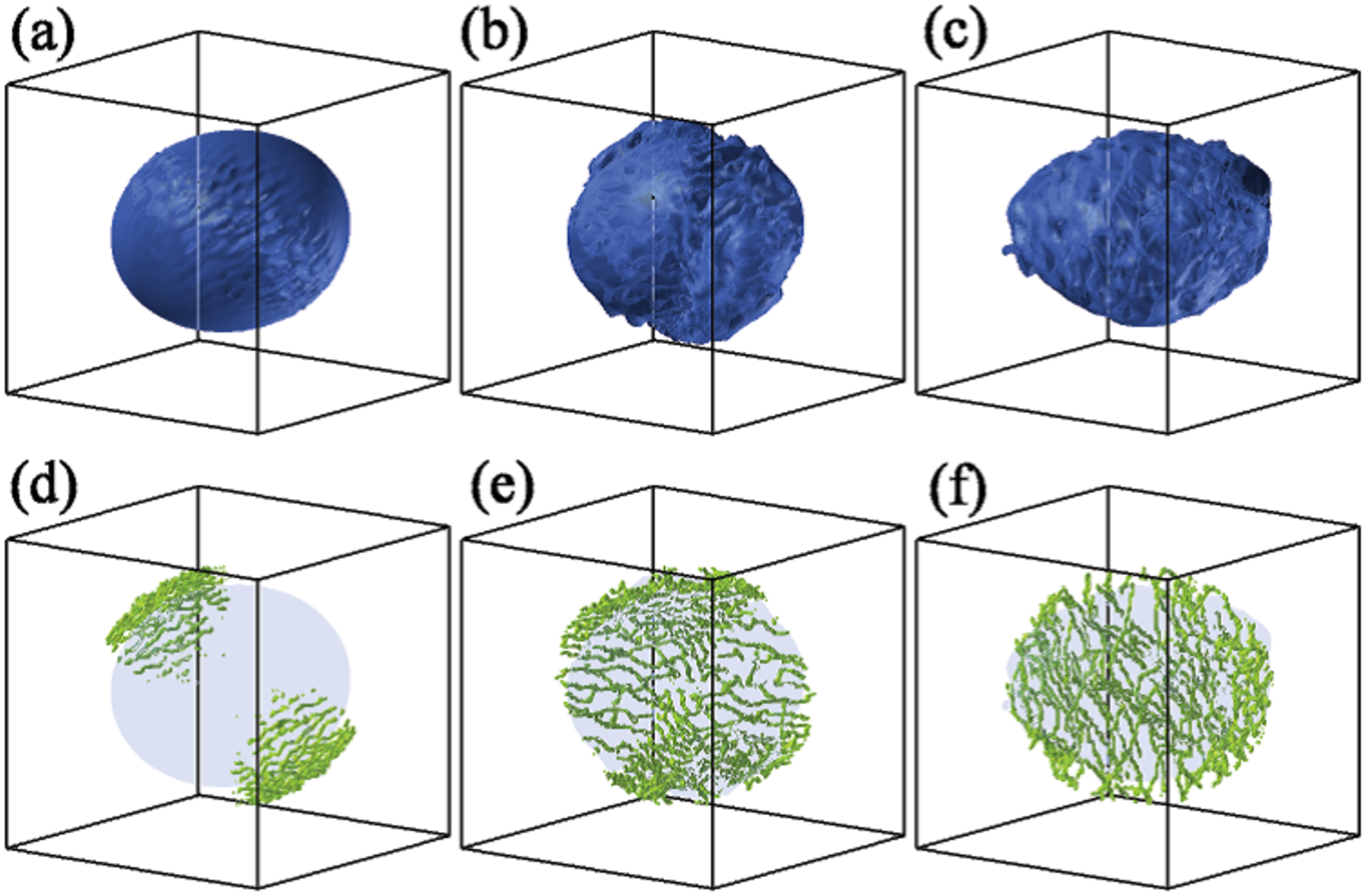}
\caption{\label{fig-BEC-turbulence} (a)--(c) Isosurface plots of $5 \%$ of the maximum condensate density $\rho$.
(d)--(f) Configuration of quantized vortices inside the Thomas--Fermi radius $R\sub{TF}$.
(a), (d) $t \omega = 10$, (b), (e) $t \omega = 50$, (c), (f) $t \omega = 300$.
The method for identifying vortices in (d)--(f) is the same as that in Fig. \ref{fig-turbulence-snapshot}. [Kobayashi and Tsubota: Phys. Rev. A {\bf 76} (2007) 045603, reproduced with permission. Copyright 2007 the American Physical Society.]}
\end{figure}

To visualize the  turbulence, we plot the isosurface of $\rho$ and the spatial distribution of the vortices inside the condensate in Figs \ref{fig-BEC-turbulence} (a)--(f).
At $t \omega = 10$, the surface of the BEC becomes unstable (Figs. \ref{fig-BEC-turbulence} (a) and (d)), and vortices appear in the BEC at $t \omega = 50$ (Figs. \ref{fig-BEC-turbulence} (b) and (e)).
Figures \ref{fig-BEC-turbulence} (c) and (f) shows QT with no crystallization but with highly tangled vortices at $t \omega = 300$.

Finally, we note that the obtained energy spectrum in Fig. \ref{fig-BEC-spectrum} (a) is not so clear straight line  and its consistency with the Kolmogorov law is incomplete.
This inconsistency comes from the anisotropy of turbulence around the $y$-axis around which there is no rotation, and is improved by other simulations of QT of trapped BECs under the precessional rotation around three axes \cite{Kobayashi:2008}.

Recently, a turbulent state has been realized in atomic BECs using two methods.
Weiler {\it et al.} performed a rapid quench of an $^{87}$Rb gas through the BEC transition temperature \cite{Weiler:2008}, which is similar to the method advocated by Berloff and Svistunov \cite{Berloff:2002}.
Through the high density fluctuation regime (weak turbulence) in a short period, several vortices and anti-vortices were formed, creating the system turbulence.
As a method with better control of the turbulence, the experimental group of Begnato introduced an external oscillatory perturbation to a $^{87}$Rb BEC \cite{Henn:2009a,Henn:2009b,Seman:2011}.
This oscillating magnetic field was produced by a pair of anti-Helmholz coils which were not perfectly aligned to the vertical axis of the cigar-shaped condensate.
Additionally, the components along the two equal directions that result in the radial symmetry of the trap were slightly different.
This oscillatory field induced a coherent mode excitation in a BEC.
For small amplitudes of the oscillating field and short excitation periods, dipolar modes, quadrupolar modes, and scissor modes of the BEC were observed, but no vortices appeared.
Increasing both parameters, the vortices grew in number, eventually leading to the turbulent state.
In the turbulent regime, they observed a rapid increase in the number of vortices followed by proliferation of vortex lines in all directions, where many vortices with no preferred orientation formed a vortex tangle.
Another remarkable feature is that a completely different hydrodynamic regime followed: the suppression of aspect ratio inversion during free expansion, despite the asymmetric expansion (from a cigar-shaped to pancake-shaped) of the usual quantum gas of bosons, or the isotropic expansion of a thermal cloud.
Although the theoretical understanding of this effect remains incomplete, it represents a remarkable new effect in atomic superfluids.

For a better understanding of the experimental results, we performed numerical simulations based on the GP equation \cite{Seman:2011}.
The net potential $V$ acting on the atoms is the sum of the harmonic magnetic trap and the oscillatory field, and can be approximately expressed by
\begin{align}
\begin{split}
V = \frac{M}{2} &\big[ \omega_x^2 \left\{ x \cos \theta_1 + y \sin \theta_1 - z \sin \theta_2 - \delta_1 ( 1 - \cos \Omega_0 t) \right\}^2 \\
&+ \omega_r^2 \left\{ y \cos \theta_1 - x \sin \theta_1 - \delta_2 (1 - \cos \Omega_0 t) \right\}^2 \\
&+ \omega_r^2 \left\{ z \cos \theta_2 + x \sin \theta_2 - \delta_3 (1 - \cos \Omega_0 t) \right\}^2 \big],
\end{split}
\end{align}
where $\theta_i = A_i (1 - \cos \Omega_0 t)$ are time dependent angles.
For the experimental conditions, $N = 3 \times 10^5$, $\omega_x = 2 \pi \times 23$Hz, $\omega_r = 2 \pi \times 210$Hz, $\Omega_0 = 2 \pi \times 200$Hz, $A_1 \simeq \pi / 60$, and $A_2 \simeq \pi / 120$ were used.
The amplitudes for the translational oscillation of the potential minimum are $( \delta_1, \delta_2, \delta_3) = \alpha a_r (2,  5, 3)$$\mu$m, where $a_r = \sqrt{\hbar/(M \omega_r)}$ and $\alpha$ is a variable parameter that represents the amplitude of the center-of-mass oscillation, being proportional to the amplitude of the excitation.
For simplicity, we employ $\Psi = \Psi_r(y,z) \Psi_x(x)$ and consider 2D simulations in $y$--$z$ space.
Here, we consider the BEC surrounded by a thermal cloud and use the constant $\gamma$ for the GP equation \eqref{eq-dissipated-GP}.
Since the thermal atoms, which are the origin of the dissipation, move together with the potential, we also have to consider the reference frame co-moving with the potential.
In this frame, the GP equation becomes
\begin{align}
(i - \gamma) \hbar \frac{\partial \Psi_r}{\partial t} = \left[ - \frac{\hbar^2 \nabla_r^2}{2 M} + V_r - \mu + g_r |\Psi_r|^2 - \Vec{\Omega}(t) \cdot \Vec{L} - \Vec{v}(t) \cdot \Vec{P} \right] \Psi_r,
\end{align}
with momentum $\Vec{P} = - i \hbar \nabla$ and angular momentum $\Vec{L} = - i \hbar \Vec{r} \times \nabla$.
Here $\nabla_r^2 = \partial^2 / \partial y^2 + \partial^2 / \partial z^2$ and $g_r = 4 \pi a / R_x$, with $R_x$ being the characteristic size of the condensate along $x$ axis.
We consider $\Vec{L} = (\Omega_x, 0, 0) \sin \Omega_0 t$ and $\Vec{v} = (0, v_y, 0) \sin \Omega_0 t$.
Using half of the oscillation period $T = \pi / \Omega_0$, we obtain $v_y \simeq 2 \delta_2 / T = 2 \Omega_0 \delta_2 / \pi$.
The rotation frequency $\Omega_x$ is also estimated as $\Omega_x \simeq 2 A_2 / T = \Omega_0 / 60$, providing a very small contribution.

\begin{figure}[htb]
\centering
\includegraphics[width=0.99\linewidth]{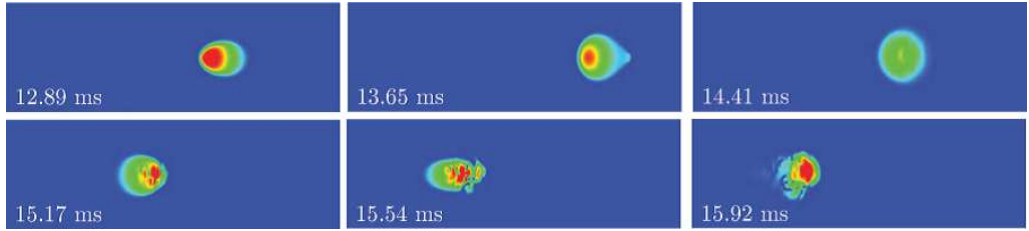}
\caption{\label{fig-turbulence-experiment} Snapshots of the BEC after different times of excitation \cite{Seman:2011}. Figure shows the 2D plot of the density profile. The colors range from red (high density) to blue (low density). [Seman, Henn, Shiozaki, Roati, Poveda-Cuevas, Magalh{\~a}es, Yukalov, Tsubota, Kobayashi, Kasamatsu, Bagnato: Laser Phys. Lett. {\bf 8} (2011) 691, reproduced with permission. Copyright 2011 WILEY-VCH Verlag GmbH \& Co. KGaA.]}
\end{figure}

Figure \ref{fig-turbulence-experiment} shows snapshots of the density profile for different excitation times ranging from $13$ to $17$ms.
Additionally, we have calculated the mean angular momentum per atom, $\langle L_x \rangle = \int d\Vec{r} \: \Psi_r^\ast L_x \Psi_r$, as a function of the excitation time.
Using $\alpha = 1.6$ and $\gamma = 0.02$, our simulation shows that $\langle L_x \rangle$ blows up after $15$ ms of excitation.
At this point, the condensate forms wavy patterns which develop to dark solitary waves which subsequently decay into several vortex pairs via the snake instability \cite{Feder:2000}.
A more complex dynamic takes place after the first events of vortex formation, consisting of the generation of an undetermined number of vortices, characterizing the emergence of the turbulent regime.
As $\alpha$ increases, the nucleation of vortices occurs at earlier times, with a faster evolution to QT.
This agrees well with the observations.
Furthermore, the time scale of the vortex events in the simulation, which is of the order of $10$ ms, is consistent with the times observed in the experiment.
These results demonstrate that the combination of rotation and translation is essential to produce vortices.
The dynamics of the BEC strongly depend on the strength of the dissipation $\gamma$.
If the dissipation is absent, no instability associated with soliton creation occurs.
Values of $\gamma$ between $0.015$ and $0.025$ are optimal for the generation of vortices and QT.
The simulations cannot reproduce the full experimental results, since the experimental system is a 3D gas.
Nevertheless, good qualitative agreement with the experiment has been achieved.

\subsection{Cascade process in quantum turbulence}

The most important concept for understanding QT is the cascade process of the energy and vortices as well as CT.
In QT, there are two regions of the cascade process in wavenumber space \cite{Vinen:2007}.
The first region is called the classical region below the inverse of the mean intervortex spacing.
The dynamics of vortices in the classical region are dominated by the Richardson cascade, in which large vortices are broken up self-similarly into smaller ones, or the collective dynamics of aggregated quantized vortices at scales larger than the intervortex spacing.
Such behavior of vortices supports the analogy of QT to CT, namely the Kolmogorov energy spectrum \cite{Vinen:2002}.
The second region is called the quantum region, in which vortex dynamics are dominated by the effects of the quantized circulation, specifically the Kelvin wave cascade of vortices, which does not appear in CT \cite{Svistunov:1995,Vinen:2001}.
The Kelvin wave cascade is also a very important concept in understanding the dissipation mechanism of QT at very low temperatures.

\begin{figure}[htb]
\centering
\includegraphics[width=0.95\linewidth]{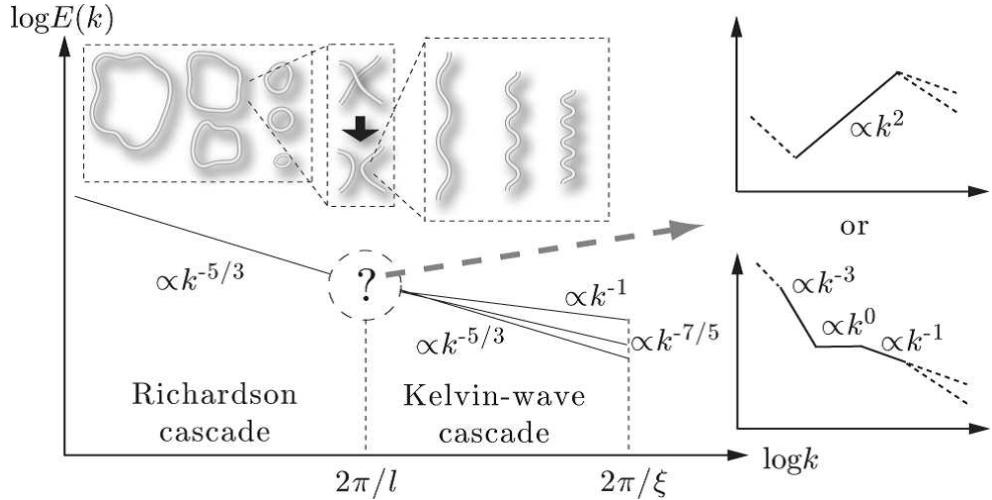}
\caption{\label{fig-overall-cascade} Overall picture of the energy spectrum of QT at zero temperature.
The energy spectrum depends on the scale and its properties change at about the scale of the mean intervortex spacing $l$.
When $k < k_l = 2 \pi / l$, a Richardson cascade of quantized vortices transfers energy from large to small scales, maintaining the Kolmogorov spectrum $E(k) = C \varepsilon^{2/3} k^{-5/3}$.
When $k > k_l$, energy is transferred by the Kelvin-wave cascade, which is a nonlinear interaction between Kelvin waves of different wavenumbers.
In this region, the energy spectrum also takes the power-law structure $E(k) \propto k^{-\eta}$ and the three values $\eta = - 7 / 5$, $\eta = - 5 / 3$, and $\eta = -1$ have been predicted.
Eventually, energy is dissipated at scales of $\xi$ by the radiation of elementary excitations.
The crossover region between Richardson cascading and Kelvin-wave cascading regions remains an open question, and two scenarios, $E(k) \propto k^2$ and $E(k) \propto k^{-3}$--$k^0$--$k^{-1}$, are proposed (see text).}
\end{figure}

Here, we briefly summarize the overall picture of the energy spectrum of QT at zero temperature \cite{Halperin:2009} based on theoretical and numerical studies (Fig. \ref{fig-overall-cascade}).
If a vortex tangle in QT is homogeneous and isotropic, there are two characteristic length scales: the mean intervortex spacing $l = L^{-1/2}$ with a vortex line length density $L$, and the healing length $\xi$ corresponding to the size of the vortex core.
$l$ is much larger than $\xi$ in the case of superfluid helium, while both are usually of the same order for atomic BEC.
Here, we consider the former case for simplicity.
Using $l$ and $\xi$, we can define the corresponding wavenumbers $k_l = 2\pi / l$ and $k_\xi = 2 \pi / \xi$.
At length scales larger than $l$, the dynamics of QT are dominated by a tangled structure of many vortices.
Because vortex dynamics become collective at large scales, quantization of the circulation is not relevant and the dynamics are similar to those of eddies in CT.
This is why this region can be referred to as the classical region.
As a result, the energy spectrum $E(k)$ in the range $k < k_l$ obeys the Kolmogorov law.
In the classical region, vortices sustain a Richardson cascade that transfers energy from smaller to larger wavenumbers without dissipation.
The Richardson cascade can be understood as large vortices breaking up into smaller ones in real space.

Vortices in QT can reconnect many times (see Fig. \ref{fig-reconnection}), which is the dominant dynamics at length scales comparable to $l$.
Through the reconnections, small cusps or distortion waves are formed on the vortex lines, which are regarded as the primary source of Kelvin waves in QT \cite{Tsubota:2000,Svistunov:1995,Kivotides:2001}.
The wavelength of the created Kelvin waves is of the order of $l$.

At length scales smaller than $l$, which is referred to as the quantum region, the Richardson cascade is no longer dominant and the quantized circulation of vortices and motion of each vortex line become significant \cite{Vinen:2000,Svistunov:1995,Samuels:1990,Vinen:2003,Kozik:2004,Kozik:2005,Nazarenko:2006,Boffetta:2009,Lvov:2010,Baggaley:2011}.
In this range, vortex dynamics are characterized by the cascade process of the Kelvin waves formed by reconnection.
The nonlinear interaction of the Kelvin waves is the origin of the cascade from small to large wavenumbers.
The energy spectrum in the quantum region $k_l < k < k_\xi$ is theoretically predicted to obey a Kolmogorov-like power law: $E(k) \propto k^\eta$.
Three theoretical values of $\eta$ have been predicted: $\eta = -7/5$ \cite{Vinen:2000,Kozik:2004,Kozik:2005,Nazarenko:2006,Boffetta:2009}, $\eta = - 5/3$ \cite{Lvov:2010,Boue:2011}, and $\eta = -1$ \cite{Kivotides:2001,Vinen:2003,Nazarenko:2006,Boffetta:2009}.
At finite temperatures where the mutual friction between superfluid and normal fluid is effective, Kelvin waves and relevant turbulent flow are strongly dissipated by the viscosity of the normal fluid, and the cascade of Kelvin waves turns off.

In the region $k \sim k_\xi$, the Kelvin waves with wavelength $\xi$ change to elementary excitations, such as phonons and rotons, in the primary decay process of QT near zero temperature \cite{Vinen:2001}.

There is one open question regarding the energy spectrum in the region $k \sim k_l$, namely the transitional region between the Richardson cascade and the Kelvin-wave cascade, referred to as the classical--quantum crossover.
A theoretical study proposed a bottleneck effect connecting the spectrum satisfying the thermalization spectrum $E(k) \propto k^2$ \cite{Lvov:2007}.
Another theoretical prediction is based on vortex reconnection dynamics on the scale $\sim l$, in which the transitional cascade process was predicted to occur by reconnection of the vortex bundle \cite{Kozik:2008}.
The crossover range is divided into three subranges, giving $E(k) \propto k^{-3}$, $E(k) \propto k^{0}$, and $E(k) \propto k^{-1}$ in each region.
We discuss this spectrum in Sec. \ref{subsubsec-classical-quantum}

\subsubsection{Classical region} \label{subsubsec-classical}

There is a significant issue regarding the energy spectrum in the classical region $k < k_l$ in  terms of the analogy of QT to CT.
In this region, the energy spectrum is determined by the collective behavior of many vortices, such as the vortex tangle and the aggregated bundle structure at scales larger than $l$.
Several numerical studies have calculated the energy spectrum in this region by simulating QT at zero temperature.
We have numerically found and already discussed the Kolmogorov energy spectrum \eqref{eq-Kolmogorov-energy-spectrum} by using the GP model in Sec. \ref{subsec-GP} (Sec. \ref{subsubsec-turbulence-zero-temperature}).
Araki {\it et al.} also found the Kolmogorov energy spectrum through the vortex-filament model \cite{Araki:2002}.

In our numerical studies discussed in Sec. \ref{subsubsec-turbulence-zero-temperature}, however, the system size was not so large and the inertial range was less than one order in wavenumber space.
Furthermore, the mean intervortex spacing $l$ was close to the healing length $\xi$, being too short to study the Kelvin wave cascade.
To obtain the energy spectrum of a wider range of wavenumber space, Yepez {\it et al.} performed a large-scale simulation of the GP model by using a novel unitary quantum lattice gas algorithm \cite{Yepez:2009}.
They found that the incompressible kinetic spectrum $E\sub{kin}\up{i}$ had three distinct power-law $k^{-\alpha}$ regions that ranged from the classical turbulent regime of Kolmogorov $\alpha = 5/3$ at large scales $k < (\sqrt{3} / (2 \pi)) L / \xi$ to the quantum Kelvin-wave cascades $\alpha = 3$ at small scales $k > (\sqrt{3} / 2) L / \xi$.
There was a semiclassical region $6.34 \lesssim k \lesssim 7.11$ connecting the Kolmogorov and Kelvin-wave spectra $(\sqrt{3} / (2 \pi)) L / \xi < (\sqrt{3} / 2) L / \xi$.
Compared with our simulations, this simulation supplied the Kolmogorov spectrum over a much wider inertial range, with about two orders in wavenumber space.
Although they related the $k^{-3}$ spectrum at small scales $k > (\sqrt{3} / 2) L / \xi$ to the Kelvin-wave cascade, the length scale in this region is smaller than the vortex core size.
Hence the $k^{-3}$ spectrum comes most probably not from the Kelvin-wave cascade but from the velocity profile abour a vortex, as pointed by several authors \cite{LN:2010,KB:2010,Yepez:2010}.

To clarify the Kolmogorov energy spectrum in the classical region and the energy spectrum in the classical--quantum crossover, Sasa {\it et al.} also performed a simulation of the GP equation at much larger scales than that discussed in Sec. \ref{subsubsec-turbulence-zero-temperature}, {\it i.e.}, from $L = 128 \xi$ to $512 \xi$ \cite{Sasa:2011}.
In contrast to the work by Yepez {\it et al.}, Sasa {\it et al.} focused on the classical region (and classical--quantum crossover)of $k < l^{-1}$.
We numerically solved the GP equation \eqref{eq-dissipated-GP-Fourier} with the dissipation term $\tilde{\gamma}$ of the step function form \eqref{eq-step-function-dissipation} with $\gamma_0 = 1$.
We investigated the decaying turbulence without energy injection, and a uniform density $\rho = 1$ and a random $\phi$ as the initial wave function to create our system turbulence.
The random phase is arranged in the same way as in our previous work discussed in Sec. \ref{subsubsec-turbulence-zero-temperature} for decaying turbulence, {\it i.e.}, placing random numbers between $-\alpha \pi$ to $\alpha \pi$ at every distance $\lambda$ and connecting them smoothly, where $\alpha$ is the control parameter for the energy injection.

\begin{figure}
\centering
\includegraphics[width=0.95\linewidth]{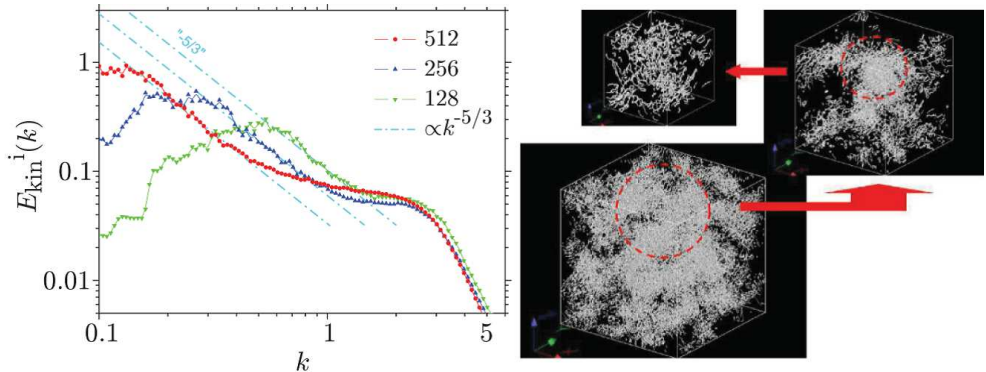}
\caption{\label{fig-Big-turbulence-spectrum} Left: Simulation results for the incompressible kinetic energy spectra $E\sub{kin}\up{i}(k)$. $\Lambda$ varies from $\Lambda \simeq 1.5$ for $L = 128$ to $\Lambda \simeq 2.2$ for $L = 512$.
Dot-dashed line: Kolmogorov spectrum.
In the figure, length and energy are normalized by $\xi$ and $\hbar^2 / (4 M \xi^2)$.
Right: A snapshot of vortex lines at the fully developed turbulent state of $L = 512$ demonstrating the self-similarity of the bundle structure (see the dotted circles representing the zoomed regions whose vortex distributions are shown subsequently), typical for fully developed turbulence. [Sasa, Kano, Machida, L'vov, Rudenko, and Tsubota: Phys. Rev. B {\bf 84} (2011) 054525, reproduced with permission. Copyright 2011 the American Physical Society.]}
\end{figure}

The main numerical results are shown in Fig. \ref{fig-Big-turbulence-spectrum}.
The left panel shows the incompressible kinetic energy spectrum $E\sub{kin}\up{i}(k)$.
The Kolmogorov spectrum extends to the lower-$k$ range and increases with the system size.
The visible extent of the Kolmogorov spectrum is much larger than that in all the results including our previous simulations.
The right panel of Fig. \ref{fig-Big-turbulence-spectrum} displays large self-similar structures of tangled vortices in the fully turbulent state: large-scale vortex bundles in the maximum size, $512 \xi$, and smaller self-similar tangled structures inside this cubic region in the subsequent insets.
The visualization of vortices clearly shows the bundle structure, which has never been confirmed in GP simulations in smaller boxes.

An important observation is a plateau-like region for $k \gtrsim 1.5 \xi$, which is a definite pileup over the Kolmogorov spectrum; this is a clear manifestation of energy stagnation.
We expect this plateau-like structure to be an indication of the bottleneck effect \cite{Lvov:2007} proposed by L'vov {\it et al.} in the classical--quantum crossover region, which will be discussed in Sec. \ref{subsubsec-classical-quantum}.
Detailed comparisons are discussed in Ref. \cite{Sasa:2011}.

\subsubsection{Quantum region}

In the region of $k > k_l$, the picture of aggregated vortices is no longer effective and the motion of each vortex line becomes essential.
The most probable dominant dynamics of vortices are Kelvin waves, which originate from distortion waves on the vortex lines after their reconnection.
A Kelvin wave is a transverse, circularly polarized wave motion, with the approximate dispersion relation for a rectilinear vortex:
\begin{align}
\omega_k = \frac{\kappa k^2}{4 \pi} \left\{ \log \left(\frac{1}{k \xi}\right) + c \right\} \label{eq-Kelvin-wave-dispersion}
\end{align}
with a dimensionless constant $c \sim 1$.
$k$ is the wavenumber of the Kelvin wave, and is different from that used for the energy spectrum.
Kelvin waves were first observed by inducing torsional oscillations in a rotating superfluid $^{4}$He \cite{Hall:1958,Hall:1960}.

Although the Kelvin-wave cascade seems to be a very important mechanism in QT at scales smaller than $l$ at very low temperatures, it is a non-trivial problem regarding the actual cascade process.
At finite temperatures where there is a significant fraction of normal fluid, Kelvin waves are damped by mutual friction.
On the other hand, at very low temperatures they can be damped only by the radiation of phonons.
Vinen estimated the rate of radiation, and found that it is extremely low unless the frequency is very high \cite{Vinen:2001}.
In QT, the main origin of the Kelvin-wave nucleation is the vortex reconnection.
When two vortices reconnect, they twist to become locally antiparallel at the reconnection point and create small cusps or kinks after the reconnection which were confirmed numerically \cite{Schwarz:1985,Schwarz:1988,Tsubota:2000}.
Svistunov suggested that the relaxation process of these cusps or kinks causes the emission of Kelvin waves, and plays an important role in the decay of QT at low temperatures.
Following Svistunov, Vinen {\it et al.} performed a numerical simulation of a Kelvin wave excited along a single vortex line using the vortex filament model discussed in Sec. \ref{filament} \cite{Vinen:2003}.
The authors consider a model system in which helium is contained in the space between two parallel sheets, separated by a distance $\ell\sub{B} = 1$cm, with a single, initially rectilinear, vortex stretched between opposite points on the two sheets.
Kelvin waves can be excited on this vortex, and periodic boundary conditions are applied at each end.
The allowed wavenumbers of the Kelvin waves are given by
\begin{align}
k = \frac{2 \pi n}{\ell\sub{B}},
\end{align}
where $n$ is a natural number.
The authors imagine that the mode with a small integer $n_0$ is continuously driven.
As the amplitude increases nonlinearly, coupling to other modes occurs and they can expect energy to flow from the mode $n_0$ to other modes with both large and small wave numbers.
To mimic the effect of phonon emission, they introduce strong damping  for all modes with $n$ exceeding a larger critical value $n\sub{c}$.
Then, they can obtain a statistically steady state in which the energy injection in the mode with $n_0$ is balanced by dissipation in the modes with $n > n\sub{c}$ and calculate the corresponding energy spectrum.
The authors observe no reconnection.

\begin{figure}[htb]
\centering
\includegraphics[width=0.95\linewidth]{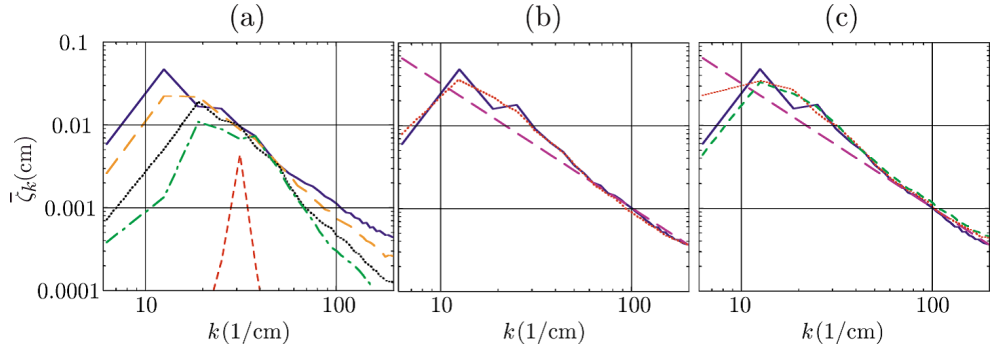}
\caption{\label{fig-Kelvin-wave-cascade} (a) Time development of $\bar{\zeta}_k(t)$ in the Kelvin-wave cascade. The short-dashed, dash-dotted, dotted, long-dashed, and solid lines refer, respectively, to averages over $0$--$800$, $10000$--$10800$, $20000$--$20800$, $40000$--$40800$, and $140000$--$140800$ s.
(b) Steady state values of $\bar{\zeta}_k$ for two different drive amplitudes. The solid line and dotted line are for, respectively, $V = 2.5 \times 10^{-5}$ cm s$^{-1}$ and $V = 2.5 \times 10^{-4}$ cm s$^{-1}$.
The long-dashed line has the form of Eq. \eqref{eq-Kelvin-wave-spectrum-Mitani}.
(c) Steady state values of $\bar{\zeta}_k$ for three different wavenumbers. The dotted, short-dashed, and solid lines refer, respectively, to $k_0 = 2 \pi $ cm$^{-1}$, $k_0 = 4 \pi$c m$^{-1}$, and $k_0 = 10 \pi$ cm$^{-1}$.
Again the long-dashed line has the form of Eq. \eqref{eq-Kelvin-wave-spectrum-Mitani}. [Vinen, Tsubota, and Mitani: Phys. Rev. Lett. {\bf 91} (2003) 135301, reproduced with permission. Copyright 2003 the American Physical Society.]}
\end{figure}

The simulations are based on the vortex filament model with the full Biot--Savart law based on Eq. \eqref{eq-s0dot}.
The force that drives one mode is of the form $V \rho \kappa \sin(k_0 z - \omega_0 t)$, where $k_0 = 2 \pi n_0 / \ell\sub{B}$, $\rho$ is the density of the helium, and $\omega_0$ is related to $k_0$ by the dispersion relation \eqref{eq-Kelvin-wave-dispersion}.
Damping at the highest wavenumber $k\sub{c} = 2 \pi n\sub{c} / \ell\sub{B} = 1/ 60$cm$^{-1}$ is applied using a periodic smoothing process.
We calculate the root mean square amplitudes $\bar{\zeta}_k(t) = \langle \zeta_k^\ast \zeta_k \rangle^{1/2}$ of the Fourier components of the displacement of the vortex from the original straight line.
Figure \ref{fig-Kelvin-wave-cascade} (a) shows how these amplitudes develop in time after the application of a drive with $V = 2.5 \times 10^{-5}$cm s$^{-1}$ and $k_0 = 10 \pi$cm$^{-1}$.
We see that initially only the mode of $k_0$ is excited.
However, as time passes, nonlinear interactions lead to excitation of all other modes.
Eventually the spectrum reaches a statistically steady state.
For large values of $k$, where the modes practically form a continuum, the steady state is observed to have, to a good approximation, a spectrum of the simple form
\begin{align}
\bar{\zeta}_k^2 = A \ell\sub{B}^{-1} k^{-3}, \label{eq-Kelvin-wave-spectrum-Mitani}
\end{align}
where the dimensionless parameter $A$ is of order unity.

Figures \ref{fig-Kelvin-wave-cascade} (b) and (c) show the effects, respectively, of increasing the drive amplitude $V$ by a factor of $10$ and of changing the drive wavenumber $k_0$.
We see that there is little effect on the steady state, within the error of the simulations, at least at the higher wavenumbers.
The steady state takes longer to be established at the lower drive amplitude, which suggests that even with a small drive amplitude, the same steady state would be established after a sufficiently large time.

The mean energy per unit length of a vortex in mode $k$ is related to $\bar{\zeta}_k$ by the equation
\begin{align}
E\sub{K}(k) = \epsilon\sub{K} k^2 \bar{\zeta}_k^2, \label{eq-Kelvin-wave-energy}
\end{align}
where $\epsilon\sub{K}$ is an effective energy per unit length of vortex, given by
\begin{align}
\epsilon\sub{K} = \frac{\rho \kappa^2}{4 \pi} \left\{ \log \left(\frac{1}{k a}\right) + c \right\}.
\end{align}
It follows from Eqs. \eqref{eq-Kelvin-wave-spectrum-Mitani} and \eqref{eq-Kelvin-wave-energy} that
\begin{align}
E\sub{K}(k) = A \epsilon\sub{K} (k \ell\sub{B})^{-1}. \label{eq-Kelvin-wave-energy-spectrum-Mitani}
\end{align}
Thus the steady state is characterized by the energy spectrum \eqref{eq-Kelvin-wave-energy-spectrum-Mitani}, and this spectrum is insensitive to the frequency and amplitude of the drive and to the power input at the drive frequency.

What happens to this kind of Kelvin-wave cascade in QT?
When two vortices reconnect, they twist to become locally antiparallel at the reconnection point and create small cusps or kinks after the reconnection.
Svistunov suggested that the relaxation of these cusps or kinks causes the emission of Kelvin waves \cite{Svistunov:1995}.
Following this suggestion, Vinen analyzed the energy spectrum of the Kelvin-wave cascade by introducing a smoothed length of vortex line per unit volume after all the Kelvin waves were removed, and considered $E\sub{K}(k) dk$, the energy per unit length of the smoothed vortex lines associated with Kelvin waves in the range $k$ to $k + dk$ \cite{Vinen:2001}.
By the dimensional analysis, $E\sub{K}(k)$ was estimated as
\begin{align}
E\sub{K}(k) = A \rho \kappa^2 k^{-1}, \label{eq-Kelvin-wave-energy-spectrum-Vinen}
\end{align}
with a constant $A$ of order unity.
This form is consistent with Eq. \eqref{eq-Kelvin-wave-energy-spectrum-Mitani}.

Kivotides {\it et al.} numerically confirmed the generation of Kelvin waves through reconnections using the vortex filament model \cite{Kivotides:2001}.
They also calculated the energy spectrum $E(k)$ defined by \eqref{eq-energy-spectrum-definition} for the superfluid velocity, and found that $E(k)$ developed approximately a $k^{-1}$ form.
Kivotides suggested that the fluctuations of the superfluid velocity field were induced by the Kelvin waves on the filament, {\it i.e.}, $E(k) \sim E\sub{K}(k)$, and their result was consistent with Vinen's analysis of Eq. \eqref{eq-Kelvin-wave-energy-spectrum-Vinen}.

The result of Eq. \eqref{eq-Kelvin-wave-energy-spectrum-Mitani}, Vinen's analysis \eqref{eq-Kelvin-wave-energy-spectrum-Vinen}, and Kivotides's result show the energy spectrum in the quantum regime to be $E(k) \propto k^{-1}$.
However, several forms of $E(k)$ have been theoretically predicted by various approaches.
Kozik and Svistunov analyzed the Kelvin-wave cascade considering the three-kelvon scattering process within the weak-turbulence theory \cite{Kozik:2004,Kozik:2005} and obtain the spectrum $E(k) \propto k^{-7/5}$.
Nazarenko and Boffetta {\it et al.} presented a nonlinear differential equation model, pointing out that turbulence displays a dual cascade behavior of both the direct energy cascade supporting $E(k) \propto k^{-7/5}$ and the inverse cascade supporting $E(k) \propto k^{-1}$ of wave action \cite{Boffetta:2009,Nazarenko:2006}.
L'vov and Nazarenko also derive the new cascade scenario due to the four-Kelvin-wave scattering process showing the energy spectrum $E(k) \propto k^{-5/3}$ \cite{Lvov:2010}.
Yepez {\it et al.} suggest the $E(k) \propto k^{-3}$ from the numerical result of the GP model as discussed in the previous section \cite{Yepez:2009}.
Bou\'e {\it et al.} discuss the discrepancies of these spectrum, especially, the spectra proposed by Kozik and Svistunov, and L'vov and Nazarenko, and point out that it comes from the difference between local (Kozik and Svistunov) and nonlocal (L'vov and Nazarenko) theories for Kelvin-wave dynamics \cite{Boue:2011}.
They also perform the numerical simulation showing an agreement with the nonlocal predictions.
Kozik and Svistunov also discuss the importance of the symmetry and related Noether's constants of motion in the Kelvin-waves when discussing the locality or its absence \cite{Kozik:2010}.

\subsubsection{Classical--quantum crossover} \label{subsubsec-classical-quantum}

As discussed in the previous sections, there are two different types of energy spectra in the classical $(k < k_l)$ and quantum $(k_l < k < k_\xi)$ regions.
We have addressed an important question: How do these two energy spectra connect to each other at the length scale $l$?
Although there are several theoretical and numerical reports on this region, consistency among these works has not yet been obtained.
In the analysis of this classical--quantum crossover, $\Lambda = \log(l/\xi)$ appears to be an important parameter.
In typical $^4$He experiments, $\Lambda$ is about $15$.

Kozik and Svistunov suggested a picture for the crossover region, in which the locally induced motion of the vortex lines emerges at the scale of $r_0 \sim \Lambda^{1/2} l$, and the crossover range is divided into three subranges, $r_0^{-1} < k < \lambda\sub{b}^{-1}$, $\lambda\sub{b}^{-1} < k < \lambda\sub{c}^{-1}$, and $\lambda\sub{c}^{-1} < k < \lambda_\ast^{-1}$, where $\lambda\sub{b} \sim \Lambda^{1/4} l$, $\lambda\sub{c} \sim l / \Lambda^{1/4} $, and $\lambda_\ast \sim l / \Lambda^{1/2}$ \cite{Kozik:2008}.
In the first region, $r_0^{-1} < k < \lambda\sub{b}^{-1}$, polarized vortex lines are organized in bundles and reconnect with other bundles to form Kelvin waves with amplitude $\bar{\zeta}_k \sim r_0 k^{-1}$.
In the second region, $\lambda\sub{b}^{-1} < k < \lambda\sub{c}^{-1}$, the cascade is supported by nearest neighbor reconnections in a bundle, and $\bar{\zeta}_k \sim l (\lambda\sub{b} k)^{-1/2}$.
In the third range, $\lambda\sub{c}^{-1} < k < \lambda_\ast^{-1}$, the cascade is driven by self-reconnection of vortex lines, giving $\bar{\zeta}_k \sim k^{-1}$.
The Kelvin-wave spectrum $\bar{\zeta}_k$ smoothly connects these ranges.
Although they emphasized that the energy spectrum $E(k)$ is practically meaningful only in the classical region, we can estimate $E(k)$ from their model: $E(k) \propto k^{-3}$ in $r_0^{-1} < k < \lambda\sub{b}^{-1}$, $E(k) \propto k^{0}$ in $\lambda\sub{b}^{-1} < k < \lambda\sub{c}^{-1}$, and $E(k) \propto k^{-1}$ in $\lambda\sub{c}^{-1} < k < \lambda_\ast^{-1}$.

L'vov {\it et al.} suggested another scenario for the classical--quantum crossover: bottleneck crossover between the two regions \cite{Lvov:2007}.
For $k \sim l^{-1}$ and $\Lambda \gg 1$, the energy of Kelvin waves is much larger than the hydrodynamic energy due to superfluid velocity by vortices at the same energy flux.
As a result, there is a bottleneck energy accumulation around $k \sim l^{-1}$ and the energy spectrum becomes $E(k) \propto k^2$ for superfluid velocity and $E\sub{K}(k) \propto k^{0}$ for Kelvin waves, followed by equipartition of the hydrodynamic energy and the energy of Kelvin waves, respectively.
This scenario is completely different from that suggested by Kozik and Svistunov because there is no energy stagnation in the model by Kozik and Svistunov.

\begin{figure}[hbt]
\centering
\includegraphics[width=0.5\linewidth]{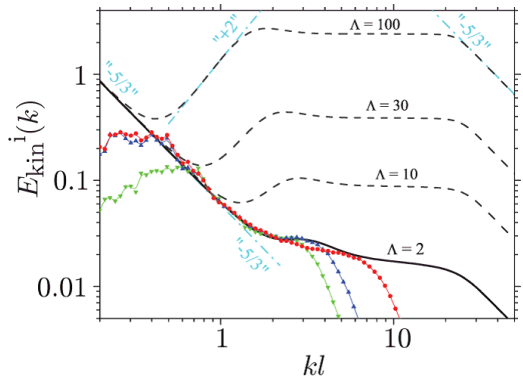}
\caption{\label{fig-Big-turbulence-comparison} Incompressible energy spectra plotted vs. $k l$.
The simulation results (the same symbols as in Fig. \ref{fig-Big-turbulence-spectrum}) and the model by L'vov {\it et al.} for $\Lambda = 10$, $30$, $100$ (dashed curves) are brought together to the theoretical (solid) curve with $\Lambda = 2$ by superposing the Kolmogorov spectrum (for both simulations and model) and plateau regions (only for simulations.
The dot-dashed lines show different scaling asymptotes. [Sasa, Kano, Machida, L'vov, Rudenko, and Tsubota: Phys. Rev. B {\bf 84} (2011) 054525, reproduced with permission. Copyright 2011 the American Physical Society.]}
\end{figure}
As discussed in Sec. \ref{subsubsec-classical}, the large-scale numerical simulation of the GP model quantitatively supports the existence of a bottleneck effect in the crossover region \cite{Sasa:2011}.
Figure \ref{fig-Big-turbulence-comparison} shows a comparison between the numerical result and the theoretical prediction for different $\Lambda$. (For the sake of a better comparison we replotted the simulation data, bringing them all together to the model by L'vov {\it et al.} with $\Lambda = 2$ by superposing the Kolmogorov spectrum and plateau regions.)
However, $\Lambda$ is too small to predict the exact mechanism and the comparison shown in Fig. \ref{fig-Big-turbulence-comparison} remains problematic.
More theoretical studies and numerical and laboratory experiments are required to fully understand the vortex dynamics in the scale crossover region.

\section{Quantum hydrodynamic instability in two-component Bose--Einstein condensates}
\label{two-component}

Hydrodynamic instability is of fundamental importance in classical fluid dynamics \cite{Chandrasekhar:1981, Kundu:2008}.  
 The instability causes characteristic wavy patterns developing into turbulent flows from a basic flow after complex dynamics of eddies.
 Because of the universal applicability of the theory,
 hydrodynamic instability appears in different kinds of fluids.
 Superfluids are no exception.

  Although superfluid dynamics are described by hydrodynamic equations similar to those in classical fluid dynamics,
 the macroscopic quantum effects, {\it i.e.}, superfluidity and vortex quantization,
 can cause significant differences between the hydrodynamic instabilities in quantum and classical fluid systems.
 Superfluidity, {\it i.e.} flow without friction, enables us to study different hydrodynamic instability arising from frictionless basic flows.
 Such an instability has no classical counterparts because frictionless flows
 cannot be achieved in classical fluid systems because of the  presence of viscosity.
 On the other hand, the appearance of a quantized vortex should cause a significant difference between the quantum and classical fluids at least in the nonlinear stage of instability where vortices are relevant.
 Therefore, quantum effects cause differences in both linear stability and nonlinear dynamics between quantum and classical hydrodynamic instability.

 In this section, we discuss hydrodynamic instability in quantum fluids, namely, quantum hydrodynamic instability, especially in two-component BECs.
 Historically, the study of quantum hydrodynamic instability has been developed in helium superfluid systems.
 First, we shall briefly review hydrodynamic instability in superfluid systems focusing on helium superfluid systems.
Then, we introduce hydrodynamic instability in two-component BECs.

\subsection{Hydrodynamic instabilities in superfluid systems}
 The Landau instability \cite{KhalatnikovBook1965} is the fundamental mechanism for the linear stability of frictionless flows.
 The Landau instability occurs when frictionless states become unstable when the superflow velocity relative to the external environment, such as the container wall or the thermal excitations dragged by the wall, exceeds a critical value.
 This instability is a thermodynamic instability,
 where elementary excitations with negative energy are spontaneously amplified in the relaxational process, decreasing the thermodynamic energy of the system.

 The most important example of quantum hydrodynamic instability is the instability of thermal counterflow in superfluid $^4$He.
 The thermal counterflow instability has been studied in parallel with QT in superfluid $^4$He \cite{Vinen:1957a, Vinen:1957b, Vinen:1957c, Vinen:1957d} as was introduced in Sec. \ref{cf}.
 The thermal counterflow instability occurs when the relative velocity between the normal fluid and superfluid components exceeds a critical value.
 Then, remnant vortices attached to the container wall are stretched by the mutual friction,
 and the stretched vortices repeat reconnections, leading to QT.

 The stretch of remnant vortices in the thermal counterflow instability may be understood more fundamentally as an instability of single quantized vortices, the Kelvin-wave instability,
sometimes called the Donnelly--Glaberson instability \cite{Glaberson:1974, Ostermeyer:1975, 1965DonnellyPRL}.
 Kelvin-wave instability can occur when there is a relative helical flow along a vortex line between normal fluid and superfluid components,
 which is realized by injecting a heat current along the quantized vortices.
The instability leads to amplification of the Kelvin waves and helical deformations of the vortex line.
 When the Kelvin-wave instability is induced in rotating superfluids with vortex lattices,
 the instability can develop into QT \cite{RotatingTurbExp,RotatingTurbNum}.
The Kelvin-wave instability and its counterpart have been discussed for other superfluids and superconductors.
 The instability can occur in superfluid $^3$He-B in a rotating cylinder,
 where the rotating counterflow of a vortex-free superfluid component and a rotating normal fluid component is realized \cite{KWI3He}.
 In atomic BECs, it was proposed that the Kelvin-wave instability can be induced as a spontaneous excitation of a {\it kelvon}, a quantum of a Kelvin wave due to the Landau instability \cite{Takeuchi2009PRA79}.
  The Kelvin-wave instability has also been discussed for quantized vortices in a rotating neutron star \cite{Peralta:2006,Glampedakis:2008}.
 In type-II superconductors,
 the counterpart of the Kelvin-wave instability is the spiral-vortex expansion instability \cite{Clem1977PRL38},
 where a flux vortex becomes unstable against the growth of helical perturbations in the presence of a sufficiently large current density applied parallel to its axis. 

 The above examples of quantum hydrodynamic instability are phenomena that do not occur in classical fluids.
 Also of interest is the study of the quantum counterpart of classical hydrodynamic instability.
 The Kelvin--Helmholtz instability, one of the most fundamental instabilities in classical fluid dynamics, was first studied experimentally in a superfluid system at the interface between the A and B phases of superfluid ${}^3$He \cite{Helsinki}. 
  When the relative velocity between the two phases exceeds a critical value, the penetration of quantized vortices across the interface was detected by counting the number of vortices by NMR before and after the instability\footnote{The instability observed in the experiment \cite{Helsinki} is the thermodynamic instability triggered by the Landau instability of ripplons of the interface between the A and B phases.
 Strictly speaking, this instability is not the counterpart of the classical Kelvin--Helmholtz instability,
 which is the dynamic instability of ripplons. See the following section.}.
 The Kelvin--Helmholtz instability in superfluid systems has been discussed on different kinds of interfaces,
 such as normal--superfluid interfaces \cite{Henn:2009a,Korshunov:2002}, nuclear--nuclear superfluid interfaces \cite{KHI-neutron-star}, and superfluid--superfluid interface in atomic two-component BECs \cite{KHItwoBEC}.

 Recently, there has been growing interest in hydrodynamic instability in atomic BECs.
 The unique dynamics due to quantized vortices have been reported on the quantum counterparts of classical hydrodynamic instabilities,
the Kelvin--Helmholtz instability \cite{KHItwoBEC}, Rayleigh--Taylor instability \cite{Sasaki:2009, Gautam:2010},  Strouhal instability \cite{SIinBEC}, Richtmyer--Meshkov instability \cite{RMIinBEC}, and Plateau--Rayleigh (capillary) instability \cite{PRIinBEC}, among others.
 There are several merits in considering an atomic BEC system to study quantum hydrodynamic instability.
 The most important advantage of such a system over other superfluid systems is that we can visualize directly the whole time development of the order parameters in the instability dynamics,
 from the linear stage to the nonlinear development including vortex nucleation \cite{Neely2010}.
 In addition, since the system is less dissipative at ultra low temperatures,
 we can observe the hydrodynamic instabilities, which are obscured by dominant thermodynamic instability in dissipative systems.
 The theoretical advantage of this system is that we can predict quantitatively the detailed dynamics of quantum hydrodynamic instability within the mean field approximation.

 Multi-component atomic BECs provides an ideal ground to study novel hydrodynamic phenomena of multi-superfluid systems.
 Two-component atomic BECs are the simplest systems of multi-component superfluids \cite{KTUreview},
 which can be created in cold-atom systems with multiple hyperfine spin states or a mixture of different atomic species.
 Recent experimental advances enable us to study a variety of superfluid dynamics in two-component BECs in a more controllable manner.
 The intra- and inter-component interactions can be tuned with the help of the the Feshbach resonance \cite{Thalhammer:2008, Papp:2008,Tojo:2010},
 and the external potentials are controllable independently on both components by utilizing the difference between the Zeeman shifts of the two components.
 Recently, hydrodynamic instability of counterflows in miscible two superfluids, called countersuperflow instability, was observed for the first time in two-component BECs by Hamner {\it et. al.} \cite{HamnerCEH}.
 Coutersuperflow instability is unique to multi-component superfluid systems.
 It was also suggested that countersuperflow instability can develop into a binary QT composed of two superfluids \cite{TakeuchiIT2010PRL105}.

 In the following subsection, we introduce our recent work on quantum hydrodynamic instability in two-component BECs.
 The next subsection is devoted to introducing the hydrodynamic formalism for two-component BECs to make it easier to understand the subsequent subsections.
 Then we develop the discussion into concrete problems, countersuperflow instability \cite{TakeuchiIT2010PRL105, 2011IshinoPRA} and quantum Kelvin-Helmholtz instability \cite{KHItwoBEC, 2010SuzukiPRA},
 which are the most fundamental hydrodynamic instabilities in two-component BECs.


\subsection{Linear stability and hydrodynamic formalism}


 Let us start with the Lagrangian of two-component BECs \cite{Pethick:2008},
\begin{eqnarray}
L=i\frac{\hbar}{2}\int dV\sum_j\left(\Psi_j^*\partial_t \Psi_j-\Psi_j\partial_t \Psi_j^*\right)-K
,
\label{eq:Lagrangian_2cmp}
\end{eqnarray}
where the index $j$ refers to the $j$th component.
If we take into account the motion of an external environment with
the translational velocity ${\Vec V}_{ex}$ and the angular velocity ${\Vec \Omega}_{ex}$,
 the thermodynamic energy $K$ is generally written as \cite{Landau_Stat}
\begin{eqnarray}
K=\int dV\left[{\cal K} -({\Vec V}_{ex}+{\Vec \Omega}_{ex}\times{\Vec r})\cdot{\Vec J}\right],
\end{eqnarray}
 where
${\cal K}$ is the thermodynamic energy density for ${\Vec V}_{ex}={\Vec \Omega}_{ex}=0$, and
 ${\Vec J}= \frac{\hbar}{2i}\sum_j(\Psi_j^*{\Vec \nabla}\Psi_j-\Psi_j{\Vec \nabla}\Psi_j^*)$ is the total momentum density of the two components.
 The energy density ${\cal K}$ is written as ${\cal K}={\cal K}_1+{\cal K}_2$ with
\begin{eqnarray}
{\cal K}_j=\frac{\hbar^2}{2m_j}|{\Vec \nabla}\Psi_j|^2+(U_j-\mu_j)|\Psi_j|^2
+\frac{1}{2}\sum_{k}g_{jk}|\Psi_j|^2|\Psi_k|^2,
\end{eqnarray}
 where $m_j$, $U_j({\Vec r})$, and $\mu_j$ are the particle mass, the external potential, and the chemical potential of the $j$th component, respectively.
 The inter- and intracomponent interaction constants $g_{jk}$ have the form $g_{jk}=2\pi\hbar^2a_{jk}(m_j^{-1}+m_k^{-1})$, where $a_{jk}$ is the $s$-wave scattering length between the $j$th and $k$th components.
For simplicity, the environment is supposed to be at rest in the laboratory frame throughout our discussion, namely ${\Vec V}_{ex}={\Vec \Omega}_{ex}=0$.

 From the Lagrangian (\ref{eq:Lagrangian_2cmp}), one obtain the coupled GP equations,
\begin{eqnarray}
i\hbar\partial_t\Psi_j
=\left[-\frac{\hbar^2}{2m_j}{\Vec \nabla}^2+U_j-\mu_j
+\sum_k g_{jk}|\Psi_k|^2\right]\Psi_j.
\label{eq:GP2cmp}
\end{eqnarray}
 Steady superflows are realized as a stationary solution $\Psi_j({\Vec r},t)=\Phi_j({\Vec r})$ of the GP equations (\ref{eq:GP2cmp}).
 For example, without external potential $U_j=0$,
 the GP equations (\ref{eq:GP2cmp}) have the stationary solutions $\Phi_j({\Vec r})\propto e^{im_j{\Vec V}_j\cdot{\Vec r}/\hbar}$ of steady uniform superflows with arbitrary velocity ${\Vec V}_j$.

To clarify the relation to fluid dynamics, we introduce a hydrodynamic formalism.
 By inserting $\Psi_j=f_j e^{i\theta_j}$ into the Lagrangian density ${\cal L}$,
 we obtain
\begin{eqnarray}
{\cal L}
&=&
-\sum_j\Biggl\{
\hbar  n_j \partial_t \theta_j
+\frac{m_jn_j}{2}{\Vec v}_j^2
\nonumber \\
&& \ \ \ \ \ \ \ \ \
+\frac{\hbar^2}{2m_j}\left({\Vec \nabla}\sqrt{n_j}\right)^2
+(U_j-\mu_j)n_j
+\frac{1}{2}\sum_{k}g_{jk}n_kn_j
\Biggr\},
\end{eqnarray}
 where  $n_j=f_j^2$ and ${\Vec v}_j=\frac{\hbar}{m_j}{\Vec \nabla}\theta_j$ is the density and the superfluid velocity of the $j$th component, respectively.
 The variation with respect to $f_j$ and $\theta_j$ yields a set of hydrodynamic equations
\begin{eqnarray}
&&
\partial_t n_j+{\Vec \nabla}\cdot (n_j{\Vec v}_j)=0,
\label{eq:continum} \\
&&
m_j\partial_t  {\Vec v}_j
=
-{\Vec \nabla}\left[\frac{m_j}{2}{\Vec v}_j^2+q_j+U_j+\mu^h_j \right],
\label{eq:motion}
\end{eqnarray}
 where 
$q_j=-\frac{\hbar^2}{2m_j}({\Vec \nabla}^2f_j)/f_j$ is the quantum pressure term,
 and $\mu^h_j=\sum_k g_{jk}n_k$ is the hydrostatic chemical potential,
 which is named after the hydrostatic pressure in fluid dynamics.
 These hydrodynamic equations are analogs of those of multi-phase fluids in classical fluid dynamics.
 The first equation represents the conservation law of density $n_j$
 and the second has a similar form to the Euler equation of irrotational flows.
 If we neglect the inter-component interaction, $g_{12}=0$,
 the term ${\Vec \nabla}\mu_j$ reduces to $\frac{1}{n_j}{\Vec \nabla}p_j^h
$ with the hydrostatic pressure
\begin{eqnarray}
p_j^h=\frac{1}{2}g_{jj}n_j^2
\label{eq:p_hyd}
\end{eqnarray}
 of the $j$th component.
In a stationary state, the second equation reduces to
\begin{eqnarray}
\frac{m_j}{2}{\Vec v}_j^2+q_j+U_j+\mu^h_j =\mu_j={\rm const.} ,
\label{eq:Bernoulli}
\end{eqnarray}
which is the counterpart of the Bernoulli theorem.

 The interaction between different components comes from the term $g_{jk}n_k~(j\neq k)$ in the hydrostatic chemical potential $\mu^h_j$.
 The force density ${\Vec f}_{jk}$ on the $j$th component by the $k$th component is written as
\begin{eqnarray}
{\Vec f}_{jk}=-g_{jk}n_j{\Vec \nabla}n_k.
\end{eqnarray}
The inter-component force ${\Vec F}_{jk}=\int dV {\Vec f}_{jk}$ obeys the principle of action and reaction, ${\Vec F}_{jk}=-{\Vec F}_{kj}$.
 The force vanishes for uniform density profiles $n_j={\rm const.}$,
 which makes it possible to realize a stationary relative flow ${\Vec v}_1 \neq {\Vec v}_2 $ between different components, namely, a countersuperflow state.

Linear stability analysis can be done in a manner similar to that in classical fluid dynamics.
 To perform the linear stability analysis, we introduce small perturbations
\begin{eqnarray}
n_j ({\Vec r}, t) &=& \bar{n}_j ({\Vec r})+\delta{n}_j ({\Vec r}, t),
\label{eq:linear_continuity}\\
\theta_j({\Vec r}, t) &=& \bar{\theta}_j({\Vec r})+\delta{\theta}_j ({\Vec r}, t),
\\
{\Vec v}_j ({\Vec r}, t) &=& \bar{\Vec v}_j ({\Vec r})+\delta{\Vec v}_j ({\Vec r}, t),
\end{eqnarray}
 where $\bar{\Vec v}_j =\frac{\hbar}{m_j}{\Vec \nabla}\bar{\theta}_j$ and $\delta{\Vec v}_j=\frac{\hbar}{m_j}{\Vec \nabla} \delta{\theta}_j$.
By linearizing Eqs. (\ref{eq:continum}) and (\ref{eq:motion}), we obtain the linearized equations
\begin{eqnarray}
&&
\partial_t \delta n_j+{\Vec \nabla}\cdot(\bar{n}_j\delta {\Vec v}_j+\delta{n}_j\bar{\Vec v}_j)
=0
\label{eq:linear_continue}\\
&&
m_j\partial_t\delta{\Vec v}_j=-{\Vec \nabla}\left[
m_j\bar{\Vec v}_j\cdot\delta{\Vec v}_j+\delta q_j+\delta \mu_j^h
\right],
\label{eq:linear_motion}
\end{eqnarray}
where 
\begin{eqnarray}
\delta q_j=\frac{\hbar^2}{4m_j\bar{n}_j^3}
[
\bar{n}_j({\Vec \nabla}^2\bar{n}_j)
+\bar{n}_j({\Vec \nabla}\bar{n}_j)\cdot{\Vec \nabla}
-({\Vec \nabla}\bar{n}_j)^2
-\bar{n}_j^2{\Vec \nabla}^2]\delta n_j
\end{eqnarray}
 and
$\delta \mu_j^h= \sum_k g_{jk}\delta n_k$. 
 These equations determine the linear stability of the stationary state.

 As the first example we consider the linear stability of uniform superflows without relative velocity:
 $\bar{\Vec v}=\bar{\Vec v}_1=\bar{\Vec v}_2={\rm const.}$, $V_j({\Vec r})={\rm const.}$ and $\bar{n}_j({\Vec r})={\rm const.}$
 The perturbations may be written as $\delta n_j \propto \cos({\Vec q}\cdot{\Vec r}-\omega t),~\delta \theta_j \propto \sin({\Vec q}\cdot{\Vec r}-\omega t)$ with the wave number ${\Vec q}$ and the frequency $\omega$.
The linearized equations (\ref{eq:linear_continue}) and (\ref{eq:linear_motion}) are reduced to the eigenvalue equations for the eigenvalue $\omega_0\equiv \omega-\bar{\Vec v}\cdot{\Vec q}$,
 and we obtain the dispersion
\begin{eqnarray}
\omega=\bar{\Vec v}\cdot{\Vec q}
\pm\sqrt{\frac{1}{2}(\omega _{1}^2+\omega _{2}^2)\pm\frac{1}{2}\sqrt{(\omega_{1}^2-\omega_{2}^2)^2+4c_{12}^4q^4}},
\label{eq:dispersion_nonrel}
\end{eqnarray}
with $\omega_j^2=c_{jj}^2q^2+\frac{\hbar^2}{4m_j^2}q^4$ and
$c_{jk}^2=\sqrt{g_{jk}^2\frac{\bar{n}_j\bar{n}_k}{m_jm_k}}$.
 The first term comes from the Doppler shift due to the background superflow $\bar{\Vec v}$ and the second term refers to the eigenvalue $\omega_0$.

 The thermodynamic stability of superflows or the Landau instability is investigated from the deviation $\delta K$ of the thermodynamic energy $K$ due to perturbations.
 By using the equations (\ref{eq:linear_continue}) and (\ref{eq:linear_motion}), the deviation energy $\delta K$ is written as
\begin{eqnarray}
\delta K=\frac{\hbar}{2}\int dV\sum_j
\left(
\delta \theta_j\partial_t\delta n_j -\delta n_j\partial_t\delta \theta_j
\right).
\end{eqnarray}
 The deviation energy $\delta K$ due to the perturbation is proportional to the frequency $\omega$ (\ref{eq:dispersion_nonrel}).
 Since $\omega$ changes its sign by increasing the velocity $\bar{v}=|\bar{\Vec v}|$ with the eigenvalue $\omega_0$ fixed,
 $\delta K$ becomes negative for a perturbation when $\bar{v}$ exceeds a critical velocity $v_L$, called the Landau critical velocity.
Then the perturbation is amplified to decrease the thermodynamic energy of the system.
 From the dispersion (\ref{eq:dispersion_nonrel}), the Landau critical velocity is given by 
 \begin{eqnarray}
 v_L =\sqrt{\frac{1}{2}(c_{11}^2+c_{22}^2)-\frac{1}{2}\sqrt{(c_{11}^2-c_{22}^2)^2+4c_{12}^4}}.
 \end{eqnarray}

 The frequency can take a complex value, where the perturbations are exponentially amplified as $\propto e^{\sigma t}$ with $\sigma= {\rm Im}~\omega>0$.
 Then the system is called dynamically unstable.
 The dynamic instability occurs when $g_{12}^2>g_{11}g_{22}$ for the dispersion (\ref{eq:dispersion_nonrel}).
 The instability means that homogeneous condensates are unstable;
 if $g_{12}>\sqrt{g_{11}g_{22}}$, the two components will undergo phase separations due to the strong inter-component interaction,
 while if $-g_{12} > \sqrt{g_{11}g_{22}}$, the condensates are unstable to the formation of a denser droplet containing both components.
 These dynamic instabilities are classified as {\it hydrostatic} instability rather than hydrodynamic instability in the sense that they occur without superflow, $\bar{\Vec u}_j=0$.

 Dynamic instability is purely an internal instability in isolated systems without external environments,
 where the particle numbers $N_j=\int dV n_j$ and the energy $E=K+\sum_j\mu_jN_j$ are conserved.
On the other hand, the Landau instability in this case is a hydrodynamic instability induced by the frictional dissipation between the condensates and the external environment.
 If the system under consideration is dissipative,
rather than dynamically unstable, the Landau instability is dominant on hydrodynamic instability.
 The frictional dissipation can be small and thus dynamic instability can be dominant in atomic BECs when condensates are trapped in `a smoothed-wall container' made by electromagnetic fields, at low temperatures, where there is a small amount of the normal fluid component.

  In the following subsections,
 we discuss the countersuperflow instability and the Kelvin--Helmholtz instability,
 which are fundamental instabilities in the presence of relative velocity between different components.
 These instabilities belong to the dynamic instability attributed to the relative flow of two condensates,
 although the Landau instability occurs due to the relative motion between the environment and condensates.

\subsection{Countersuperflow instability in miscible two-component BECs}
 It has already been mentioned that countersuperflow states can be realized as a basic flow in two-component BECs. 
It is interesting to compare countersuperflow with thermal counterflow of superfluid $^4$He.
 Both states are a counterflow of two miscible fluid components, which has no analog in classical fluid dynamics. 
 The latter is a counterflow of normal fluid and superfluid components,
 which becomes thermodynamically unstable when the relative velocity exceeds a critical value.
 The former is a counterflow of two superfluids.
 At first sight, it seems that the countersuperflow state can be stable for arbitrary relative velocities
 since each component is a superfluid by itself.
 However, the countersuperflows in two-component BECs become dynamically unstable over a critical relative velocity \cite{2001LawPRA63,Yukalov:2004}.
 In this subsection, we discuss the linear stability and nonlinear dynamics of countersuperflow instability in two-component BECs.

\subsubsection{Linear stability of countersuperflows}
Let us consider uniform counterflows in two-component BECs in an isolated homogeneous system,
$n_j=\bar{n}_j={\rm const.}$ and ${\Vec v}_j=\bar{\Vec v}_j={\rm const.}$ with relative velocity ${\Vec v}_R\equiv \bar{\Vec v}_2-\bar{\Vec v}_1\neq 0$.
Because of the Galilean invariance,
 we can neglect the translational motion of the whole system without loss of generality,
$({\Vec P}_1+{\Vec P}_2)=M_1\bar{\Vec v}_1+M_2\bar{\Vec v}_2=0$ with the total momentum ${\Vec P}_j=m_j\int dV n_j{\Vec v}_j$  and the total mass $M_j=m_jN_j$  of the $j$th component.

 The linear stability of the countersuperflows is evaluated using
 the coupled linearized equations (\ref{eq:linear_continue}) and (\ref{eq:linear_motion}).
 The problem is reduced to solving the equations
\begin{eqnarray}
(\partial_t+\bar{\Vec v}_j\cdot{\Vec \nabla})^2\delta \bar{n}_j
=\frac{\bar{n}_j}{m_j}
{\Vec \nabla}^2(\delta \mu_j+\delta q_j).
\label{eq:linear_homo}
\end{eqnarray}
By substituting $\delta n_j \propto \cos({\Vec q}\cdot{\Vec r}-\omega t)$ into Eq. (\ref{eq:linear_homo}),
we obtain
\begin{eqnarray}
[(\omega-\bar{\Vec v}_1\cdot{\Vec q})^2-\omega_1^2][(\omega-\bar{\Vec v}_2\cdot{\Vec q})^2-\omega_2^2]=c_{12}^4q^4.
\end{eqnarray}
 Although the stability is investigated by solving this quartic equation,
 the eigenvalue has a complicated form in general.
 However, for a symmetric case $m_{11}=m_{22}=m$, $n_{1}=n_{2}=n$, and $g_{11}=g_{22}=g$, which is a reasonable approximation, {\it e.g.}, for two-component BECs of $^{87}$Rb atoms \cite{Tojo:2010, HamnerCEH},
the dispersion relation is reduced to a simple form
\begin{eqnarray}
\varepsilon'^2
=\frac{1}{4}{\Vec q}'^4+{\Vec q}'^2+\frac{1}{4}q_{\parallel}'^2V_{R}'^2 
\pm\sqrt{\bigl(\frac{1}{4}{\Vec q}'^4+{\Vec q}'^2\bigr)q_{\parallel}'^2V_{R}'^2+{\Vec q}'^4\gamma^2},
\label{eq:omega}
\end{eqnarray}
 where $\varepsilon'=\hbar\omega/gn$, ${\Vec q}'={\Vec q}\xi$ with $\xi=\hbar/\sqrt{mgn}$, $V'_{R}=|{\Vec V}'_{R}|=|{\Vec v}_{R}|/c$ with $c=\sqrt{gn/m}$, and $\gamma = g_{12}/g$.
 Here ${\Vec q}'^2=q_{\parallel}'^2+q_{\perp}'^2$ with $q_{\parallel}'=|{\Vec q}'\cdot{\Vec V}'_{R}/V'_{R}|$ and $q_{\perp}' \geq 0$.
 By comparing the last term with the sum of the other terms on the right hand side in Eq. (\ref{eq:omega}),
the condition ${\rm Im}~\varepsilon' \neq 0$ for the dynamic instability is found to be
 \begin{eqnarray}
   \sqrt{\frac{1}{4}{\Vec q}'^4+{\Vec q}'^2(1-\gamma)}<\frac{1}{2}q'_{\parallel}V'_{R}<\sqrt{\frac{1}{4}{\Vec q}'^4+{\Vec q}'^2(1+\gamma)}.
   \label{eq:condition}
 \end{eqnarray}
 It is reasonable to expect that this inequality is never satisfied without the inter-component interaction ($\gamma=0$).

 \begin{figure}
   \centering
   \includegraphics[width=.7 \linewidth]{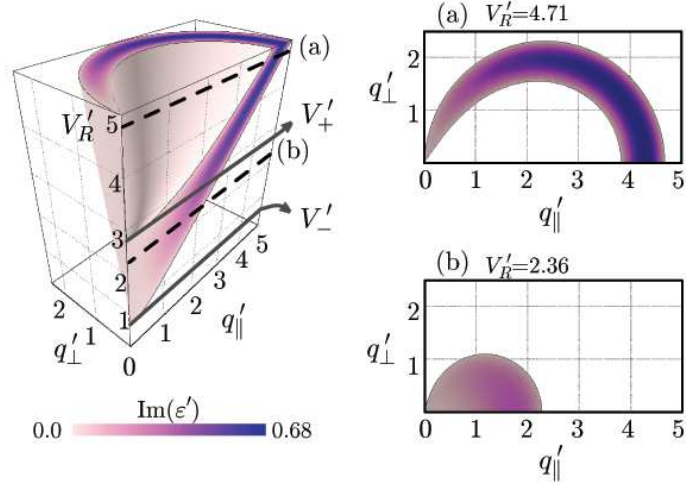}
   \caption{
     Phase diagram of the countersuperflow instability for $\gamma=0.9$.
     The dashed lines represent (a) $V'_{R}=4.71$ and (b) $2.36$ and the solid lines show the critical velocities $V_{\pm}'=2\sqrt{1\pm \gamma}$. 
 The two right-hand plots show the cross-section surfaces of $V'_{R}=4.71$ and $2.36$ of the phase diagrams. [Ishino, Tsubota and Takeuchi: Phys. Rev. A {\bf 83} (2011) 063602, reproduced with permission. Copyright 2011 the American Physical Society.]
   }
   \label{fig:souzu}
 \end{figure}%

 Figure \ref{fig:souzu} shows the phase diagram of the countersuperflow instability for $\gamma=0.9$.
 The unstable region is characterized by the lower and upper critical velocities
 \begin{eqnarray}
   V'_{\pm}=2\sqrt{(1\pm|\gamma|)}.
   \label{eq:Vc}
 \end{eqnarray}
 The unstable region appears when $V'_{R}$ exceeds the lower critical velocity $V'_{-}$.
 The distribution of the unstable modes, which have larger values of $|{\rm Im}~\varepsilon'|$, depends on the relative velocity.
 For sufficiently large relative velocity with $V'_{R}>V'_{+}$,
 the cross section of the unstable region has a crescent-like form [Figs. \ref{fig:souzu} (a)].
In this case, we find that the unstable mode with large values of $|{\rm Im}~\varepsilon'|$ are distributed in the region with higher wave number $q'_{\perp}$.
 On the other hand, if $V'_{R}$ decreases below $V'_{+}$,
 the unstable region is broadly distributed around $q'_{\perp}= 0$ [Fig. \ref{fig:souzu} (b)].

 \begin{figure*}[htb]
   \centering
   \includegraphics[width=1. \linewidth]{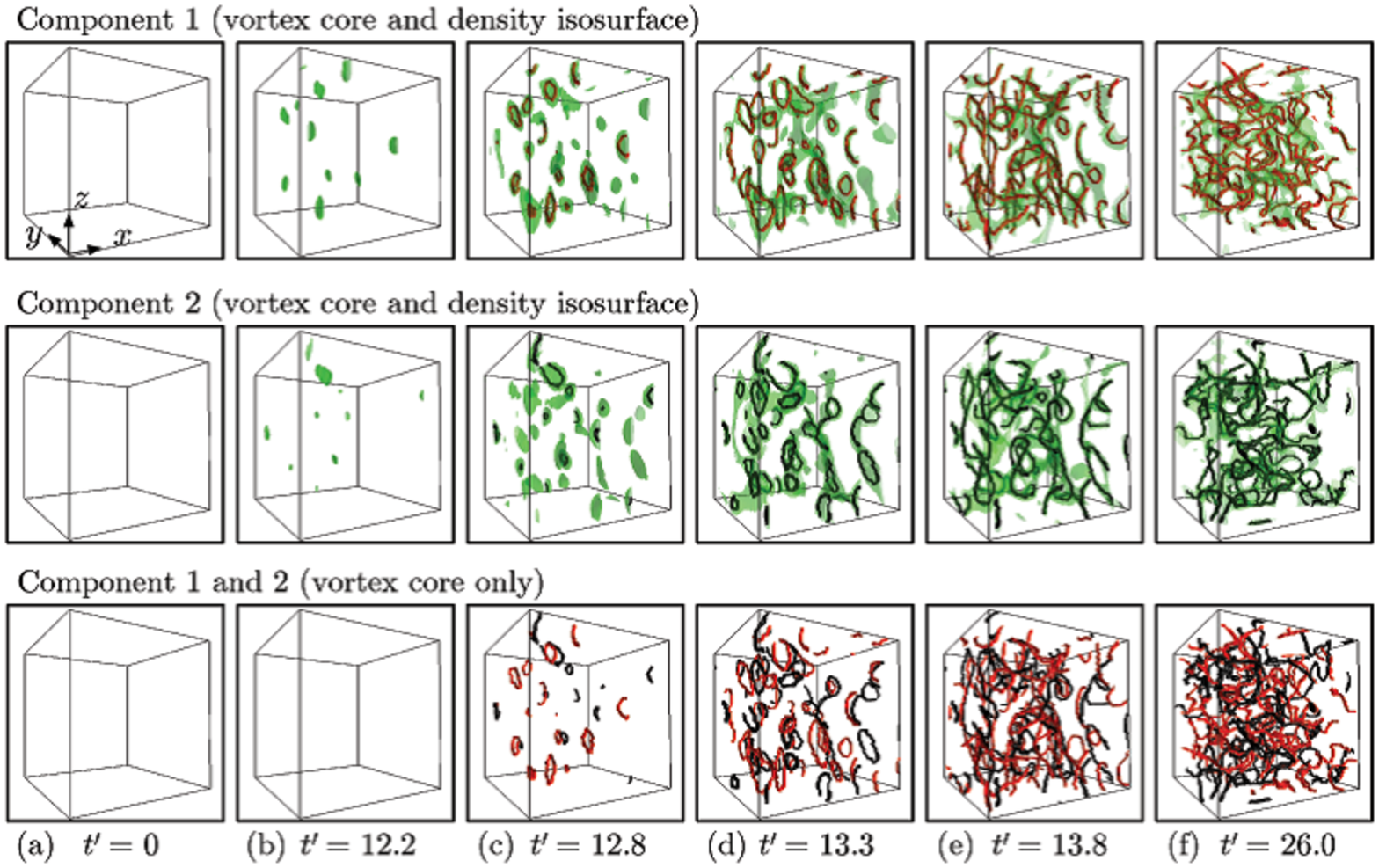}
   \caption{
 Nonlinear dynamics of vortex cores in countersuperflow instability.
 The top and middle panels show the vortex dynamics in the first and second components, respectively.
 The surface plots represent the low density isosurface of $|\Psi_1|^2/n=0.1$ (top) and $|\Psi_2|^2/n=0.1$ (middle).  [Ishino, Tsubota and Takeuchi: Phys. Rev. A {\bf 83} (2011) 063602, reproduced with permission. Copyright 2011 the American Physical Society.]
 }
   \label{fig:3D}
 \end{figure*}%

\subsubsection{Nonlinear development of countersuperflow instability}

 The distribution of the unstable modes in the wave number space strongly affects the nonlinear dynamics of the vortex nucleation after the linear amplification of the modes.
 Figure \ref{fig:3D} shows a typical time development of the countersuperflow instability,
 obtained by numerical simulation of the GP equations (\ref{eq:GP2cmp}).
 The numerical simulations were done in a three-dimensional box under periodic boundary conditions with the parameter settings $m_1=m_2=m$, $g_{11}=g_{22}=g$, $n_1=n_2=n$, $\bar{\Vec v}_1=-\bar{\Vec v}_2$, $\mu_j=m{\Vec v}_R^2/8+(g+g_{12})n$, $\gamma=0.9$, and $V'_R=4.71$.
 We add a small amount of white noise in the initial countersuperflow state to trigger the instability.
Because of the symmetric parameters between the two components,
 the nonlinear dynamics are similar for both components.

 After the exponential amplification of the unstable modes,
 disk-shaped low-density regions appear in both components,
 which face in the direction parallel to the initial relative velocity [Fig. \ref{fig:3D}(b)],
 and then vortex rings are nucleated inside the regions  [Fig. \ref{fig:3D}(c)].
 Since the size of the vortex rings is similar to those of the low density regions,
 the vortex distribution is characterized by the density pattern emerging after the onset of the instability.
 Thus, if the instability is so strong that the density pattern grows soon into vortex rings,
 the vortex number density immediately after the instability is estimated by the wavelength of the most unstable modes in the phase diagram.
 For the limit of large relative velocity $V_R'\gg V_+'$,
 the unstable region determined by the inequality (\ref{eq:condition}) is reduced to
$ (q'_{\parallel}-\frac{1}{2}V'_{R})^2+q_{\perp}'^2=\frac{1}{4}V_{R}'^2$,
 and then the vortex line density is estimated to be $\sim {\Vec v}_{R}^2/\kappa^2$.

For the case of small relative velocity, e.g. $V'_{-}<V'_{R}<V'_{+}$, the size of the nucleated vortex rings becomes large because the unstable modes with larger imaginary part are distributed mainly around $q_\bot=0$.
 If the perpendicular wavelength $2\pi/q_\bot$ of the unstable modes is similar to or larger than the system size normal to the relative flow,
 the instability does not cause nucleation of vortex rings,
 but rather makes vortex lines across the system or distorted stripe patterns, as observed by Hamner {\it et al.} \cite{HamnerCEH}. 

 Typical dynamics after the vortex ring nucleation are demonstrated in Figs. \ref{fig:3D}(d)--(f).
 The vortex rings propagate along the initial relative velocity in the opposite direction between the two components, and the ring size increases with time.
 When vortex rings come close to each other,
 the rings are distorted by the interaction and make reconnections [Fig. \ref{fig:3D}(e)].
 The distortion and the reconnection mainly occur between the vortices in the same components
 because the vortex--vortex interaction between the same component is larger than that between different components \cite{2011EtoPRA}; reconnection does not occur between vortices in the different components.
 Quantized vortices in both components become tangled due to the distortion,
 forming binary QT [Fig. \ref{fig:3D}(f)].

 \begin{figure*}[htb]
   \centering
   \includegraphics[width=.5 \linewidth]{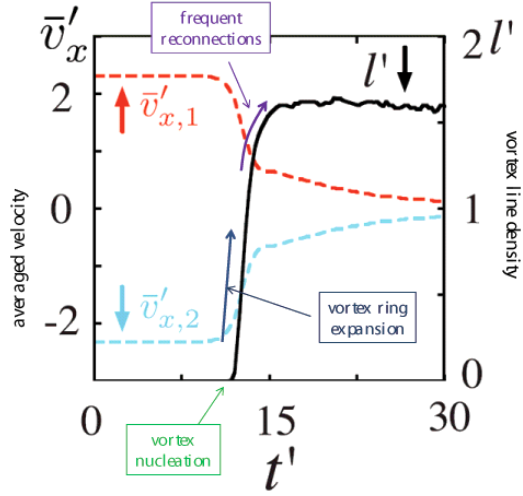}
   \caption{
Time evolution of the total vortex line density $l=l'\xi^{-2}$ and the velocity $V_{x,j}=v_{x,j}' \xi/\tau$ of the $j$th component along the initial relative velocity. The time and length are scaled by $\xi=\hbar/\sqrt{mgn}$ and $\tau=\hbar/\mu$.
 }
   \label{fig:exchange}
 \end{figure*}%

 The time development of the countersuperflow instability can be considered as the frictional relaxation of the relative motion of the two condensates.
 In general, a frictional force between two interacting objects causes a decrease in their relative motion and their kinetic energy is dissipated in various forms of energy, {\it i.e.}, the internal energy such as heat.
 Here, we define the macroscopic kinetic energy $E_{kin}$ of the two condensates as
\begin{eqnarray}
E_{kin}=\frac{{\Vec P}_2^2}{2M_1}+\frac{{\Vec P}_2^2}{2M_2}=\frac{M'}{2}{\Vec V}_R^2
\end{eqnarray} 
 with the macroscopic relative velocity ${\Vec V}_R\equiv{\Vec P}_2/M_2-{\Vec P}_1/M_1$  and the macroscopic reduced mass $M'=(M_1^{-1}+M_2^{-1})^{-1}$.
 The corresponding `internal energy' $E_{int}$ is defined by subtracting $E_{kin}$ from the total energy $K$; $E_{int}\equiv K-E_{kin}$.
 Then, the frictional force ${\Vec F}_R$ between the two components can be defined from the time variation of the relative kinetic energy $E_R\equiv\frac{1}{2}M'{\Vec V}_R^2$:
\begin{eqnarray}
\frac{d}{dt}E_R={\Vec F}_R\cdot{\Vec V}_R,
\end{eqnarray}
 with the frictional force
\begin{eqnarray}
{\Vec F}_R\equiv M' \frac{d}{dt}{\Vec V}_R.
\label{eq:FrictionalForce}
\end{eqnarray}
 Since $M'$ is the conserved quantity,
 ${\Vec F}_R$ is proportional to the time variation of the relative velocity ${\Vec V}_R$.

 Figure \ref{fig:exchange} shows the time evolutions of
 the averaged velocity ${\Vec V}_j={\Vec P}_j/M_j$ of the $j$th component and the sum of the vortex line density of the two components.
 Throughout the development,
 the velocity ${\Vec V}_j$ is almost perpendicular to the initial velocity $\bar{\Vec v}_j \parallel \hat{\Vec x}$ with the unit vector $\hat{\Vec x}$  along the $x$-axis,
 and only the parallel component $V_{x,j}={\Vec V}_j\cdot \hat{\Vec x}$ is plotted in Fig. \ref{fig:exchange}.
 From the relation (\ref{eq:FrictionalForce}),
 the reduction rate of $|V_{x, 2}-V_{x, 1}|$ is proportional to the frictional force between the two components.

 The friction between the two components is small in the linear stage of the instability, but grows drastically as the vortex line density increases.
 The decrease in the kinetic energy $E_R$ by the friction is offset with the increase in the `internal energy' by nucleating and expanding vortex rings.
 On the other hand, the vortex reconnection suppresses the friction
 since the distortion of the vortex ring configuration due to the reconnection disturbs the free expansion of the vortex rings.
 In addition,
 the vortex line length can be decreased after vortex reconnections
 since some of the energy is dissipated for phonon emission.
 The vortex line density increases to a maximum value
 when the two effects, the vortex ring expansion and the vortex reconnection, are balanced in the vortex-tangled state.
 The frictional relaxation continues but its rate decreases in the tangled state.
 The total length starts to decrease when the relative velocity becomes almost zero,
 and then the binary QT will decay.

 The dynamics of the turbulence transition in countersuperflow instability is similar to that in thermal counterflow instability.
 Recall that in the thermal counterflow instability,
 quantized vortices are stretched by the mutual friction between the superfluid and normal fluid components.
 The steady QT developed from the thermal counterflow instability is anisotropic since the relative velocity between the two components is sustained externally by applying a temperature gradient through the system.
On the other hand,
 the counterpart of the mutual friction is caused by
 the friction between the two condensates in the countersuperflow system.
 Quasi-steady turbulence is realized temporarily when the vortex line density approaches the maximum value.
 It is numerically shown that the maximum vortex line density is proportional to the square of the initial relative velocity,
 similarly to the relation of Eq. (\ref{Lvns}),
 but the vortex tangle can be isotropic when the momentum exchange is completed and the relative velocity vanishes \cite{2011IshinoPRA}.

\subsubsection{Conclusion}
 Countersuperflow states in miscible two-component BECs become dynamically unstable when the relative velocity between the different components exceeds a critical value.
 The countersuperflow instability causes vortex nucleation and stretching vortices leading to isotropic binary QT.
 The time development of the countersuperflow instability is interpreted as frictional reduction of the relative motion of two condensates.
 These phenomena are interesting in two senses.
 One is that the relative superflows decay due to the mutual friction between two superfluids,
 each of which consists of the `frictionless' superfluid component by itself.
 The other is that the QT of multi-component BECs can be realized from the countersuperflow instability.
 These phenomena can be observed with current experimental techniques \cite{HamnerCEH, Hoefer:2011} if the size of the condensates is sufficiently large in the direction perpendicular to the relative velocity.
 We hope that these phenomena will be observed in future experiments.


\subsection{Kelvin--Helmholtz instability in immiscible two-component BECs}
 In classical fluids,
 the Kelvin--Helmholtz instability can occur when there is a sufficient velocity difference across the interface between two fluids with different mass densities \cite{Chandrasekhar:1981, Kundu:2008}.
 A vortex sheet exists along the interface due to the velocity difference and
the instability induces exponential amplification of the oscillating modes of the vortex sheet.
 The instability typically develops into roll-up patterns of the interface in the nonlinear stage.
 The Kelvin--Helmholtz instability is one of the most fundamental hydrodynamic instabilities in fluid dynamics, related to several familiar phenomena such as wind-generated ocean waves, flapping flags, billow clouds, and sand dunes.
 In this subsection,  we discuss hydrodynamic instability in phase-separated condensates,
 where the inter-component interaction parameter satisfies the immiscible condition $g_{12}>\sqrt{g_{11}g_{22}}$.
 We shall show that,
 in the presence of the relative velocity between the phase-separated condensates,
 the instability is related to the Kelvin--Helmholtz instability in classical fluid dynamics.

\subsubsection{Linear stability of a flat interface}
\begin{figure} [htpb] \centering
  \includegraphics[width=.6 \linewidth]{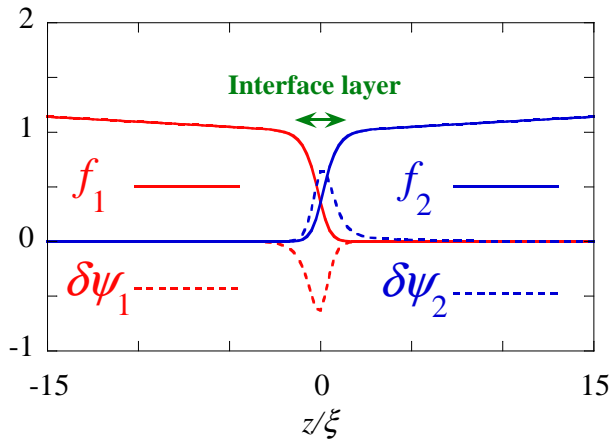}
  \caption{Profiles of the order parameter amplitudes $f_j=\sqrt{n_j}$ of phase separated two-component BECs under external potentials $U_j(z)=G_j z$ for the $j$th component.
 The real part of the excitation function $\delta\psi_j=u_j(z)-v_j(z)^*$ of a typical interface mode is plotted with broken lines.
 The parameters are set as $\mu=\mu_j-\frac{m_j}{2}v_j^2$, $m=m_j$, $g=g_{jj}$, $g_{12}=10 g$, and $G_j=-(-1)^jG/2$ with $G=0.02 \mu/\xi$.
 The amplitude and the length are scaled by $\mu/g$ and $\xi=\hbar/\sqrt{m\mu}$, respectively. [Takeuchi, Suzuki, Kasamatsu, Saito and Tsubota: Phys. Rev. B {\bf 81} (2010) 094517, reproduced with permission. Copyright 2010 the American Physical Society.]
}
\label{fig:profile}
\end{figure}
 We consider a flat interface in phase-separated condensates.
 The interface is located at $z=0$, where $f_1 > f_2$ for $z<0$ and $f_2 < f_1$ for $z>0$ as shown in Fig. \ref{fig:profile}.
 The position of the interface is stabilized at  $z=0$ under the small potential gradient $\partial_z U_j=G_j={\rm const.}$ with $G_1>0$ and $G_2<0$.
 The stationary solution has a form $\Phi_j({\Vec r})=f_j(z)e^{im_j{\Vec V}_j\cdot{\Vec r}/\hbar}$, where the superfluid velocity ${\Vec V}_j$ is parallel to the interface.
 The interface layer can be defined as the region sandwiched between the regions $n_1\sim n_1^T$ and $n_2\sim n_2^T$,
 where $n_j^T(z)=(\mu_j-U_j(z)-\frac{m_j}{2}\bar{\Vec v}_j^2)/g_{jj}$ is the bulk density obtained by neglecting the quantum pressure and the density of the different components $n_k=0~(k\neq j)$ in the Bernoulli equations (\ref{eq:Bernoulli}).
 The thickness of the layer decreases when the inter-component interaction becomes large.
 For simplicity, we consider strong phase separation with sufficiently large $g_{12}$,
 where the thickness is minimized to the order of $\xi_j\sim \hbar/\sqrt{m_jg_{jj}n_j^T(0)}$, the healing length of a single condensate.

The linear stability of the stationary states $\Phi_j$ is investigated by linearizing the GP equations with respect to a collective excitation $\delta\Psi_j({\Vec r},t)=\Psi_j({\Vec r},t)-\Phi_j({\Vec r})$.
 An oscillating perturbation of frequency $\omega$ is conventionally described with the Bogoliubov formalism 
 \begin{eqnarray}
 \delta \Psi_j=e^{im_j{\Vec V}_j\cdot{\Vec r}/\hbar}\left\{u_j(z)e^{i{\Vec q}\cdot{\Vec r}-i\omega t}-[v_j(z)e^{i{\Vec q}\cdot{\Vec r}-i\omega t}]^*\right\},
 \end {eqnarray}
 where the wave number ${\Vec q}$ is parallel to the interface.
 The functions $u_j$ and $v_j$ obey the reduced Bogoliubov--de Gennes (BdG) equations,
\begin{eqnarray}
&&
\left[\frac{\hbar^2}{2m_j}\left({\Vec q}+\frac{m_j}{\hbar}{\Vec V}_j\right)^2-\frac{\hbar^2}{2m_j}\frac{d^2}{dz^2}+U_j-\mu_j\right]u_j
\nonumber\\
&& \ \ \ \ \ \ \ \ \ +\sum_k g_{jk}\left(f_k^2u_j+f_jf_ku_k-f_jf_kv_k \right)=\hbar\omega u_j
,
\label{eq:BdG2cmp1}
\\
&&
-\left[\frac{\hbar^2}{2m_j}\left({\Vec q}-\frac{m_j}{\hbar}{\Vec V}_j\right)^2-\frac{\hbar^2}{2m_j}\frac{d^2}{dz^2}+U_j-\mu_j\right]v_j
\nonumber\\
&& \ \ \ \ \ \ \ \ \ -\sum_k g_{jk}\left(f_k^2v_j+f_jf_kv_k-f_jf_ku_k \right)=\hbar\omega v_j.
\label{eq:BdG2cmp2}
\end{eqnarray}
 These equations determine the linear stability of the stationary states.

 Excitations in the phase-separated states are generally classified into two types.
 One is bulk modes such as phonons, which can propagate in the bulk far from the interface.
 The other is localized modes, which disturb the order parameters only locally around the interface and decay exponentially in the bulk (see Fig. \ref{fig:profile}).
 The simplest example of localized modes is a transverse shift of the interface in the $z$ direction.
 In a manner similar to the linear stability analysis of the Kelvin--Helmholtz instability in hydrodynamics,  
 we consider only oscillations of such interface modes, neglecting the bulk modes and the internal structure of the interface.

If the thickness of the interface is neglected and the interface position is represented by the single-valued function $z=\eta(x,y,t)$,
 the interface modes may be approximately described by the effective Lagrangian
\begin{eqnarray}
L_{\rm eff}=\int dxdy \left[ \int^{\eta}_{-\infty}dz {\cal P}_1+\int_{\eta}^{\infty}dz {\cal P}_2-\alpha {\cal S}\right],
\label{eq:Leff}
\end{eqnarray}
where we used
\begin{eqnarray}
&&
{\cal P}_j=
-\hbar  n_j \partial_t \theta_j
-\frac{m_jn_j}{2}{\Vec v}_j^2
+\frac{\hbar^2}{2m_j}f_j{\Vec \nabla}^2f_j
-(U_j-\mu_j)n_j
-\frac{1}{2}g_{jj}n_j^2
\\
&&{\cal S}=\sqrt{1+\left(\partial_x\eta\right)^2+\left(\partial_y\eta\right)^2}
\end{eqnarray}
 and the interface tension coefficient $\alpha$ was introduced.
 Let us consider a flat interface $\eta=0$ in a stationary state in a homogeneous system along the $x$ and $y$ axes.
 For a small perturbation, we obtain the equation of motion for the interface position $\eta$,
\begin{eqnarray}
{\cal P}_1(\eta)-{\cal P}_2(\eta)+\alpha{\Vec \nabla}^2\eta=0.
\label{eq:KHI_Bernoulli}
\end{eqnarray}
 This equation is an analogue of the Bernoulli theorem on the interface.
 In the stationary state $\eta=0$,
 the term ${\cal P}_1$ is reduced to the hydrostatic pressure $p_j^h$ (\ref{eq:p_hyd}) of the $j$th component at the interface; we then obtain $p_1^h=p_2^h$.

 We can employ the kinematic boundary condition on the interface, similar to the discussion for hydrodynamics,
\begin{eqnarray}
(\delta {\Vec v}_{j})_z=\partial_t\eta+{\Vec v}_{j}\cdot{\Vec \nabla} \eta.
\label{eq:bound1}
\end{eqnarray}
Based on the assumption that the density perturbation $\delta n_j$ at $z\sim0$ is caused by the local transverse shift of the density profile,
 we can write the density perturbation along the interface as 
\begin{eqnarray}
\delta n_j=-\eta\partial_z n_j.
\label{eq:bound2}
\end{eqnarray}

 To obtain the dispersion of the interface modes, we assume the localized perturbations with a form $\eta \propto \sin({\Vec q}\cdot{\Vec r}-\omega t)$, $\delta n_j \propto e^{-(-1)^ja_j z}\sin({\Vec q}\cdot{\Vec r}-\omega t)$ and $\delta \theta_j \propto e^{-(-1)^ja_j z}\cos({\Vec q}\cdot{\Vec r}-\omega t)$.
 We obtain $a_j=q=|{\Vec q}|$ in the approximation neglecting the quantum pressure term, $\partial_z \bar{n}_j\approx\partial_z n_j^T=-(-1)^jG/2g_{jj}$, by linearizing Eqs. (\ref{eq:linear_continuity}), (\ref{eq:bound1}), and (\ref{eq:bound2}) with respect to the perturbations.
 It is straightforward to derive the dispersion relation for the interface mode,
\begin{eqnarray}
\omega({\Vec q})={\Vec q}\cdot{\Vec v}_G\pm\frac{1}{\sqrt{\rho_1+\rho_2}}\sqrt{q(F+\alpha q^2)-\frac{\rho_1\rho_2}{\rho_1+\rho_2}({\Vec v}_R\cdot{\Vec q})^2},
\label{eq:dispersion}
\end{eqnarray}
 where $\rho_j = m_jn_j^T(0)$, $F=G_1n_1^T(0)-G_2n_2^T(0)$, ${\Vec v}_R={\Vec v}_2-{\Vec v}_1$, and ${\Vec v}_G=\frac{\rho_1\bar{\Vec v}_1+\rho_2\bar{\Vec v}_2}{\rho_1+\rho_2}$.


 Dynamic instability occurs when the imaginary part ${\rm Im}~\omega$ becomes nonzero for $\frac{F+\alpha q^2}{q}<\frac{\rho_1\rho_2}{\rho_1+\rho_2}v_R^2$ with $v_R=|{\Vec v}_R|$.
 This instability is the counterpart of the Kelvin--Helmholtz instability in classical fluid dynamics.
 The critical relative velocity $V_{D}$ for the dynamic instability is given by
\begin{eqnarray*}
V_{\rm D}=\sqrt{2\frac{\rho_1+\rho_2}{\rho_1\rho_2}\sqrt{F\alpha}}.
\end{eqnarray*}
 Note that the interface is hydrostatically unstable without the relative velocity for $F < 0$.
 This is an analog of the Rayleigh--Taylor instability,
 where a heavier fluid is located above a lighter fluid and the interface between the two fluids is unbalanced under gravity.

 The Landau instability is evaluated from Eq. (\ref{eq:dispersion}).
 We obtain the Landau critical velocity $V_L$ for the velocity $v_G=|{\Vec V}_G|$ from the condition $\omega=0$,
\begin{eqnarray}
V_{L}=\frac{1}{\sqrt{\rho_1+\rho_2}}\sqrt{2\sqrt{F\alpha}-\frac{\rho_1\rho_2}{\rho_1+\rho_2}v_R^2}.
\label{eq:VL}
\end{eqnarray}
 The Landau critical velocity $V_L$ depends on the relative velocity $v_R$.
 When $v_R>V_D$, the system becomes thermodynamically unstable for an arbitrarily small velocity $v_G$.
 On the other hand, even when $v_G>V_L$, the system can be still dynamically stable with ${\rm Im}~\omega=0$.
 Therefore, the Landau instability can occur in general before the onset of dynamic instability in a dissipative system.
 The Landau instability has been experimentally observed at the interface between the A and B phases of $^3$He \cite{Helsinki, Volovik:2002}.

\begin{figure} [htpb] \centering
  \includegraphics[width=.5 \linewidth]{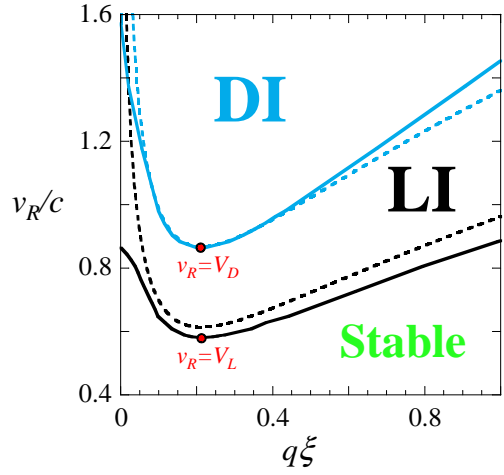}
  \caption{
 Phase diagram of the dynamic instability (DI) and the Landau instability (DI) for $V_1=0$ and $V_2<0$.
 The curves are the boundaries of the DI and LI regions obtained from numerical calculation of the BdG equations (solid curves) and from the dispersion of Eq. (\ref{eq:dispersion}) (broken curves).
 The parameters are the same as those in Fig. \ref{fig:profile}.
 The interface tension is calculated as $\alpha=0.886 \mu^2 \xi/g$ according to Ref. \cite{2008SchaeybroeckPRA}.
 The relative velocity $v_R$ and the wave number $q$ are scaled by $c=\sqrt{g\rho/m^2}$ and $\xi=\hbar/\sqrt{g\rho}$ with $\rho=\rho_j=m\mu/g$.[Takeuchi, Suzuki, Kasamatsu, Saito and Tsubota: Phys. Rev. B {\bf 81} (2010) 094517, reproduced with permission. Copyright 2010 the American Physical Society.]
}
\label{fig:phase_diagram}
\end{figure}

 Figure \ref{fig:phase_diagram} shows the phase diagram of the dynamic instability (DI) and the Landau instability (LI),
 obtained by the dispersion (\ref{eq:dispersion}).
 Here, we considered ${\Vec q}\parallel {\Vec v}_R$ and $V_1=0$ with the parameters described in the caption of Fig. \ref{fig:profile}.
 The results are compared with those obtained by the direct numerical computations of the BdG equations (\ref{eq:BdG2cmp1}) and (\ref{eq:BdG2cmp2}).
 The analytic results are in good agreement with the numerical results.
 Note that the approximation demonstrated here is not applied to the perturbation with $q \gtrsim 1/\xi$ since the interface thickness $\sim \xi$ is neglected in this model.
The difference for small $q$ comes from the inadequate treatment of the density and phase perturbation when the penetration depth $a^{-1}=q^{-1}$ of the interface modes becomes comparable to the system size.

\subsubsection{Nonlinear development of dynamic Kelvin--Helmholtz instability}
 The nonlinear time development of the dynamic and Landau instabilities is investigated numerically.
 We shall demonstrate typical developments of instabilities in quasi-two-dimensional systems neglecting the $y$ coordinate.

\begin{figure} [hbtp] \centering
  \includegraphics[width=.99 \linewidth]{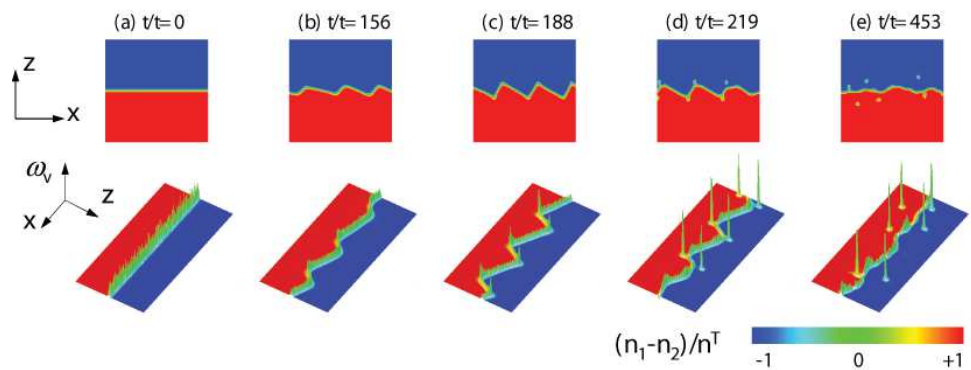}
  \caption{
 Time development of the dynamic Kelvin--Helmholtz instability for $v_R=0.98c>V_D$.
 The density difference is scaled by $n^T=\mu/g$.
 The height in the lower figures represents the vorticity $\omega_v$.
 The numerical simulation was done under the periodic boundary condition along the $x$ axis.
 The system size is $64 \xi\times 64 \xi$ with $\xi=\hbar/\sqrt{g\rho}$.
 The time is scaled by $\tau=\hbar/\mu$.[Takeuchi, Suzuki, Kasamatsu, Saito and Tsubota: Phys. Rev. B {\bf 81} (2010) 094517, reproduced with permission. Copyright 2010 the American Physical Society.]
}
\label{fig:dynamics_N}
\end{figure}

 We first discuss the nonlinear time development of the dynamic instability (dynamic Kelvin--Helmholtz instability) for $v_R>V_D$.
 The nonlinear dynamics of the Kelvin--Helmholtz instability are obtained by numerically solving the GP equations (\ref{eq:GP2cmp}) with a quasi-two-dimensional system periodic along the relative velocity.
 Figure \ref{fig:dynamics_N} shows a typical time development of the density difference $n_1-n_2$ in the dynamic Kelvin--Helmholtz instability.
 The vorticity $\omega_v$ and the mass current velocity ${\Vec v}$,
\begin{eqnarray}
\omega_v\equiv {\rm rot}~{\Vec v},~~~~{\Vec v}=\frac{{\Vec j}_1+{\Vec j}_2}{m_1 n_1+m_2 n_2},
\end{eqnarray}
 are useful for understanding the phenomena.
 When there is a velocity difference between components across the interface,
 the vorticity $\omega_v$ is distributed along the interface.
 A quantized vortex in the bulk far from the interface has a singular peak at its core in the vorticity distribution.

 In the linear stage of the instability,
 a random seed, added in the initial stationary state,
 grows into a sinusoidal interface wave.
 The wave number of the sinusoidal wave corresponds to that of the unstable mode with the largest imaginary part $|{\rm Im}~\omega|$.
 As the amplitude of the wave becomes large,
 the sine wave is distorted [Fig.~\ref{fig:dynamics_N}~(b)],
 and deforms into a sawtooth wave [Fig.~\ref{fig:dynamics_N}~(c)].
 Then, quantized vortices are released from the edges of the sawtooth waves.
 The vorticity $\omega_v$ is localized on the edges of the sawtooth waves and creates singular peaks [Fig.~\ref{fig:dynamics_N}~(d)].
 These peaks are released from the vortex sheet, becoming a singly quantized vortex with a circulation of $\kappa=h/m$ [Fig.~\ref{fig:dynamics_N}~(e)].
 The vorticity of the vortex sheet on the interface is reduced after the release of vortices,
 and then the velocity difference across the interface decreases locally.
 The relative velocity across the interface after the vortex nucleation is roughly estimated to be the total vorticity on the interface divided by the length of the sheet.
 Since six quantized vortices are released from the interface [Fig.~\ref{fig:dynamics_N}~(f)],
 the relative velocity decreases by about $6\kappa/L\sim 0.6 c$ with the system size $L=64\xi$ below the threshold $V_D$.
 Then the instability stops,  and vortex nucleation occurs no more.
 The released vortices continue to drift along the interface and the system does not recover the initial flat interface.

 Since the Kelvin--Helmholtz instability occurs locally around the interface,
 the relative motion of the two components does not ultimately vanish, in contrast to the countersuperflow instability demonstrated in the previous subsection.
 The instability can occur globally if the interface thickness becomes comparable to the system size for a small inter-component interaction with $g_{12}\sim g$.
 The instability phenomenon then becomes similar to that of countersuperflow instability.
 The crossover between counterflow instability and Kelvin--Helmholtz instability is investigated in Ref. \cite{2010SuzukiPRA}.

\subsubsection{Nonlinear development of thermodynamic Kelvin--Helmholtz instability}
 We next discuss the nonlinear time development of the Landau instability (thermodynamic Kelvin--Helmholtz instability) for $v_R>V_L$.
 The dissipative dynamics can be qualitatively investigated by solving the dissipative GP equations,
 which are obtained by replacing the time-derivative term $i \partial_t$ by $(i-\gamma)\partial_t$ in the GP equations (\ref{eq:GP2cmp}) \cite{2003KasamatsuPRA}.

\begin{figure} [hbtp] \centering
  \includegraphics[width=.99 \linewidth]{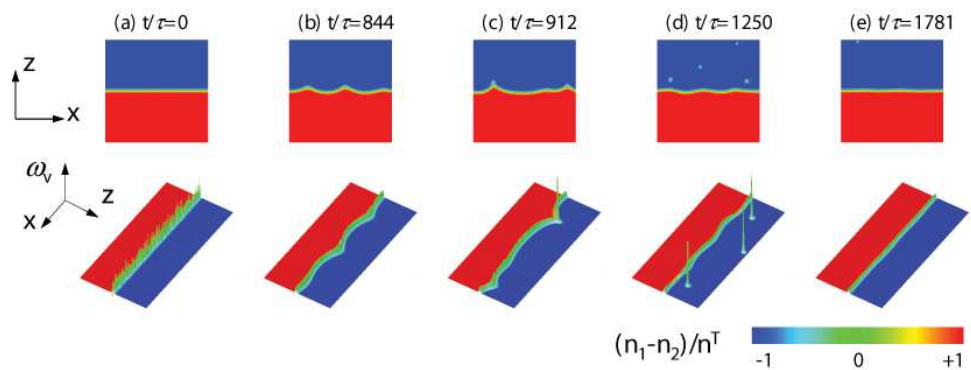}
  \caption{
 Time development of the thermodynamic Kelvin--Helmholtz instability for $v_R=0.79c>V_L$.
 We set the dissipation coefficient to be $\gamma=0.03$.
}
\label{fig:dynamics_D}
\end{figure}

 Figure \ref{fig:dynamics_D} shows the time development of the instability obtained by solving the dissipative GP equations. 
 In the nonlinear stage,  the interface has flattened troughs and peaked crests [Fig. \ref{fig:dynamics_D}(b)].
 The patterns are diphycercally asymmetric in contrast with the patterns of the dynamic Kelvin--Helmholtz instability.
 In this case, the vorticity is localized at the crests [Fig. \ref{fig:dynamics_D}(c)] and a single quantized vortex is nucleated from each crest only into the upper side [Fig. \ref{fig:dynamics_D}(d)].

 The nucleated vortices are dragged away from the interface to the upper direction due to the dissipation,
 which decreases the flow velocity of the 2nd component (the blue region in Fig. \ref{fig:dynamics_D}) at the rest frame.
 The event of the vortex dragging is interpreted as phase slippage \cite{1985AvenelPRL}.
 The vortex nucleation stops after four vortices are nucleated,
 where the relative velocity across the interface decreases by $4\kappa/L\sim 0.4 c$ below the threshold $V_L$ of the Landau instability.
 Then the interface recovers a flat form with less vorticity (relative velocity) than the initial state.

\subsubsection{Conclusion}
 The interface modes are amplified due to dynamic instability when the relative velocity across the interface exceeds a critical value in phase-separated two-component BECs.
 The instability is interpreted as the quantum counterpart of the Kelvin--Helmholtz instability when the interface motion is represented using the hydrodynamic formalism.
 The nonlinear dynamics of the quantum Kelvin--Helmholtz instability are quite different from those of classical Kelvin--Helmholtz instability governed by quantized vortices.
 The Landau instability for the interface modes in the presence of relative velocity,
 called the thermodynamic Kelvin--Helmholtz instability, causes vortex nucleation and phase slippage from the interface, which has no analog in classical fluid dynamics.
 The Kelvin--Helmholtz instability in trapped systems and the possibility of its experimental realization was discussed in detail in Ref. \cite{2010SuzukiPRA}.
 We believe that the first observation of the dynamic Kelvin--Helmholtz instability, the quantum counterpart of the classical Kelvin--Helmholtz instability, will soon be realized in future experiments.

\section{Conclusions}
We have reviewed recent topics on quantum hydrodynamics (QHD) in superfluid helium and atomic BECs, chiefly focusing on the activity of our group.
Quantized vortices were discovered in superfluid $^4$He in the 1950s. However, they have recently grown in importance, for two reasons. 
The first reason is that the research of QT entered a new era since the mid 1990s leaving the previous studies almost limited to thermal counterflow. 
The second reason is the realization of atomic BECs in 1995. Modern optical techniques have enabled the direct visualization of quantized vortices, and multi-component BECs have further enriched the world of quantized vortices.
In this concluding section, we describe the main motivation of this research and the interesting topics that are not addressed in the text.
The discussions in this section are limited to the case at zero temperature where the normal fluid component is negligible. 

Comparing QT and CT reminds us of the motivation of studying QHD. 
Turbulence in a classical viscous fluid appears to be comprised of eddies. 
However, these eddies are unstable and not well defined. 
The circulation is not conserved and is not identical for each eddy. 
QT consists of a tangle of quantized vortices that have the same conserved circulation.  
Looking back at the history of science, {\it reductionism}, which tries to understand the nature of complex things by reducing them to the interactions of their parts, has played an extremely important role. 
The success of solid-state physics owes much to  {\it reductionism}. 
In contrast, conventional fluid physics is not reducible to elements, and thus does not enjoy the benefits of {\it reductionism}. 
However, QT is different, being reduced to quantized vortices; {\it reductionism} is applicable to quantum turbulence.  
The main interests would be quantum hydrodynamic instability and QT beyond the instability. 
How can we approach these problems from  {\it reductionism}?

We should reveal the transition to QT and the nature of QT. 
Statistical quantities are useful in order to investigate the transition and the nature of QT.
Here we list the possible statistical quantities.
\begin{enumerate}
\item Energy spectra. The energy spectrum of fully developed QT is expected to obey the Kolmogorov law.
 However, the present understanding of the story is not so simple \cite{Vinen:2010}. 
It is generally believed that there are two kinds of QT, namely quasi-classical turbulence and ultra-quantum turbulence. 
In quasi-classical turbulence, most of the turbulent energy is concentrated in the large-scaled eddies, typically on scales larger than the mean inter-vortex distance $\ell$. 
This is quite similar to the case of classical homogeneous turbulence. 
The discreteness of each quantized vortex is not so relevant; they are expected to make some coherent structure (vortex bundles). 
Then the energy spectra would follow the Kolmogorov law.
If we switch off the energy injection sustaining the turbulence, the vortex line density (VLD) decays as $L \propto t^{-3/2}$.
On the other hand, ultra-quantum turbulence has no quasi-classical motion on scales greater than $\ell$.
Most of energy is concentrated on scales smaller than $\ell$.
When it decays, the VLD reduces as $L \propto t^{-1}$.
There are no direct observations of the energy spectra at such low temperatures.
However, the two types of decay $L \propto t^{-3/2}$ and $t^{-1}$ are observed experimentally \cite{Bradley:2006,Walmsley:2007,Walmsley:2008}.
The essential point is how to connect the configuration of quantized vortices with energy spectra.
If a configuration of vortices is given, the energy spectrum is determined uniquely.
Then, what types of vortex configuration can make quasi-classical or ultra-quantum turbulence? 

\item PDF(Probability density function) of superfluid velocity $\mathbf{v}_s$. This was discussed in Sec. \ref{velocity statistics}. The PDF shows classical Gaussian in low velocity and non-classical power-law in high velocity. However, there are only a few theoretical works on this problem \cite{White:2010, Adachi:2011}, no systematic studies in connection with energy spectra and the configuration of vortices.

\item Vortex length distribution.  If self-similar Richardson cascade appears, it is expected to yield some self-similar power-law in the vortex length distribution. There are still a few numerical works \cite{Araki:2002, Kobayashi:2007,Mitani:2006}. It is impossible to observe the distribution in superfluid helium, but possible in atomic BECs.

\item Drag coefficient. This quantity was discussed in Sec. \ref{vibrating structure}.  
The drag coefficient $C_D$ is inversely proportional to the velocity in laminar flow and of order unity  in turbulent flow. This is a well-known story in CT, and confirmed in QT too \cite{Skrbek:2009}. This change in $C_D$ can be another sign of transition to QT.

\end{enumerate}

 It is possible and important to consider QT as a transient state in the relaxation process far from thermal equilibrium \cite{Nowak:2011,Nowak:2012} .  QT corresponds to nonthermal fixed points in a nonperturbative quantum-field theoretic approach, following some scaling law characteristic of dynamical critical phenomena. This kind of approach should make progress in near future.  

Section \ref{two-component} described chiefly hydrodynamics and QT in two-component BECs.
A spinor BEC is another important system for hydrodynamics \cite{Kurnreview}.
Fujimoto and Tsubota investigated theoretically and numerically the GP model of spin-1 spinor BECs.
They considered the two cases: one is the counterflow of two components with different magnetic quantum numbers in a uniform system\cite{Fujimoto:2012a} and the other is starting from a helical spin structure in a trapped system \cite{Fujimoto:2012b} .
When the interaction is ferromagnetic, the instability is amplified to spin turbulence in both cases, where the spectrum of the spin-dependent interaction energy exhibits a -7/3 power law, different from the Kolmogorov -5/3 law.
This power law is understood from some scaling argument for the equation of motion of the spin density vector. 
Since such spin density vector can be observed \cite{Vengalattore:08}, such spin turbulence could be realized and observed.

\bibliography{bib-turbulence.bib}

\end{document}